\documentclass{aa} 
\usepackage{natbib}
\usepackage{graphicx,color}
\usepackage{txfonts}
\usepackage{url}
\usepackage[colorlinks=true, citecolor=blue]{hyperref}
\usepackage[usenames,dvipsnames,svgnames,table]{xcolor}

\newcommand{\orcid}[1]{\href{https://orcid.org/#1}{\includegraphics[width=10pt]{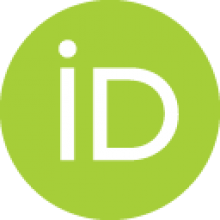}}}

\begin{document} 

\title{Galactic ArchaeoLogIcaL ExcavatiOns (GALILEO) II.\\ t-SNE Portrait of Local Fossil Relics and Structures} 

\author{
	Mario Ortigoza-Urdaneta\inst{1}\thanks{mario.ortigoza@postgrados.uda.cl}
        \and
        Katherine Vieira\inst{1}\orcid{0000-0001-5598-8720}
        \and
        Jos\'e G. Fern\'andez-Trincado\inst{2}\orcid{0000-0003-3526-5052}\thanks{jose.fernandez@ucn.cl}
        \and
        Anna. B. A. Queiroz\inst{3,4,5}
        \and
        Beatriz Barbuy\inst{6}
        \and
        Timothy C. Beers\inst{7}
        \and
        Cristina Chiappini\inst{3,5}
        \and 
        Friedrich Anders\inst{8,9,10}
        \and
        Dante Minniti\inst{11,12}\orcid{0000-0002-7064-099X}
        \and 
        Baitian Tang\inst{13}
	   }
	
\authorrunning{Mario Ortigoza-Urdaneta et al.} 
	
\institute{
	Instituto de Astronom\'ia y Ciencias Planetarias, Universidad de Atacama, Copayapu 485, Copiap\'o, Chile
		\and 
		Instituto de Astronom\'ia, Universidad Cat\'olica del Norte, Av. Angamos 0610, Antofagasta, Chile
        \and 
        Leibniz-Institut f$\ddot{u}$r Astrophysik Potsdam (AIP), An der Sternwarte 16, 14482 Potsdam, Germany
        \and
        Institut f\"{u}r Physik und Astronomie, Universit\"{a}t Potsdam, Haus 28 Karl-Liebknecht-Str. 24/25, 14476 Golm, Germany
        \and 
        Laborat\'orio Interinstitucional de e-Astronomia - LIneA, RJ 20921-400, Rio de Janeiro, Brasil
        \and
        Universidade de S\~ao Paulo, IAG, Rua do Mat\~ao 1226, Cidade Universit\'aria, S\~ao Paulo 05508-900, Brazil
        \and
		Department of Physics and Astronomy and JINA Center for the Evolution of the Elements, University of Notre Dame, Notre Dame, IN 46556, USA
        \and
        Departament de F\'isica Qu\'antica i Astrof\'isica (FQA), Universitat de Barcelona (UB),  c. Mart\'i i Franqu\`es, 1, 08028 Barcelona, Spain
        \and
        Institut de Ci\`encies del Cosmos (ICCUB), Universitat de Barcelona (UB), c. Mart\'i i Franqu\`es, 1, 08028 Barcelona, Spain
        \and 
        Institut d'Estudis Espacials de Catalunya (IEEC), c. Gran Capit\`a, 2-4, 08034 Barcelona, Spain
        \and 
     	Depto. de Cs. F\'isicas, Facultad de Ciencias Exactas, Universidad Andr\'es Bello, Av. Fern\'andez Concha 700, Las Condes, Santiago, Chile
	   \and
    	Vatican Observatory, V00120 Vatican City State, Italy
        \and 
        School of Physics and Astronomy, Sun Yat-sen University, Zhuhai 519082, China 
        \and 
        Centro de Investigaci\'on en Astronom\'ia, Universidad Bernardo O'Higgins, Avenida Viel 1497, Santiago, Chile
    }
	
\date{Received ...; Accepted ...}
\titlerunning{Local hotter structures using t-SNE, APOGEE and Gaia}
	
	
	\abstract
	{Based on high-quality APOGEE DR17 and Gaia DR3 data for 1,742 red giants stars within 5 kpc of the Sun and not rotating with the Galactic disc ($V_\phi <$ 100 km s$^{-1}$), we use the nonlinear technique of unsupervised analysis t-SNE to detect coherent structures in the space of ten chemical-abundance ratios: [Fe/H], [O/Fe], [Mg/Fe], [Si/Fe], [Ca/Fe], [C/Fe], [N/Fe], [Al/Fe], [Mn/Fe], and [Ni/Fe]. Additionally, we obtain orbital parameters for each star using the non-axisymmetric gravitational potential {\tt GravPot16}. Seven structures are detected, including the Splash, Gaia-Sausage-Enceladus (GSE), the high-$\alpha$ heated-disc population, N-C-O peculiar stars, and inner disk-like stars, plus two other groups that did not match anything previously reported in the literature, here named Galileo 5 and Galileo 6 (G5 and G6). These two groups overlap with Splash in [Fe/H], G5 being lower metallicity than G6, both between GSE and Splash in the [Mg/Mn] versus [Al/Fe] plane, G5 in the $\alpha$-rich in-situ locus, and G6 on the border of the $\alpha$-poor in-situ one; nonetheless their low [Ni/Fe] hints to a possible ex-situ origin. Their orbital energy distributions are between the Splash and GSE, with G5 being slightly more energetic than G6. We verified the robustness of all the obtained groups by exploring a large range of t-SNE parameters, applying it to various subsets of data, and also measuring the effect of abundance errors through Monte Carlo tests.}
	
	\maketitle
	
\section{Introduction}\label{sec_intro}

Numerous studies have revealed that the various structures that make up the Milky Way (MW) have been affected by both intrinsic (e.g., secular evolution, \citealt{2016A&A...595A..63V,2017sf2a.conf..223C}) and extrinsic (e.g., merger events, \citealt{Kruijssen2020}) processes. As a consequence, our galaxy has mixed populations comprising both in-situ and ex-situ star formation. The identification of these structures is fundamental to understanding the formation and evolutionary history of our galaxy.

In some cases, these structures still retain traces of their original kinematics, and can be identified by their peculiar orbits as compared to canonical populations of the MW \citep{Belokurov2018,2019A&A...631L...9K}, but in many other cases, their original motion is erased by their prolonged interaction with the MW. On the other hand, most chemical abundances in stellar atmospheres are unaffected by the dynamical history of stars. Thus, a chemical signature becomes a valuable piece of information to determine the presence and origin of a given population. Taken together, kinematics, dynamics, and chemical abundances are extremely useful to decode the history of such population \citep[e.g.,][]{Freeman2002, 2021ApJ...923..172H, 2023MNRAS.520.5671H}.

Thanks to the data provided by Gaia \citep{Brown2021} and spectroscopic surveys \citep[e.g., APOGEE, GALAH][]{Majewski2017, Buder2022}, and their unprecedented precision, great advances have been made in the detection and understanding of structures such as Gaia-Sausage Enceladus (GSE) \citep{Belokurov2018, 2018ApJ...863..113H, 2018Natur.563...85H, 2018ApJ...860L..11K,2018ApJ...863L..28M} or the Helmi stream(s) \citep{Helmi1999Natur, 2000AJ....119.2843C, 2019A&A...625A...5K}. In both cases, an ex-situ origin is attributed, coming from past mergers with the MW. On the other hand, we have the ``Splash'', a population thought to have likely arisen from the heating of the primordial disc by an early merger \citep{2017ApJ...845..101B, 2020ApJ...897L..18B, 2018ApJ...863..113H, 2019A&A...632A...4D, 2020MNRAS.494.3880B}. The Splash has also been interpreted as an outcome of clumpy star formation of the early disc \citep{2020ApJ...891L..30A, 2021MNRAS.503.1418F}.

While in the past century astronomical data was often scarce, particularly regarding chemical abundances, parallaxes, and proper motions, nowadays the situation is the opposite. This large influx of data allows for a more detailed analysis of known structures and the discovery of new populations, although the increased density of data points hampers a clean separation of the populations, which usually overlap in the many different measurements available. More data - traditionally analyzed - means it is more difficult to separate the (probable) members individually. 

The tsunami of astronomical data has driven the application of statistical techniques that allow for automatic searches to identify potential associations of stars by specific properties. It is necessary to understand how these methods work and to what extent they can help us analyze the data, as there are differences in data quality due to both instrumental and natural factors. Among the widely used techniques is Principal Component analysis (PCA), which has been successfully applied in many kinds of data with excellent results \citep{rebonato2011most,2012MNRAS.421.1231T}.

In this paper, we present the use of t-SNE (t-Distributed Stochastic Stochastic Neighbor Embedding  \citep{tSNE2003, ref_tsne} dimensionality reduction technique for a study of local low-velocity structures ($V_\phi<100$ km s$^{-1}$, less than 5 kpc from the Sun), from input data provided by APOGEE+Gaia. t-SNE has been used in several areas of astronomy, such as pre-main sequence identification 
\citep[e.g.,][]{Rim2022}, spectral classification \citep[e.g.,][]{Verma2021}, and 
abundance-space dissection \citep[e.g.,][]{Anders2018}, where we note the powerful applicability for automated searches of populations. We provide an analysis based on t-SNE detections obtained from ten chemical abundances of our sample and the dynamics provided by GravPot16\footnote{\url{https://gravpot.utinam.cnrs.fr}}.

In the last few years, several sub-populations have been identified in the MW halo, e.g., GSE, Splash,
Sequoia \citep{Barba2019}, Kraken \citep{Kruijssen2020}, and others, due to their peculiar kinematics, dynamics, and/or chemical abundances. It has been proposed by \citet{Naidu2020} and other authors that the MW halo is built entirely from accreted dwarfs and heating of the disc. Those substructures have been generally identified at rather large vertical distances from the MW plane, but in this study, we look for evidence of their presence closer to the Sun where they would co-locate with the disc system. 

This paper is organized as follows: data sources and selection are described in sections 2, 3, and 4, the MW dynamical model in section 5, the t-SNE method in sections 6 and 7, and our results in Section 8. Section 9 discusses the accretion origin of some of the structures detected, section 10 is dedicated to a quick analysis of the effects of the bar pattern speed in the obtained structures, and our conclusions are in Section 11. Additional figures and tables are located in the Appendix section.

\section{APOGEE-2 DR~17}\label{sec_apogee}
	
The data employed in this study were obtained as part of the second phase of the Apache Point Observatory Galactic Evolution Experiment \citep[APOGEE-2;][]{Majewski2017}, which was one of the four Sloan Digital Sky Survey IV surveys \citep[SDSS-IV;][]{Blanton2017}. APOGEE-2 was a high-resolution ($R\sim22,500$) near-infrared (NIR) spectroscopic survey containing observations of $657,135$ stars, whose spectra were obtained using the cryogenic, multi-fiber (300 fibers) APOGEE spectrograph \citep{Wilson2019} mounted on the 2.5m SDSS telescope \citep{Gunn2006} at Apache Point Observatory to observe the Northern Hemisphere (APOGEE-2N), and expanded to include a second APOGEE spectrograph on the 2.5m Ir\'en\'ee du Pont telescope \citep{Bowen1973} at Las Campanas Observatory to observe the Southern Hemisphere (APOGEE-2S). Each instrument records most of the \textit{H}-band ($1.51$ $\mu$m -- $1.70$ $\mu$m) on three detectors, with coverage gaps between $\sim$1.58 -- 1.59 $\mu$m and $\sim$1.64 -- 1.65 $\mu$m, and with each fiber subtending a $\sim$2'' diameter on-sky field of view in the northern instrument and 1.3'' in the southern.  We refer the interested reader to \citet{Zasowski2013}, \citet{Zasowski2017}, \citet{Beaton2021}, and \citet{Santana2021} for further details regarding the targeting strategy and design of the APOGEE-2 survey.

The final version of the APOGEE-2 catalog was published in December 2021 as part of the 17$^{\rm th}$ data release of the Sloan Digital Sky Survey \citep[DR17;][]{Abdurro2022} and is available publicly online through the SDSS Science Archive Server and Catalog Archive Server\footnote{SDSS DR17 data: \url{https://www.sdss.org/dr17/irspec/ spectro data/}}. The APOGEE-2 data reduction pipeline is described in \citet{Nidever2015}, while stellar parameters and chemical abundances in APOGEE-2 have been obtained within the APOGEE Stellar Parameters and Chemical Abundances Pipeline \cite[\texttt{ASPCAP};][]{Garcia2016}. \texttt{ASPCAP} derives stellar atmospheric parameters, radial velocities, and as many as 26 individual elemental abundances for each APOGEE-2 spectrum by comparing each to a multidimensional grid of theoretical \texttt{MARCS} model atmosphere grid \citep{Zamora2015}, employing a $\chi^2$ minimization routine with the code \texttt{FERRE} \citep{Allende2006} to derive the best-fit parameters for each spectrum. We used the abundances computed from the synspec\_fix spectral synthesis code.

The accuracy and precision of the atmospheric parameters and chemical abundances are extensively analyzed in \citet{Holtzman2018}, \citet{Henrik2018}, and \citep{Henrik2020}, while details regarding the customized \textit{H}-band line list are fully described in 
\citet{Shetrone2015}, \citet{Hasselquist2016}, \citet{Cunha2017}, and \citet{2021AJ....161..254S}.

\section{Input parameters}\label{sec_input}

\subsection{Elemental abundances and radial velocities}\label{subsec_abun}

We make use of high-quality elemental abundances for ten chemical species, including the light- (C, N), $\alpha$- (O, Mg, Si, Ca), odd-Z (Al), and iron-peak (Mn, Fe, Ni) elements, as well as precise ($<$1 km s$^{-1}$) APOGEE-2 spectroscopic radial velocity measurements when calculating kinematics and orbits for giant stars in the APOGEE DR~17 catalog \citep{Abdurro2022}. The other 16 abundances available from ASPCAP were not used because they are affected by telluric bands (e.g., Na), blended with other atomic/molecular lines, or have a lower signal-to-noise ratio (SNR). Also, for some elements a reduced number of stars have measurements.

\subsection{\texttt{StarHorse} distances}\label{subsec_starhorse}

We also make use of the precise spectro-photo-astrometric distances ($d_{\rm \texttt{StarHorse}}$) estimated with the Bayesian \texttt{StarHorse} code \citep[][]{Santiago2016, Queiroz2018, Anders2019, Queiroz2020}, which have been published in the form of a value-added catalog (VAC)\footnote{\url{https://data.sdss.org/datamodel/files/APOGEE_STARHORSE/}}. \texttt{StarHorse} combines high-resolution spectroscopic data from APOGEE-2 DR~17 with broad-band photometric data from several sources (\texttt{Pan-STARSS1}, \texttt{2MASS}, and \texttt{AllWISE}), as well as parallaxes from Gaia DR~3 when available, along with their associated uncertainties, in order to derive distances, extinctions, and astrophysical parameters for APOGEE-2 stars through a Bayesian isochrone-fitting procedure. These parameters are robust to changes in the Galactic priors assumed and corrections for the Gaia parallax zero-point offset. 

\subsection{Proper motions}\label{subsec_pm}

Astrometric data are taken from Gaia DR~3 \citep{Brown2021} that are also available for the APOGEE-2 sample. For this study, when calculating orbital parameters the \texttt{RUWE} (renormalized unit weight error) astrometric quality indicator was imposed to be \texttt{RUWE} $<$ 1.40, in order to have astrometrically well-behaved sources \citep[see, e.g.,][]{Lindegren2018}.

\section{Sample selection}\label{sec_sample}

The APOGEE-2 DR~17 catalog contains more than six hundred thousand entries. Several cuts were applied to refine our sample. For quality control of the stellar parameters and elemental abundances, we first cleaned the sample from sources with unreliable parameters by keeping only sources with \texttt{STARFLAG} and \texttt{ASPCAPFLAG} equal to zero \citep[see e.g.,][]{Holtzman2015, Holtzman2018}, and stars with good  spectra (SNR $>$ 70 pixel$^{-1}$) and good stellar abundances, which are flagged as \texttt{X\_FE\_FLAG}$=$0 (\texttt{X}$=$ C, N, O, Mg, Al, Si, Ca, Mn, Fe, and Ni). These cuts ensure that there are not major flagged issues, such as a low SNR, poor synthetic spectral fit, stellar parameters near grid boundaries, and potential problematic object spectra. 

We also cut in distance precision, by selecting only stars with less than 20\% error distances from $d_{\rm \texttt{StarHorse}}$. Application of these cuts yields an initial sample of 144,476 stars. Then, we make a cut to select red giants, by selecting from the initial sample only those stars with surface gravity values $\log$ \textit{g} $<$ 3.6, and stellar effective temperatures in the range T$_{\rm eff} < $ 5500K, yielding a sample of 142,305 stars. The upper effective temperature limit was imposed to remove stars at the hot edge of the main stellar-atmospheric model grids used in the APOGEE-2 analysis. The following cut is in kinematics, by choosing only stars with Galactocentric azimuthal velocity $V_\phi<100$ km s$^{-1}$, which minimizes the presence of the dominant Galactic disc population, yielding a sample of 9544 stars. 

We also checked that this sample did not have repeated Gaia DR3 source IDs, because the main APOGEE catalog does have, in a few cases, pairs of different APOGEE IDs associated with the same Gaia source ID. This is relevant in the next cut for {\tt RUWE}<1.4, for which a true Gaia match is obviously needed, since the match is done via the Gaia DR3 source ID listed in the APOGEE catalog. The cut in {\tt RUWE} yielded a sample of 9107 stars. We then cut the sample to stars located within a 5 kpc sphere around the Sun ($d_{\rm \texttt{StarHorse}} < $ 5 kpc), yielding a sample of 1,905 stars that sample a sufficiently large volume to reach the inner-halo region, but avoiding the Bulge region of influence. In this paper, we refer to inner halo as the part of the Milky Way halo dominated by debris from past events of accretion, involving massive objects such as dwarf galaxies or globular clusters \citep[e.g.,][]{Helmi2018Natur}.

Finally, we discard known globular cluster member stars as listed by \citet{Szalbolcs2020}, \citet{2021-Baumgardt-Vasiliev}, and \citet{2021MNRAS.505.5978V}. Our final sample for the analysis below contains 1742 red giant stars. It is also important to notice that we make no correction for the selection biases within the APOGEE-2 DR~17 survey. Thus, stars close to the Solar Neighborhood will be over-represented in our sample.


\section{Dynamical Model}\label{sec_dynmodel}

In order to construct a comprehensive orbital study of the local structures across of the Solar vicinity, we use a state-of-the-art orbital integration model in a non-axisymmetric gravitational potential that fits the structural and dynamical parameters based on recent knowledge of our Galaxy. 

For the computations in this work, we have employed the rotating ``boxy/peanut'' bar of the Galactic potential model called \texttt{GravPot16}, along with other composite stellar components. The considered structural parameters of our bar model, e.g., mass, present-day orientation, and pattern speeds, are within observational estimations that lie in the range of 1.1$\times$10$^{10}$ M$_{\odot}$ and 20$^{\circ}$, in line with \citet{Tang2018} and \citet{Fernandez-Trincado2020}, and 31--51 km s$^{-1}$ kpc$^{-1}$ \citep[see e.g.,][]{Bovy2019, Sanders2019} in increments of 10 km s$^{-1}$ kpc$^{-1}$, respectively. The bar scale lengths are $x_0=$1.46 kpc, $y_{0}=$ 0.49 kpc, $z_0=$0.39 kpc, and the middle region ends at the effective major semiaxis of the bar Rc $= 3.28$ kpc \citep{Robin2012}. The density profile of the adopted ``boxy/peanut'' bar is the same as in \citet{Robin2012}.

\texttt{GravPot16} considers, on a global scale, a 3D steady-state gravitational potential for the MW, modeled as the superposition of axisymmetric and non-axisymmetric components. The axisymmetric potential comprises the superposition of many composite stellar populations belonging to seven thin discs. For each \textit{i}$^{th}$ component of the thin disc, we implemented an Einasto density-profile law \citet[][]{Einasto1979}, as described in \citet[][]{Robin2003}, superposed with two thick-disc components, each one following a simple hyperbolic secant squared decreasing vertically from the Galactic plane plus an exponential profile decreasing with Galactocentric radius, as described in \citet{Robin2014}. We also implemented the density profile of the interstellar matter (ISM) component with a density mass as presented in \citet{Robin2003}. 

The model also correctly accounts for the underlying stellar halo, modeled by a Hernquist profile as already described in \citet{Robin2014}, and surrounded by a single spherical Dark Matter halo component \citet{Robin2003}; no time dependence of the density profiles is assumed. The most important limitations of our model are: 
\begin{itemize}
	\item[(i)] We ignore secular changes in the MW potential over time, which are expected although our Galaxy has had a quiet recent accretion history.
	\item[(ii)] We do not consider perturbations due to spiral arms, as an in-depth analysis is beyond the scope of this paper.  
\end{itemize}

For reference, the Galactic convention adopted by this work is: $X-$axis is oriented toward $l=$ 0$^{\circ}$ and $b=$ 0$^{\circ}$, the $Y-$axis is oriented toward $l$ = 90$^{\circ}$ and $b=$0$^{\circ}$, and the disc rotates toward $l=$ 90$^{\circ}$; the velocities are also oriented in these directions. In this convention, the Sun's orbital velocity vector is [U$_{\odot}$,V$_{\odot}$,W$_{\odot}$] = [$11.1$, $248.5$, $7.25$] km s$^{-1}$, in line with \citet{Brunthaler2011} and \citet{Reid2020}. The model has been rescaled to the Sun's Galactocentric distance, $R_{\odot} = $8.178 kpc \citep{Gravity2019} and Z$_{\odot}= 25$ pc \citep{Juric2008}. 


The orbital trajectories for our entire sample were integrated by adopting a simple Monte Carlo scheme and the Runge-Kutta algorithm of seventh-eight order elaborated by \citet{fehlberg68}, in order to construct the initial conditions for each star, taking into account the uncertainties in the radial velocities provided by the APOGEE-2 survey, the absolute proper motions provided by the Gaia DR3 catalog \citep{GaiaDR3}, and the spectro-photo-astrometric distances provided by \texttt{StarHorse} \citep{Queiroz2023}.
The uncertainties in the input data (e.g., $\alpha$, $\delta$, distance, proper motions, and line-of-sight velocity errors), were randomly propagated as 1$\sigma$ variations in a Gaussian Monte Carlo re-sampling. For each star, we computed one thousand orbits, computed backward in time over 3 Gyr. The average value of the orbital elements was found for these thousand realizations, with uncertainty ranges given by the 16$^{\rm th}$ and 84$^{\rm th}$ percentile values.

For this study, we use the average value of the amplitude of the vertical oscillation $|Z_{max}|$, the perigalactic distance $R_{peri}$, apogalactic distance $R_{apo}$, the eccentricity $e$, the orbital Jacobi constant $E_j$ computed in the reference frame of the bar and the ``characteristic'' orbital energy $E_{char}$ as envisioned by \citet{Moreno2015}. The $E_{char}$ vs. $E_J$ plane is used to broadly discriminate orbit populations, the lower the $E_{char}$ the more radially confined to the Galaxy, the lower the $E_J$ the more vertically confined it is. 

We also consider the minimum and maximum of the \textit{z}-component of the angular momentum in the inertial frame, which defines their orbital behavior. Under a non-axisymmetric potential, the angular momentum is not conserved, thus during the 3 Gyr integration time, a star can be either always prograde (P), always retrograde (R), or change from prograde to retrograde or vice versa (P-R). That is, a P orbit has $L_{min}<0$ and $L_{max}<0$, an R orbit has $L_{min}>0$ and $L_{max}>0$, 
and a P-R orbit has $L_{min}$ and $L_{max}$ with opposite signs. In general, P-R orbits are extremely eccentric, indicating that their transition from prograde to retrograde orbit, or vice versa, is achieved by following a very radial orbit, with their periastron rather close to the Galactic center. Their dynamics can be subject to chaotic effects with very sensitive orbits, and/or could be trapped in a resonance. Also, when orbits are isotropically heated, not just vertically, we suspect the mechanism keeping them that way, internal or external, is different from the one that formed the MW thick disc.

\section{t-SNE}\label{sec_tsne}
The t-Distributed Stochastic Neighbor Embedding (t-SNE) algorithm is a nonlinear technique of unsupervised analysis for the study of high-dimensional ensembles through a dimension reduction that allows its visualization (2D or 3D), as presented by \citet{ref_tsne}. Many areas of study  make use of this technique. due to its utility in finding possibly related structures automatically and transferring the information to spaces that allow us to visualize them. In other words, t-SNE is able to detect similarity among data points in high-dimensional space, by transforming them onto 2D values (the t-SNE plane), so that nearby points are considered as likely members of the same population, or at least somehow related. 

Astronomy presents an ideal scenario for the use of this technique since we have now large catalogs with a variety of information for each star. In the context of our work, t-SNE has proven useful, such as in \citet{Anders2018}, where the interested reader can find a summarized yet comprehensive description of the operation of t-SNE. In the following paragraphs, we provide an overall description of how the method works.

The t-SNE algorithm comprises two main stages. First, it constructs a probability distribution over pairs of high-dimensional objects so that the closer/farther objects are, the higher/lower probability assigned for being related. Second, starting from a Gaussian low-dimensional distribution, t-SNE iteratively constructs a probability distribution of points by minimizing the Kullback–Leibler divergence  between the high- and low-dimensional distributions using gradient descent. As a result, the final low-dimensional distribution resembles the probability of the high-dimension one, so that nearby objects in one are also close in the other.

The method has one main parameter, called the \texttt{perplexity} $p$, 
which can be thought of as a guess about the number of close neighbors each point has; the ideal value for $p$ depends on the sample size. Since a change in perplexity has in many cases a complex effect on the resulting map, it is recommended to try different values for $p$ \citep{wattenberg_how_2016}. According to \citet{Linderman2017EfficientAF}, two other hyper-parameters of t-SNE can be optimally chosen: the learning rate ($\sim$ 1) and the early exaggeration parameter (10\% of sample size). We used these recommended values as a starting point but eventually, other values were set, guided by the best separation of the detected groups in the t-SNE plane. 

We use the Python implementation of t-SNE included in the {\tt scikit-learn} package \citep{pedregosa2011scikit}. A word of caution is given: with the same data and parameter inputs, we recover the exact same t-SNE planes up to version 1.1.3 of {\tt scikit-learn}, but when using newer updates, i.e, version 1.2.0 onwards, the look of the t-SNE plane changes. We checked that our detected groups remain virtually the same, therefore our conclusions are not altered.

The distribution of points in the t-SNE plane depends not only on the multidimensional data input, but also on the order each of these vectors is entered. This means that the same data input sorted in different ways will yield different-looking t-SNE planes, but in all cases, the same points will be associated quite similarly. In our case, the stellar data were inserted in order of increasing [Fe/H], as an additional check on the obtained results.

The t-SNE algorithm does have two major weaknesses: 1) no missing individual data is allowed, meaning each star must have measured values in all the dimensions considered; and 2) it does not consider the effect of individual uncertainties. We deal with these issues by 1) using only stars with all ten chosen abundances measured and 2) performing a Monte Carlo experiment to show that our results are robust to abundance uncertainties. 

\section{Strategy adopted with t-SNE}\label{sec_strategy}

The input entries fed into t-SNE are the abundance ratios [Fe/H], [O/Fe], [Mg/Fe], [Si/Fe], [Ca/Fe], [C/Fe], [N/Fe], [Al/Fe], [Mn/Fe], and [Ni/Fe]. We also include [Mg/Mn], which has proven to be useful in separating accreted from in-situ structures \citep{Das2020, Naidu2020}, considering also that from previous work we know of structures that overlap in [Mg/Mn] but are separate in [Mg/Fe] and [Mn/Fe], or vice-versa. Our final results were obtained under \texttt{perplexity = 25}, \texttt{early\_exageration = 130}, \texttt{init = pca}, \texttt{learning\_rate = auto}, and  \texttt{n\_iter = 7000}, in which the detected associations of stars were more visibly separated (See Figure \ref{fig_main_tsne}, central panel). A total of seven structures were identified, and are presented for separate analysis in the coming sections. To choose this result as the final one, we thoroughly tested the effect of the input parameters and the abundance errors in the obtained results, as described below. Also, we checked their dynamical parameters to help confirm if the detected populations also exhibited additional differences in their orbits.

Stars in all the structures identified were persistently associated in the test runs despite changing the t-SNE parameters over a range of values: $20<$\texttt{perplexity}$<40$, and $80<$\texttt{early\_exageration}$<130$, although a few of the smaller groups were more unstable in their location in the t-SNE plane, moving from one location to other, getting closer or farther from the other groups. The diagrams generated by t-SNE can strongly vary the shape of the distribution in the plane depending on the initial parameters imposed (see Figure \ref{fig_tsne_params} in Appendix \ref{sec_tests}), but indeed some of the identified structures predominate and simply relocate, always staying as a single structure (Groups G1, G2, G5, and G7 mainly). Therefore, the final selection shown considered how their location in the t-SNE plane changed with respect to the other detected associations. Another important test was to verify the separation of adjacent or nearby structures by running t-SNE on smaller input subsets. Moreover, we used the obtained t-SNE values as additional inputs for a second t-SNE run, to check that the selected groups were ``squeezed'' further in the new t-SNE plane. These test helped us to trace the boundaries between nearby groups. Results of this test are shown in Appendix \ref{sec_tests}, Figure \ref{fig_tsne_iter}. We also found that the most extended structures (G1 and G2) kept each one cohesively in these runs, without further separation.

To understand the effect of abundance errors, we generated random samples based on the uncertainty of each abundance and then fed those as input to t-SNE. For each star, 100 random measurements were drawn from a Gaussian centered on the measured abundance and dispersion corresponding to its uncertainty. Samples were then created by randomly taking one out of 100 simultaneously in all input abundances and applying the t-SNE to each generated sample with the same parameters as the one chosen for our final results. We found that, despite significant variation in the distribution shape of the points in the t-SNE plane (see Figure \ref{fig_tsne_errors} in Appendix \ref{sec_tests}), the same structures kept together consistently and were detected in the different t-SNE realizations. Therefore, we conclude that abundance uncertainties do not lead to significant changes in the group identifications. Moreover, this reinforces the fact that the structures detected do not correspond to random associations. 


\section{Structures detected in the t-SNE projection}\label{sec_results}
\begin{figure*}[h]
\centering
\includegraphics[width=17cm]{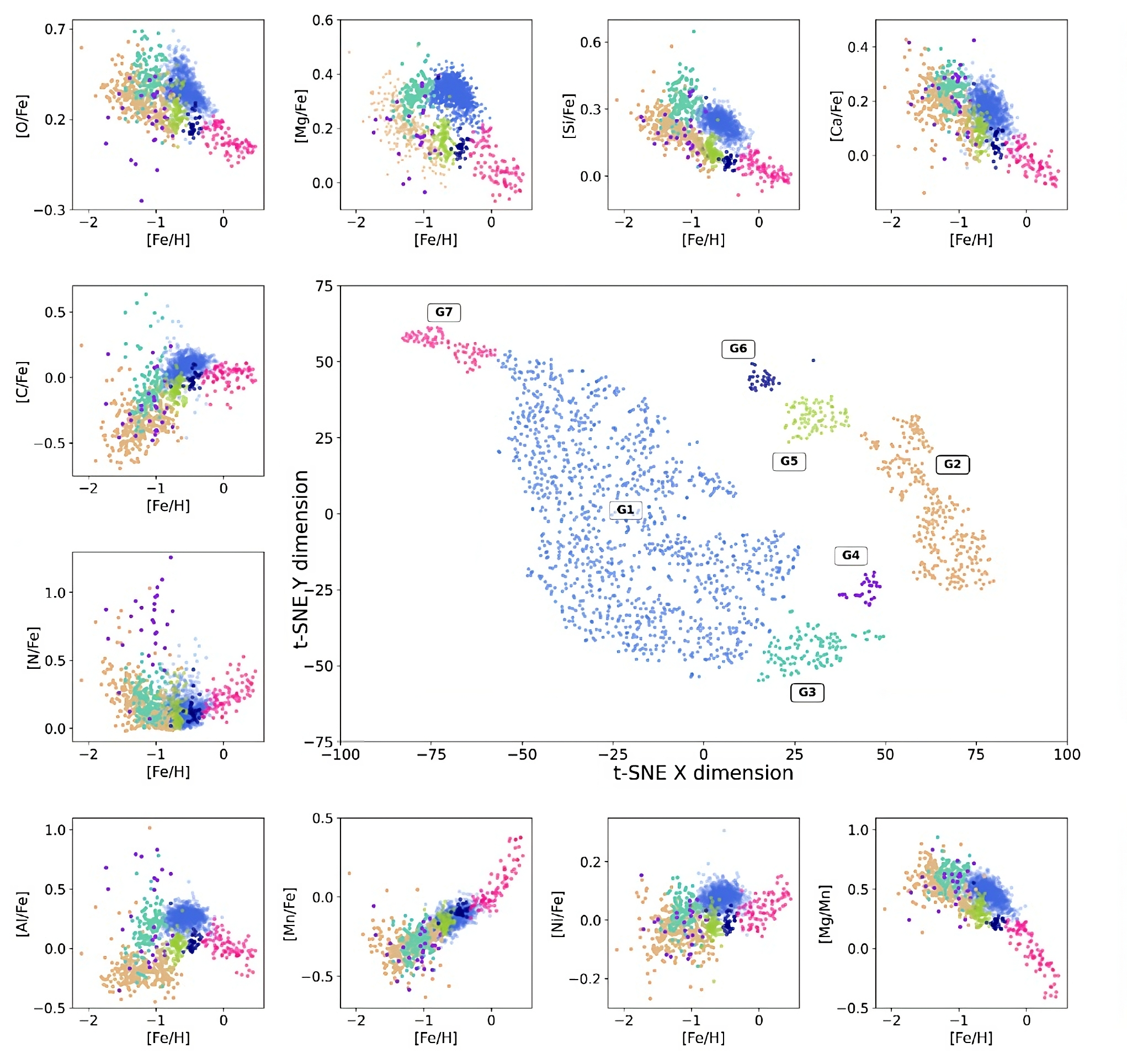}
\caption{The t-SNE plane, shown in the large main panel, with the selected groups identified by colors and labels. The  surrounding panels correspond to the ten APOGEE abundances used as inputs for t-SNE.} \label{fig_main_tsne}
\end{figure*}

\begin{figure*}[h]
\includegraphics[width=18cm]{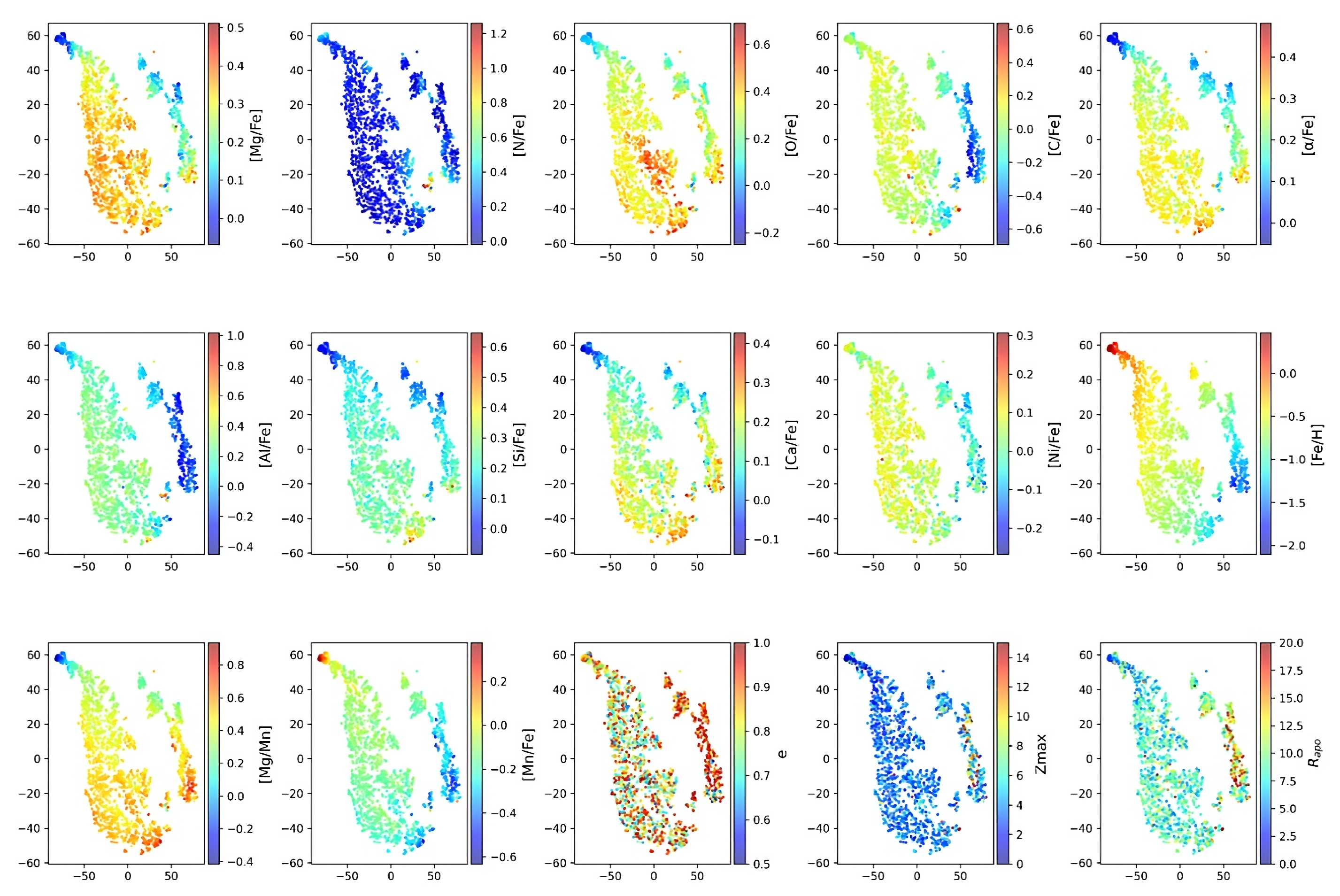}
\caption{The t-SNE plane color-coded by the indicated abundances and dynamical parameters $e$, $|Z_{max}|$ in kpc, and $R_{apo}$ in kpc.} \label{fig_tsne_color_map}
\end{figure*}

The large panel shown in Figure \ref{fig_main_tsne} is our reference t-SNE projection, 
on which we have identified and named seven substructures that clearly emerge from it. The figure also shows the distribution of these substructures in several elemental-abundance diagrams (small panels), which have been color-coded according to the substructures' names. This t-SNE plane is also shown in Figure \ref{fig_tsne_color_map}, color-coded by the indicated abundances and dynamical parameters $e$, $|Z_{max}|$ in kpc, and $R_{apo}$ in kpc.

\begin{figure}
\centering
\includegraphics[scale=0.13]{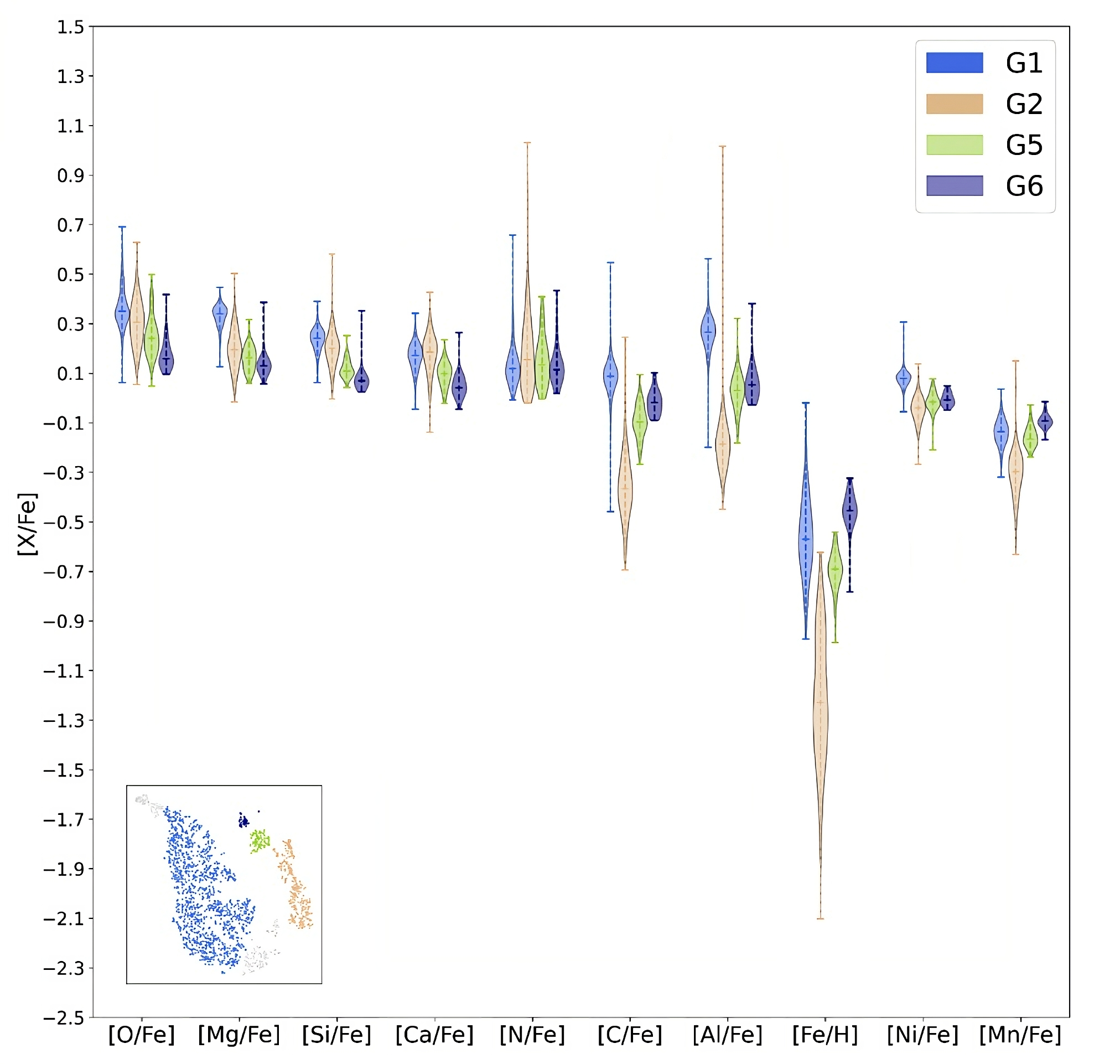}
\caption{Comparison of Groups G1 (Splash), G2 (GSE), G5 (Galileo 5), and G6 (Galileo 6), in each of the chemical abundances used in this work, displayed in the form of a violin plot. A common trend in the compared abundances for the $\alpha$-elements can be seen, as well as in the iron-peak elements. The observed distributions demonstrate that each group is different from the others in at least one chemical abundance.}
\label{fig_violin}
\end{figure}

From inspection of the t-SNE map in Figure \ref{fig_main_tsne}, we can immediately appreciate two dominant and well-populated substructures, which fit very well with the descriptions of the Splash-like population (G1, blue points, Fig. \ref{fig_main_tsne}) and a significant past merger event (G2, beige points, Fig. \ref{fig_main_tsne}) established in the literature. Five other smaller substructures were identified as well. Each substructure is described individually below, considering not only their chemical abundances but also their dynamical properties, which were not fed to the t-SNE but exhibit distinctive ranges. In all corresponding figures, $V_\phi$ is in km s$^{-1}$, distances are given in kpc, and energies in $10^5$ km$^2$ s$^{-2}$. Eccentricity $e$ and t-SNE X and t-SNE Y values are dimensionless. Figure \ref{fig_violin} shows the distribution of the chemical abundances for some of the detected groups, to prove they were all indeed different from each other in at least one of the elements. In the Appendix section \ref{app_median}, Table \ref{tab:median_values} contains the median and median absolute deviations for all the chemical abundances and dynamical parameters of each detected group. 

\subsection{Group 1: The Splash}\label{subsec_splash}

\begin{figure*}[h]
\includegraphics[width=18cm]{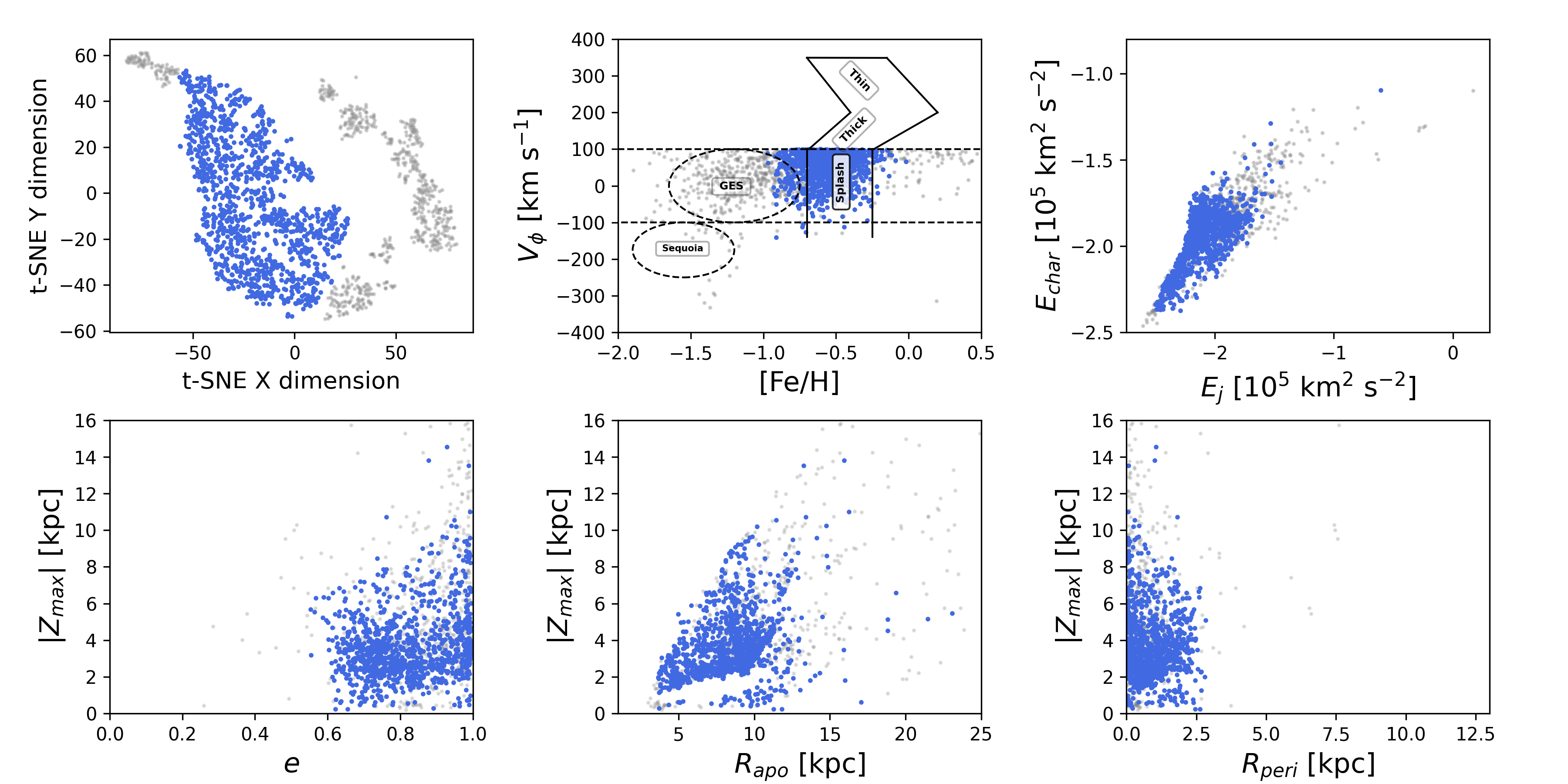}
\caption{Group 1 (G1), identified as the Splash. Energies units are $10^5$ km$^2$ s$^{-2}$. Details are explained in subsection \ref{subsec_splash}.} \label{G1}
\end{figure*}

This is the largest and most consistent structure detected by the t-SNE method, showing higher $\alpha$-element abundances $\gtrsim +0.20$ and slightly enriched levels of Al ([Al/Fe] $>+0.1$) at metallicities [Fe/H] $>-1$ (Figure \ref{fig_main_tsne}, blue points) to solar metallicity, chemistry consistent with the thick disc. Dynamically (see Figure \ref{G1}), it exhibits a wide range of $|Z_{max}|$, spanning from in-plane to 14 kpc from the Galactic plane, clearly overlapping with the inner-halo region of the Galaxy.

$R_{apo}$ spans mostly from 3 to 12 kpc, but a few stars reach up to 25 kpc, while $R_{peri}$ is mostly below 2.5 kpc. As expected from such values, these stars follow eccentric orbits, typically with $e\gtrsim0.6$ (see also the lower panels of Figure \ref{fig_tsne_color_map}, color-coded by $e, |Z_{max}|$, and $R_{apo}$). The orbital energies of these stars span a large range, from low to mid values. Much of this group fits very well with the descriptions established for the so-called Splash by different authors \citep{2017ApJ...845..101B, 2020ApJ...897L..18B, 2018ApJ...863..113H, 2019A&A...632A...4D, 2020MNRAS.494.3880B}. 

G1 is mostly (76\%) comprised of prograde orbits stars (see Table \ref{tab:orbit_percent}), which supports the idea of its origin as heated-disc stars. Nonetheless, a non-negligible portion of G1 stars (see Table \ref{tab:orbit_percent} in Appendix \ref{app_median}) are P-R and retrograde orbits, which are not only vertically but also radially heated. Therefore, these stars have the chemical signature of the MW thick disc, but are dynamically hotter, pointing at some additional mechanism (internal or external) giving these stars a higher {\it ``kick''}. 

\subsection{Group 2: \textit{Gaia}-Enceladus-Sausage-like}\label{subsec_gse}

This structure matches very well the group dubbed Gaia-Sausage-Enceladus \citep{Helmi2018Natur, Belokurov2018, 2018ApJ...863..113H, Mackereth}, showing its characteristic low concentrations of aluminum and iron: [Al/Fe] $\lesssim 0$ and [Fe/H]$<-0.6$, in its component stars (see Figure \ref{fig_violin}). Our Group 2 (G2) is deficient in carbon, with [C/Fe] $\lesssim 0$, although some outliers are seen with higher values. For the ratios [Si/Fe], [Mg/Fe], and [Ca/Fe], G2 follows the known sequence vs [Fe/H] for GSE, including the so-called ``{\it knee}'' at [Fe/H] $\sim -1$. The elements [N/Fe], [O/Fe], and [Ni/Fe] are more spread.

\begin{figure*}[h]
\includegraphics[width=18cm]{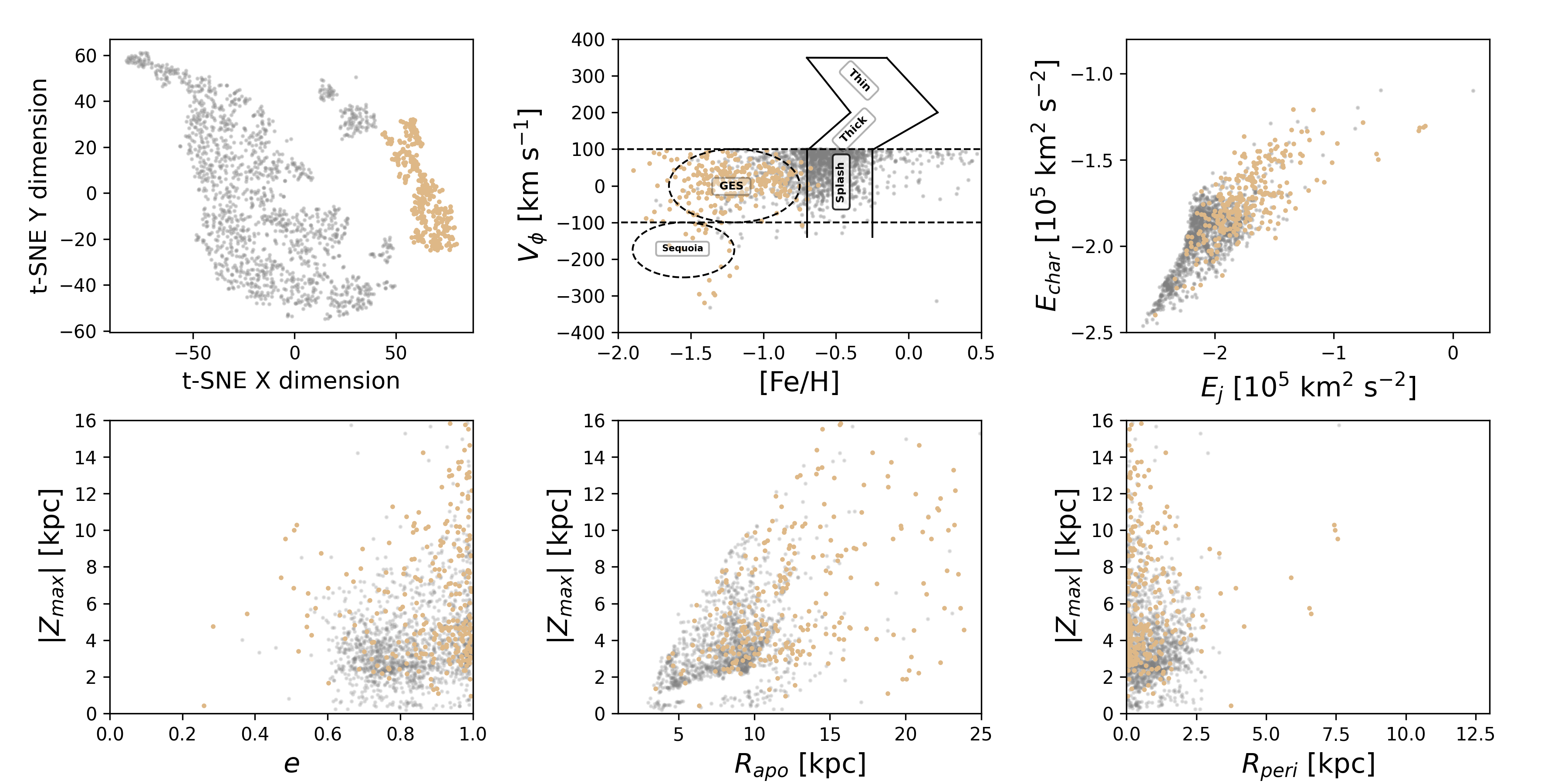}
\caption{Group 2 (G2), identified as Gaia-Sausage-Enceladus. Details are explained in subsection \ref{subsec_gse}. Panels and units are as in Figure \ref{G1}.} \label{G2}
\end{figure*}

The G2 stars (see Fig. \ref{G2}) exhibit rather eccentric orbits, spanning mainly between $0.6\leq e\leq 1$ with a visible concentration at $e>0.9$, and heights exceeding 2 kpc, reaching to 20 kpc. The energy plane $E_{char}$ versus $E_J$ exhibits high values (mainly $E_J>-2$ and $E_{char}>-2$), spread in a way consistent with the dynamics of the so-called inner-halo component, first identified by \citep{Carollo2007, Carollo2010, Beers2012d}. Most of the stars in G2 are similarly distributed between prograde and retrograde orbits with a smaller fraction in the P-R classification (See Table \ref{tab:orbit_percent}), which means G2 stars are evenly distributed in $L_Z$, not showing a preferential rotation direction. We noticed that G2 stars with the lowest $V_\phi$ values -  potentially members of Sequoia - are all also retrograde, but they spread all over the G2 locus, meaning t-SNE did not detect anything closely in common among them. Therefore, we keep those as G2 members. The periastron distances for many G2 stars are within 1 kpc of the Galactic center, with apoastron distances between 8 and 25 kpc. The highest vertical distance from the Galactic plane for these stars spans between 2 and 25 kpc. One star (APOGEE ID: 2M13393889+1836032 ) reaches $R_{apo} \sim$ 80 kpc and $|Z_{max}| \sim$ 80 kpc as well. All these features are in agreement with previous results \citep{Mackereth&Bovy2020, Naidu2020, Feuillet2021,Buder2022}.

\subsection{Group 3: High-$\alpha$ heated-disc population}\label{subsec_halpha}
From the t-SNE plane in Figure \ref{fig_main_tsne}, G3 is next and seemingly connected to G1, yet we diagnosed G3 as a separate group because in many tests (see Figs. \ref{fig_tsne_params}, \ref{fig_tsne_errors}) it always appeared at one extreme of G1, looking like an appendage connected by only a few points, and being always more metal-poor than G1 (see Fig. \ref{fig_tsne_color_map}, row 2, column 5). We also noticed that, in the t-SNE double iteration (see Fig. \ref{fig_tsne_iter}), G3 is detached from G1 while G1 itself, despite exhibiting some internal clustering, remains mostly together. Additionally, when applying t-SNE only to the stars we identified in G1 and G3 alone, under various conditions, G3 is often separated from G1, while the latter keeps its integrity.

\begin{figure*}[h]
\includegraphics[width=18cm]{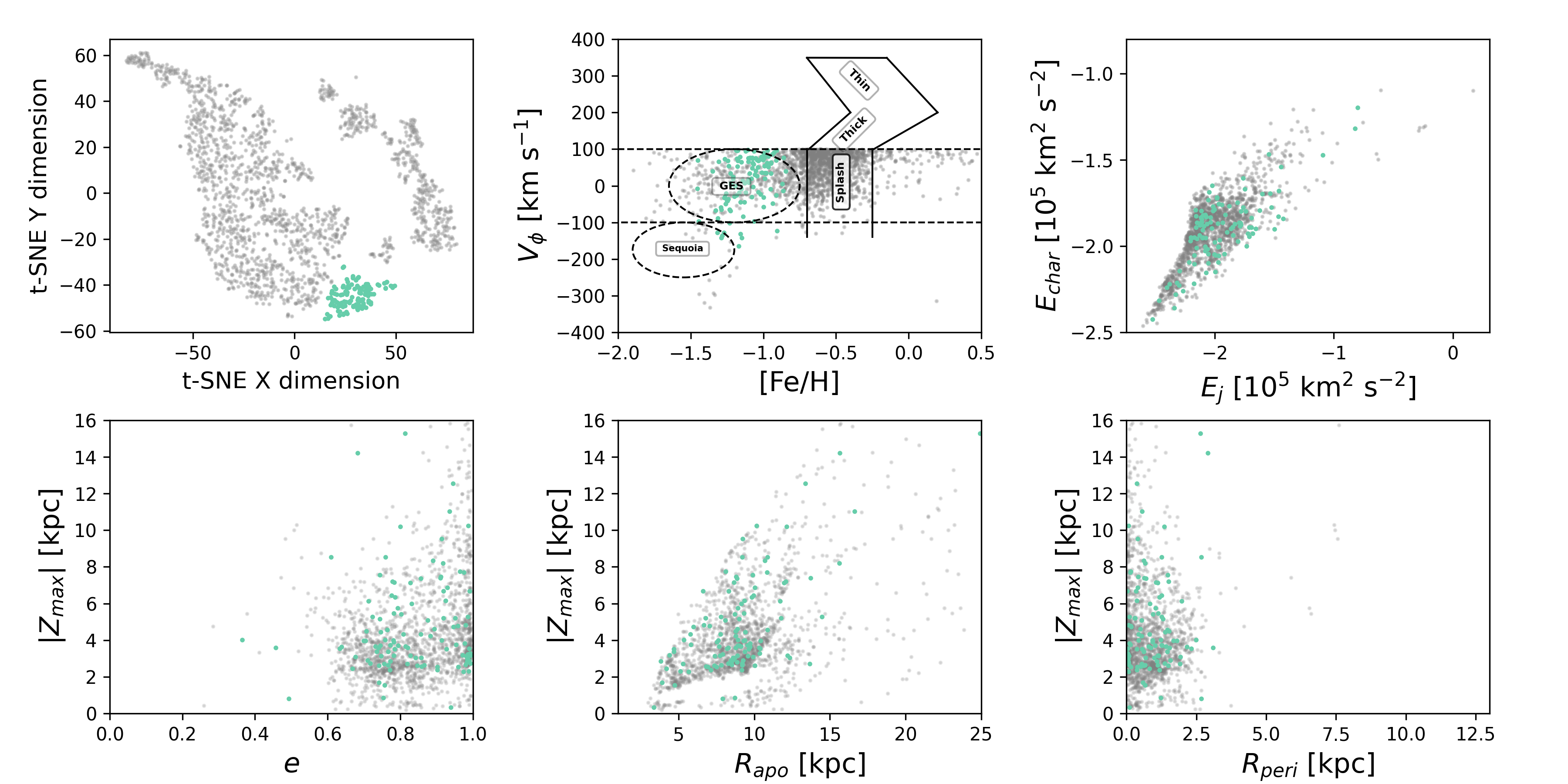}
\caption{Group 3 (G3), identified as the High-$\alpha$ heated-disc population. Details are explained in subsection \ref{subsec_halpha}. Panels and units are as in Figure \ref{G1}.} \label{G3}
\end{figure*}

Although our Group 3 (G3) stellar abundances overlap with either G1 and/or G2 in various elements, it is clear from Figure \ref{fig_main_tsne} that this structure sits between them as a separate population in the [Si/Fe] versus [Fe/H] and [Al/Fe] versus [Fe/H] planes, being the structure richest in silicon of our whole dataset. As \citet{2015MNRAS.453..758H} describe in their Figure 1, the abundance locus of canonical halo MW stars in the [$\alpha$/Fe] versus [Fe/H] plane is also fulfilled well by G3 stars.  Dynamically though, G3 star's orbital parameters spread over ranges similar to those of G1 (Splash) (see Figure \ref{G3}), although more dispersed. A population of such features has been found by \citet{2010A&A...511L..10N}:  their high-$\alpha$ {\it heated} population, and also by \citet{2018ApJ...852...49H}: their high-Mg thick disc-like population. 

\citet{2010A&A...511L..10N} suggest that these stars may be ancient disc or bulge stars “heated” to halo kinematics by merging satellite galaxies, or they could have formed as the first stars during the collapse of a proto-Galactic gas cloud, while \citet{2018ApJ...852...49H} claims that the HMg population is likely associated with in-situ formation. In Figure \ref{alvsmg_mn}, G3 sits on the high-$\alpha$ in-situ locus. In the chemical-evolution model for a Milky Way galaxy proposed by \citet{2021MNRAS.500.1385H} (their Fig. 2), this location roughly corresponds to an in-situ early disk population. We conclude that G3 corresponds to this structure, which we tag as the High-$\alpha$ population. 

\subsection{Group 4: N-C-O peculiar stars}\label{subsec_nrich}

\begin{figure*}[h]
\includegraphics[width=18cm]{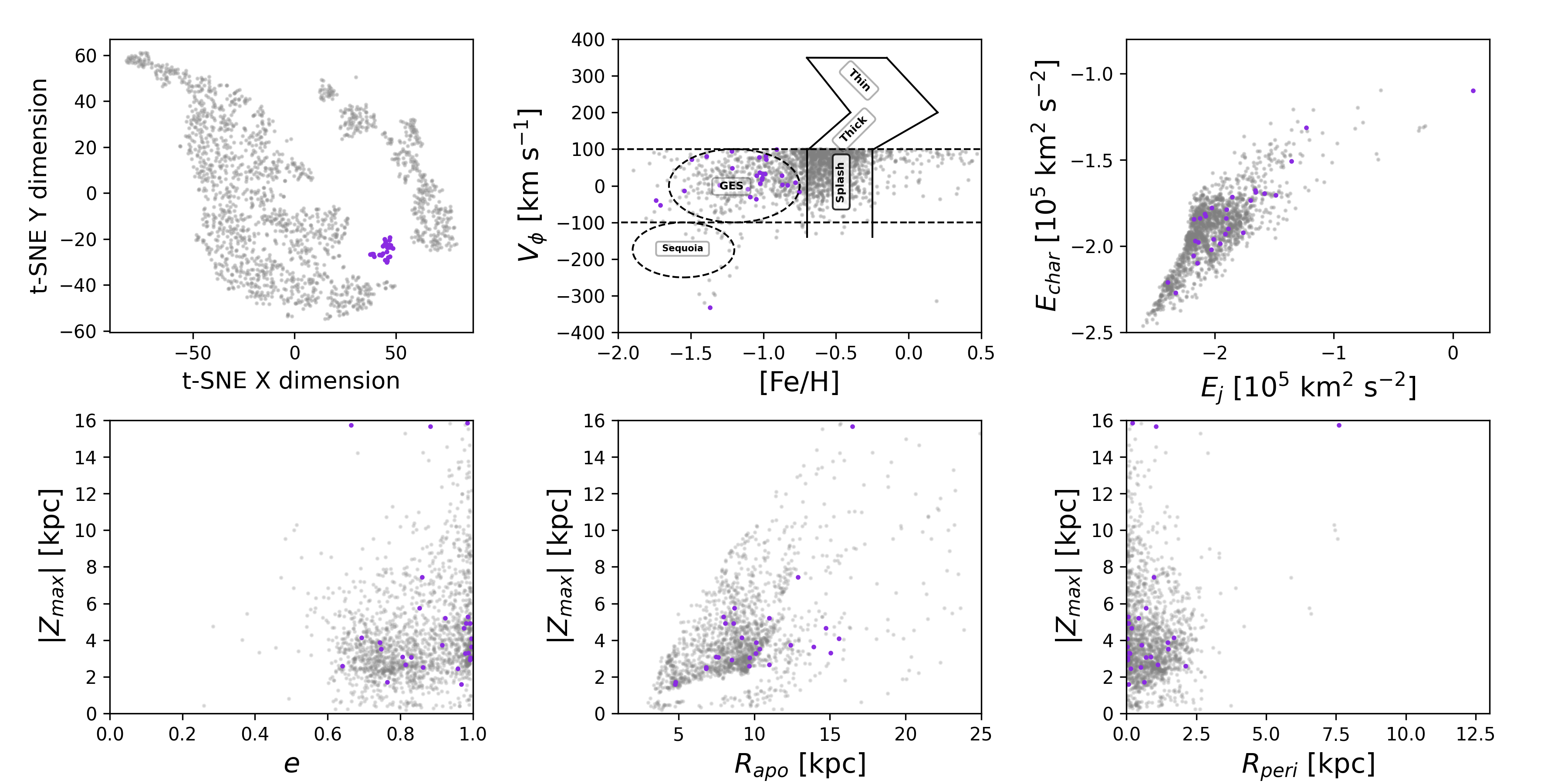}
\caption{Group 4 (G4), identified as N-C-O peculiar stars. Details are explained in subsection \ref{subsec_nrich}. Panels and units as in Figure \ref{G1}.} \label{G4}
\end{figure*}

This is a small substructure in the t-SNE space, whose abundances relative to Fe are shown in small panels in Figure \ref{fig_main_tsne}. A large fraction of the stars in this substructure is typified by near-solar and anomalously high levels of [N/Fe] that are well above ($\gtrsim +0.5$) typical Galactic levels over a range of metallicities, accompanied by decreased abundances of [C/Fe] $< +0.15$, as seen in Figures \ref{fig_main_tsne} and \ref{fig_tsne_color_map}. A portion of its stars also exhibit high [Al/Fe]. This substructure also matches very well with the atypical nitrogen-enriched population, as envisioned by a series of works \citep[see, e.g.,][]{Schiavon2017Nrich, Trincado1, Trincado2, Trincado3, Trincado4, Trincado5, Trincado6, Trincado7, Trincado8, Trincado9, Trincado10, Trincado11} that have attributed these stars to exhibit Globular Cluster (GC) \textit{second-generation}-like chemical patterns, and are likely GC debris \citep{Trincado9, Trincado8} that have now become part of the Galactic field population. The dynamics of these stars have inner-halo-like orbits, with values spreading with no particular concentration in any parameter (see Figure \ref{G4}). 

A closer examination of their abundances reveals that G4 contains three subgroups of stars:
\begin{itemize}
\item N-rich and C-poor stars: [N/Fe] $>$ +0.5 and [C/Fe] $<$ +0.15, amounting to 20 stars falling into this category, including 17 already reported by \citet{Trincado11}, plus 3 
others (APOGEE IDs: 2M06572697$+$5543115, 2M21372238$+$1244305 and 2M16254515-2620462), located within the abundance limits defined by \citet{Trincado11} for this kind of stars, and proposed by these authors to GC second-generation or chemically anomalous debris. These three stars are genuine newly identified N-rich stars, the last one (2M16254515-2620462) being a potential extra-tidal star of M~4.

\item N-poor and C-poor stars: [N/Fe] $< +0.55$ and [C/Fe] $< -0.1$, which amount to 5 stars that also exhibit sub-solar [Al/Fe] abundance ratios. When examined in the [Mg/Mn]--[Al/Fe] plane these stars clearly fall in the accreted halo region (see Figure \ref{alvsmg_mn}). We found that three of the five stars (APOGEE IDs 2M12211605-2310262, 2M23364836$-$1135404 and 2M21420647$-$3019385) are within GSE's t-SNE footprint in all the tests shown in Fig. \ref{fig_tsne_errors}. We postulate that they are part of a merger remnants (and/or likely associated with GSE, or to a number of unknown events contributing to this sub-population) present in the Milky Way.

\item N-rich, C-rich, and O-rich stars: [N/Fe] $>$ +0.4, [C/Fe] $>$ +0.15 and [O/Fe] $>$ +0.2, which amount to 3 stars. These stars also have super-solar values in Mg and Si. Stars with such enhancements have been reported by, e.g., \citet{2005ARA&A..43..531B}. It is likely that the atmosphere of these mildly carbon-enhanced stars has been contaminated either by an intrinsic process, such as self-enrichment (likely a TP-AGB), or by a past extrinsic event, that is, the mass-transfer hypothesis (binary mass transfer systems). However, with the available radial velocity scatter (\texttt{VSCATTER} $<1$ km s$^{-1}$) from three APOGEE-2 visits, it is not possible to support or reject either possibility.
\end{itemize}
In our entire sample of 1742 stars, we have 24 matches with the 412 N-rich stars listed by \citet{Trincado11}, of which the 7 not in G4 were assigned by t-SNE to G2 (GSE). Three of these latter stars have [Al/Fe] $>$ 0, which is not consistent with the main feature that distinguishes GSE, but the other four have [Al/Fe] $<$ 0, and we postulate these as second-generation or chemically anomalous debris from GSE's own GCs. This is corroborated by their locus in Figure \ref{alvsmg_mn}.

Some of the differences between the results of our work and that of \citet{Trincado11} could also be explained by two factors: (i) these authors based their analysis on a proprietary catalog internally distributed to the collaboration, which contained the resulting \texttt{ASPCAP} solutions from the ``$\dots$l33$\dots$'' runs, while our present study is based on the published \texttt{ASPCAP} run ``$\dots$synspec\_fix$\dots$'', and (ii) t-SNE does an {\it un}-supervised search for stars associated in the used abundances space, while \citet{Trincado11} does a supervised one. 

\subsection{Group 5: New structure}\label{subsec_h5}
\begin{figure*}[h]
\includegraphics[width=18cm]{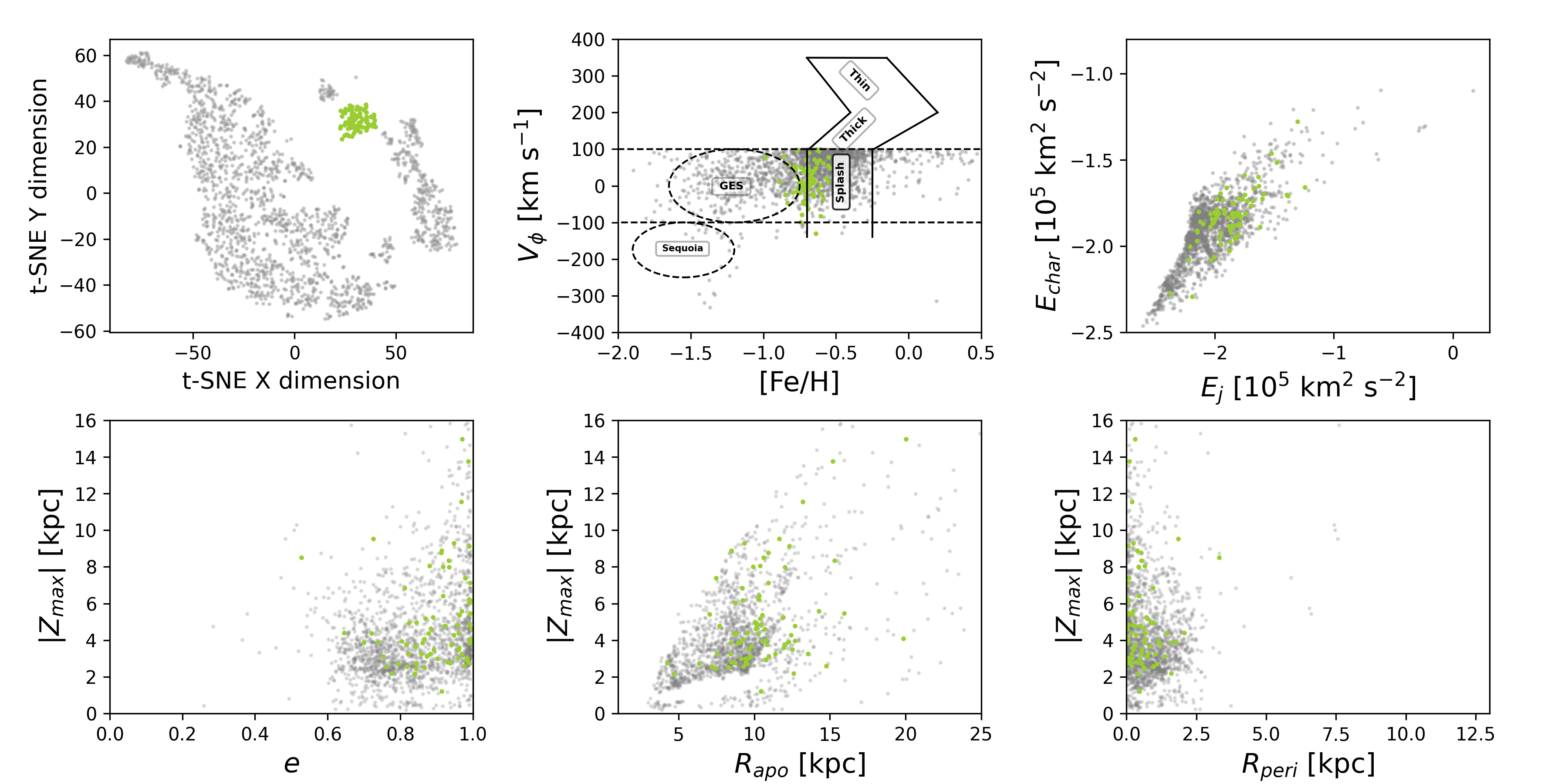}
\caption{Group 5, a new structure identified in this work and named Galileo 5 (G5). Details are explained in subsection \ref{subsec_h5}. Panels and units are as in Figure \ref{G1}.} \label{G5}
\end{figure*}

Stars in Group 5 (G5) are located next to the metal-rich border of G2 (GSE) in many abundances versus [Fe/H] in the small panels of  Figure \ref{fig_main_tsne}. This small structure appears frequently between G2 and G7 in the many t-SNE tests performed in this work. The G5 structure sits in a distinctive separate locus in the [Mg/Fe], [Si/Fe], and [Al/Fe] versus [Fe/H] planes. Compared to G1 (Splash), G5 is lower in oxygen, magnesium, silicon, carbon, and aluminum, although it has G1-like metallicity. Dynamically, the values of $e, R_{peri}, R_{apo}$, and $|Z_{max}|$ in G5 stars spread over the values spanned by the full dataset, but its energy and eccentricity value are slightly higher than G1 (see Figure \ref{fig_tsne_color_map}, lowest row 5$^{th}$ panel). The G5 structure has mostly $-2.2 \lesssim E_J \lesssim -1.5$ and $-2.1\lesssim E_{char}\lesssim -1.6$ in the energy plot of Figure \ref{G5}. We notice one star, APOGEE ID 2M15113246+4813218, which is an outlier of G5 in [Mg/Fe, [Si/Fe], [Ca/Fe], and [Al/Fe], despite its location on the t-SNE plane being well within the footprint of this group.

Chemically, there is an overlap between G5 and the metal-poor side of the structure Eos described by \citet{2022ApJ...938...21M}. Unlike these authors' Eos, our G5 group does not overlap much with either Splash (G1) or GSE (G2). Our detection also extends the eccentricity range towards slightly lower values, as we do not cut on this parameter. Like the metal-poor half of Eos, G5 is lower in nickel than the MW disc, and low values of [Ni/Fe] have been linked to accretion from present-day dwarf spheroidal galaxies \citep{2010A&A...511L..10N}, although the latter are more metal poor ([Fe/H] $< -0.8$) than G5. The investigation of \citet{2021NatAs...5..640M} carefully dated ex-situ stars with $e>0.7$ that fall below the line [Mg/Fe] $= -0.2$\,[Fe/H] $+$0.05, and found that they also exhibit sub-solar values of [Ni/Fe] and are slightly richer than GSE in [(C$+$N)/O], as seen in Figure \ref{fig_niquel}. These stars are found to be slightly younger than the high-Mg in-situ halo stars. Most of our G5 stars fall below the above-mentioned line and occupy the same locus in the plot [Ni/Fe] versus [(C+N)/O] as the confirmed ex-situ low-Mg stars. 

We did a cross-match between the 177 stars labeled as members of Eos by \citet{2022ApJ...938...21M} (private communication) with our full sample. Only 49 were in common between both studies: 11 (22\%) in G1, 7 (14\%) in G2, 1 (2\%) in G3, 0 (0\%) in G4, 19 (39\%) in G5, 8 (16\%) in G6 and 3 (6\%) in G7. Given their chemical similarity, it is not unexpected that a good portion of the common stars (55\%) belongs to G5 and G6, yet a non-negligible 22\% portion was assigned by t-SNE to G1 (Splash). Only 169 listed members of Eos have a $d_{\rm \texttt{StarHorse}}$, of which 65 are beyond 5 kpc, and therefore outside the volume sampled by our study.

While \citet{2022ApJ...938...21M} propose that Eos originated from the gas polluted by the GSE and evolved to resemble the (outer) thin disc of the MW, we instead speculate that what they detect as the single structure Eos, was separated by t-SNE in our work into two portions, the more metal-poor one being our G5 and of ex-situ origin, and the more metal-rich one which may correspond to our detected group G6, described next.
We postulate G5 as a new structure, which we name Galileo 5, after the name of the project ``Galactic ArchaeoLogIcaL ExcavatiOns'' (GALILEO), thus retaining the G5 acronym. 

\subsection{Group 6: New structure or Eos?}\label{subsec_h6}

Stars in Group 6 (G6) are more metal-rich than G5 but more metal-poor than G7, and are slightly lower than G5 in all the $\alpha$ elements, as seen in Figure \ref{fig_main_tsne}. In several t-SNE tests, G6 stars were very close to or attached to G7. Similar to the previous group, G6 also occupies a distinctive separate locus in the [Mg/Fe], [Si/Fe], and [Al/Fe] 
versus [Fe/H] planes, and is more $\alpha$-poor than G1 (Splash), overlapping partially with the abundance locus of thin-disc stars. Dynamically, like G5, G6 also exhibits high eccentricities, but is slightly less energetic than G5. 

Figure \ref{fig_violin} shows an element-by-element comparison between groups G1 (Splash), G2 (GSE), G5 (Galileo 5), and G6. For all the $\alpha$-elements, G1, G2, G5, and G6 go from richest to poorest in their mean values; for the iron-peak elements, aluminum, and carbon, G2, G5, and G6 follow an ascending trend, with G1 being either the richest or the second richest. G1 stands out as the richest structure in aluminum; G2 is easily separated from the other structures by its extremely low values in carbon, aluminum, and iron; G5 and G6 separate from each other in iron. We notice one star in G6, APOGEE ID 2M16045805-2418515, which sits a bit far from the locus of G6 in the t-SNE plane, and also looks like an outlier in several abundances. This star was  associated with G6 when running t-SNE only on the G5 and G6 data.

As commented in the previous subsection, G6 overlaps with the more metal-rich part of the reported structure Eos. Our data selection is more stringent than \citet{2022ApJ...938...21M}, and this may have helped t-SNE in separating G6 from G5 in [Fe/H]. In our data, G5 and G6 have distinctive values and distribution of metallicity, as seen in Figure \ref{fig_violin}, therefore we trust they are  separate structures. G6 abundances overlap with Aleph, as reported by \citet{2023MNRAS.520.5671H}, but Aleph clearly follows a disc-like orbit, which is not the case for G6. When studying various halo substructures in the MW halo, \citet{2023MNRAS.520.5671H} explains that halo substructures with low nickel due to a low star-formation rate can be identified by their having disc-like values of [Mn/Fe], which is the case for both G5 and G6. On the other hand, unlike G5, G6 overlaps in many abundances with the thick-disc population, although G6 stars' orbits are clearly more energetic. If we assume G6 stars were formed in the disc, a mechanism different or additional to the one that formed the thick disc is needed to keep G6 stars in those more energetic orbits. With G6 it is harder to ascertain if it has an ex-situ origin or if it matches Eos - which is proposed to be formed in situ -. We postulate G6 as a likely new ex-situ structure formed by debris from a past merger event and we name it Galileo 6, keeping the G6 acronym. Nonetheless, we recognize there is also a chance of it being the Eos structure.

\begin{figure*}[h]
\includegraphics[width=18cm]{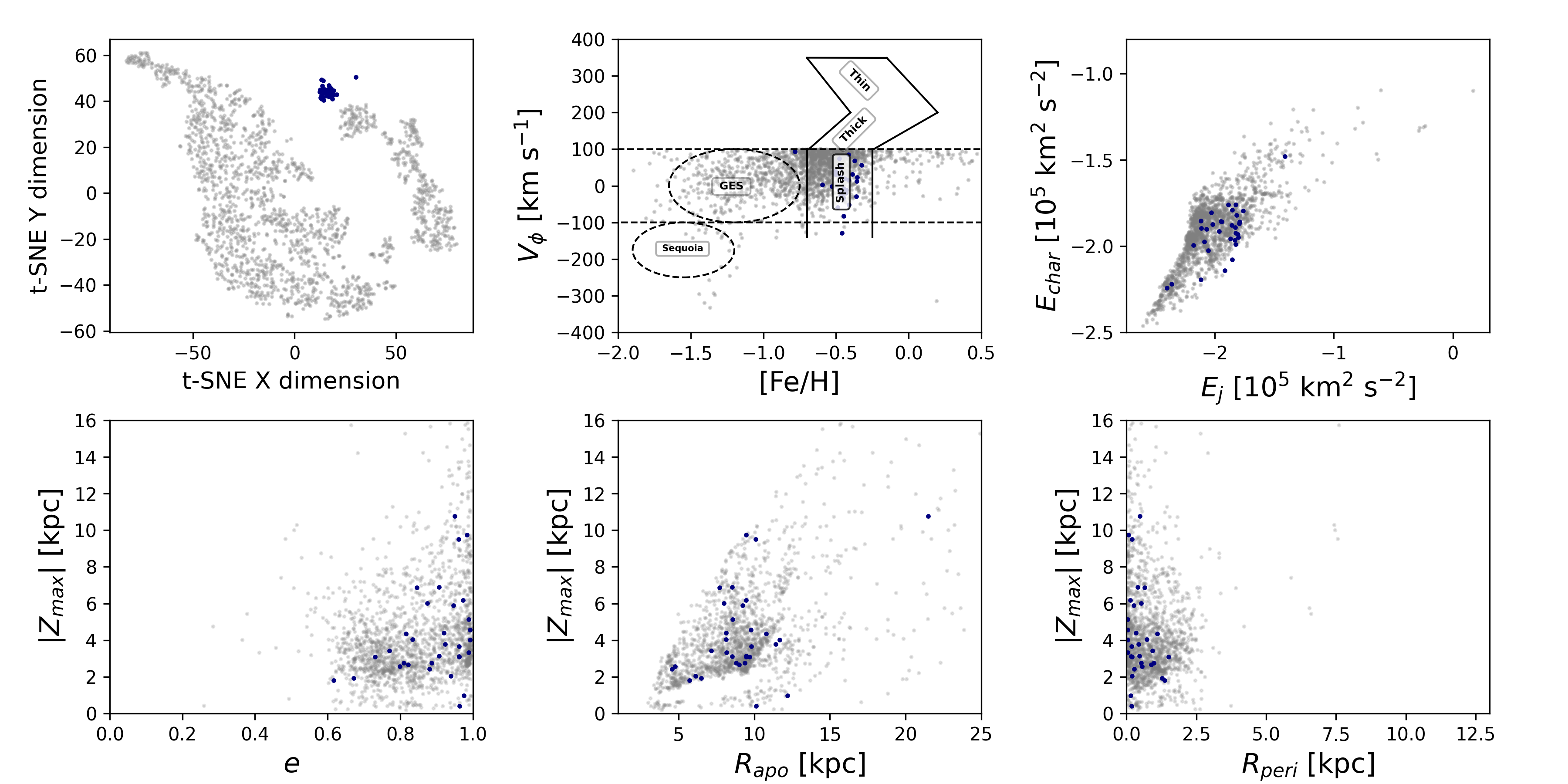}
\caption{Group 6, a new structure identified in this work and named Galileo 6 (G6). Details are explained in subsection \ref{subsec_h6}. Panels and units as in Figure \ref{G1}.} \label{G6}
\end{figure*}

 \subsection{Group 7: Inner disk-like stars}\label{subsec_bulge}
 Our Group 7 (G7) contains the richest stars of the whole dataset in the iron-peak elements: [Fe/H] $\gtrsim -0.2$ and [Mn/Fe] $\gtrsim -0.2$, and also poorer in all the $\alpha$-elements than all the other groups. G7 has the chemical signature of inner disk-like stars. About half of the G7 stars have low energy values, and this subset is also concentrated at $V_{\phi} \sim$ 80 km s$^{-1}$,  eccentricity at $e \sim 0.82, Z_{max} \lesssim 2$ kpc,  $R_{apo} < 5$ kpc, and $R_{peri} < 0.5$ kpc. Nonetheless, this subset does not cluster in a particular locus within the G7 footprint in the t-SNE plane, and not all the stars around $V_{\phi}\sim$ 80 km s$^{-1}$ belong to this subset.  A non-negligible portion of the $V_{\phi} \sim$ 80 km s$^{-1}$ stars have higher energies and halo-like orbits. The energies of G7 stars in the $E_{char}$ versus $E_J$ plane fall within the footprint of MW Bulge GCs shown in \citet{2021ApJ...908L..42F}, their Figure 4 panel d. In the t-SNE runs, we noticed that G7 kept close to G1 in several tests but it did separate sometimes, while G1 - despite some internal clustering - always retained its integrity. This was a hint of G7 being different from G1, which was independently confirmed by its dynamics.

\begin{figure*}[h]
\includegraphics[width=18cm]{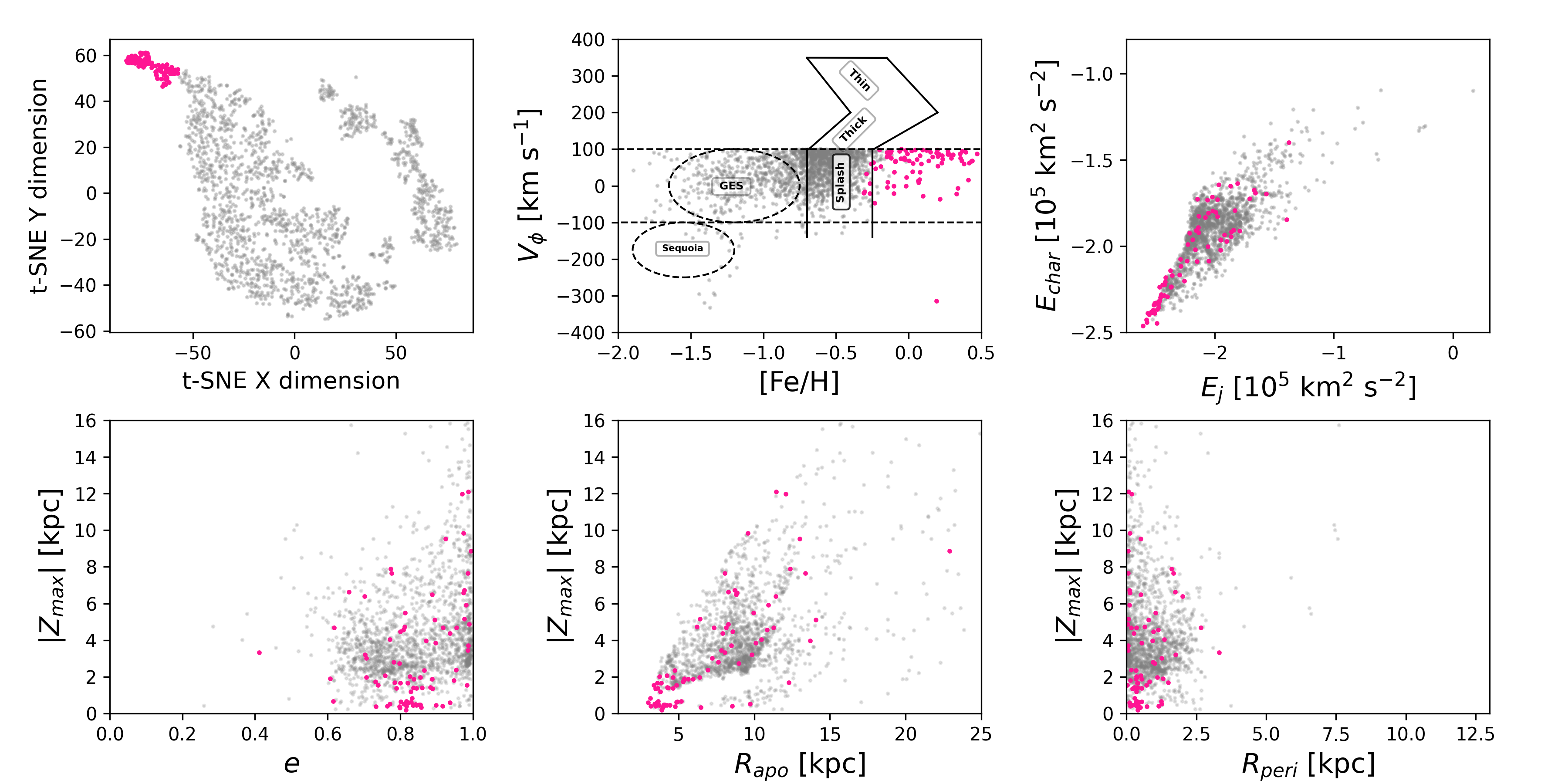}
\caption{Group 7 (G7), identified as inner disk-like stars. Details are explained in subsection \ref{subsec_bulge}. Panels and units as in Figure \ref{G1}.} \label{G7}
\end{figure*}

\section{Accretion origin for some structures}\label{sec_accretion}

\begin{figure}[h]
\includegraphics[scale=0.25]{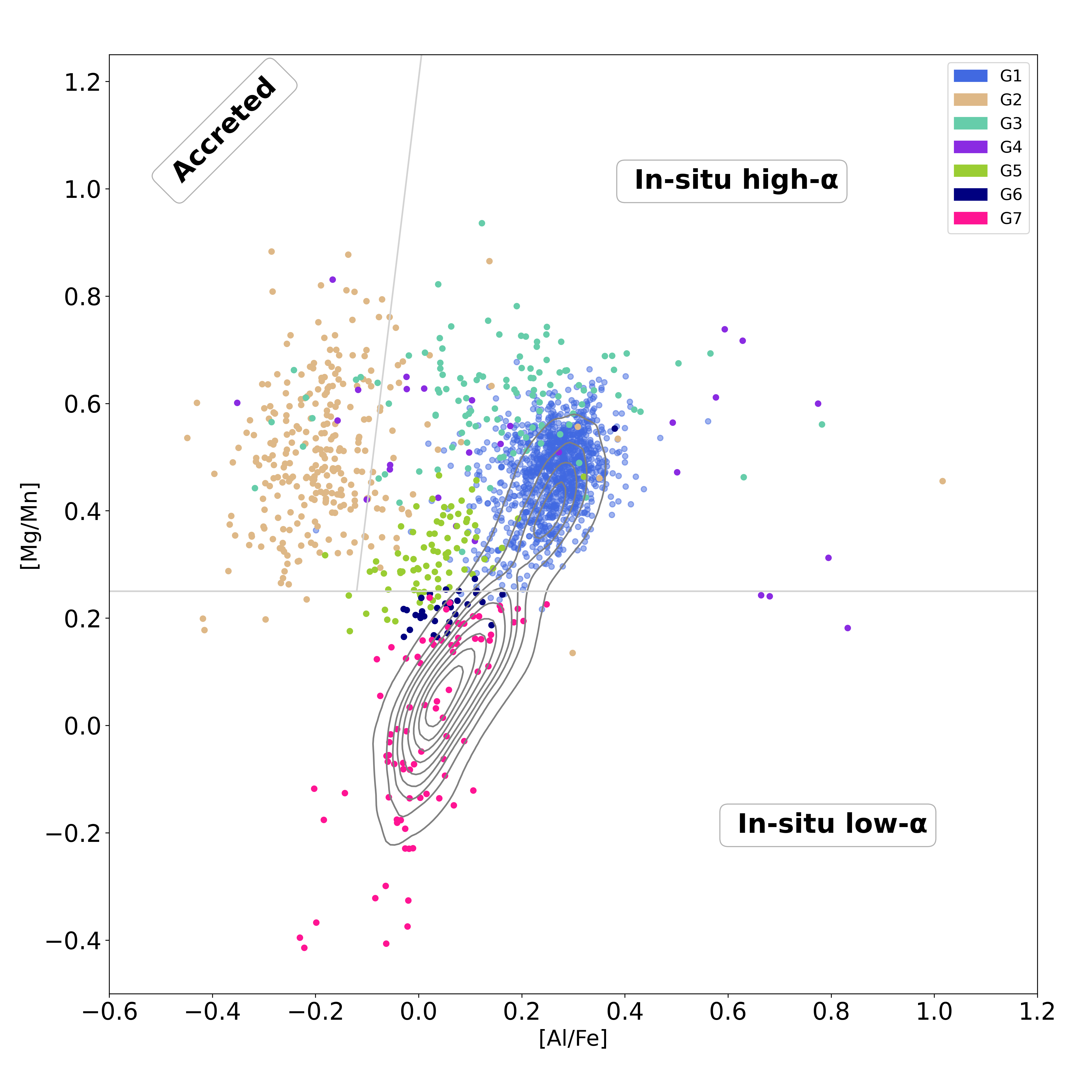}
\caption{[Mg/Mn] versus [Al/Fe] plot, color-coded for each detected group. A kernel density estimation (KDE) plot of MW disc stars is shown in grey curves. These were selected from APOGEE DR17 following the same quality criteria as our halo stars, and $V_\phi\geq 100$ km s$^{-1}$. Straight lines mark the boundaries of the accretion and in-situ high/low-$\alpha$ loci used in \citet{Das2020}. The G2 (GSE) locus is mostly limited in this plot by [Al/Fe] $<$ 0 and [Mg/Mn] $>$ +0.2.} \label{alvsmg_mn}
\end{figure}

The plot [Mg/Mn] versus [Al/Fe], shown in Figure \ref{alvsmg_mn}, has been used by several authors  \citep[e.g.,][]{Das2020, Naidu2020} as an empirical diagnostic for the origin of structures as either in-situ or accretion. Moreover, \citet{2021MNRAS.500.1385H} proves the validity of this plot with their chemical-evolution model. From the structures detected in this work, only G2 (which we identify as GSE) falls right on the locus of accreted structures in this plot, confirming its nature. Also, there are five G4 N-poor and C-poor stars that fall on this locus, which we postulate as GSE's GC first-generation debris. We also notice that, among the G2 stars, there are four N-rich and probable GSE's GC second-generation debris.

\begin{figure}[h]
\includegraphics[scale=0.25]{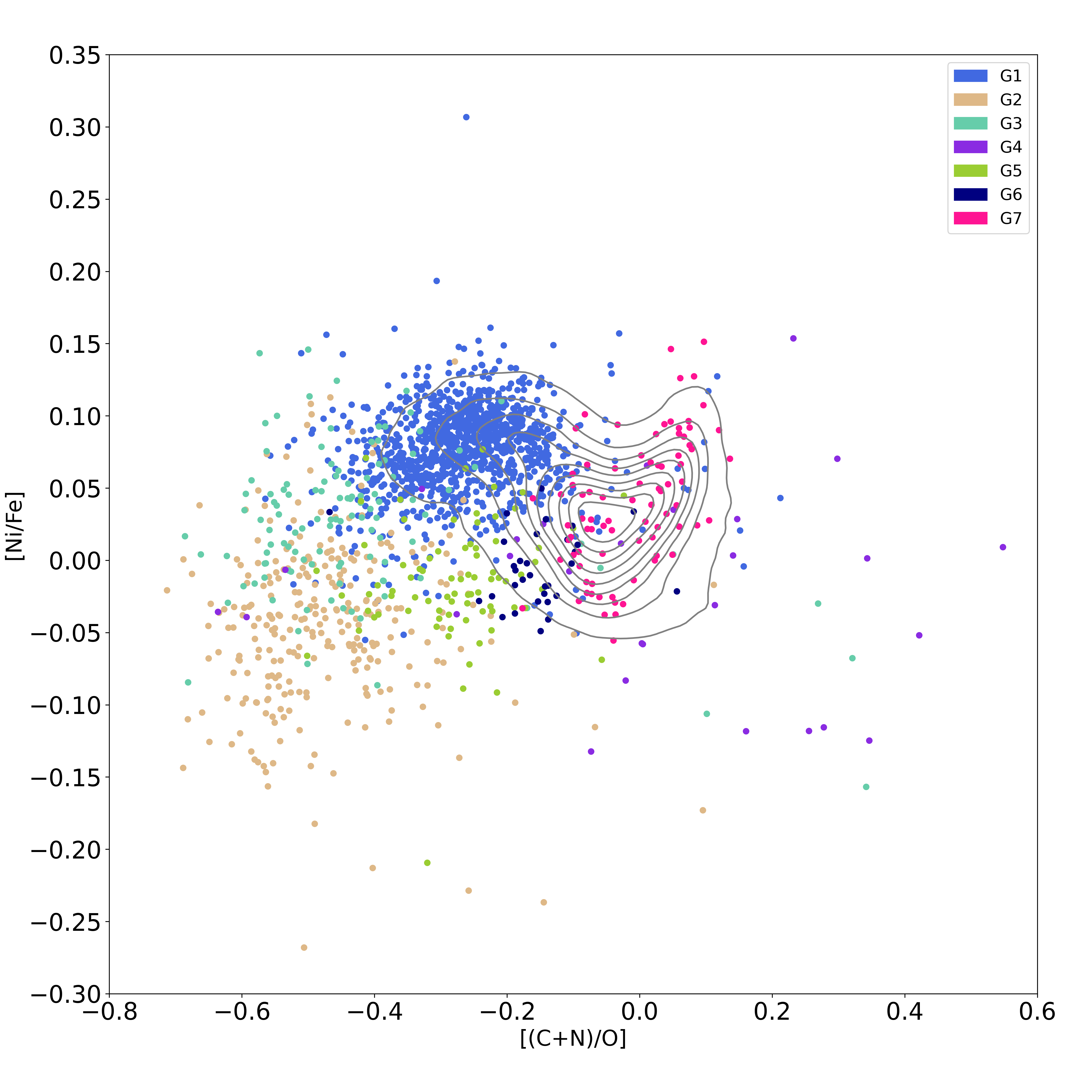}
\caption{[Ni/Fe] versus [(C$+$N)/O] plot, color-coded for each detected group. A kernel density estimation (KDE) plot of MW disc stars is shown in grey curves. These were selected from APOGEE DR17 following the same quality criteria as our halo stars, and $V_\phi\geq 100$ km s$^{-1}$. [Ni/Fe] $<$ 0 and [(C+N)/0] $< -0.2$ has been linked to ex-situ stars. In this work, G2 (Splash) and G5 (Galileo 5) fall in the locus of accreted stars. G6 (Galileo 6) overlaps with the thick-disc population.} \label{fig_niquel}
\end{figure}

As discussed in Sections \ref{subsec_h5} and \ref{subsec_h6}, the plot of [Ni/Fe] versus [(C$+$N)/O] has been used by \citet{2021NatAs...5..640M} to pinpoint the loci of identified in-situ and ex-situ stars, for which also precise seismic ages were calculated. The ex-situ stars are slightly younger than the in-situ halo stars. Low values of [Ni/Fe] with disc-like values of [Mn/Fe] have been linked to satellite galaxies of the MW with low star-formation rates, and such an abundance pattern has been observed in both G5 and G6 (see Figure \ref{fig_niquel}). Both reasons motivate us to postulate G5 and G6 as newly identified accreted structures in the MW inner halo. 

\section{The Galactic bar pattern speed $\Omega_{bar}$ role in the detected inner halo groups}\label{sec_barspeed}

In previous works studying peculiar chemical and/or kinematical structures in the solar vicinity  \citep[e.g.,][]{Naidu2020, 2022ApJ...926L..36N, 2023MNRAS.520.5671H}, the dynamical analysis was performed using axisymmetric gravitational potentials. This work is the first to analyze the role of a non-axisymmetric potential, in this case, the one induced by the bar on the dynamics of the inner-halo groups detected -- a relevant issue as they spend a good portion of their orbits in the volume occupied by the MW disc, therefore some effect is expected. Our analysis considered $\Omega_{bar}=31, 41,$ and $51$ km s$^{-1}$ kpc$^{-1}$. 
The results obtained for each of these values were very similar; we did not detect a visible or significant difference in the distribution of each of the dynamical parameters examined, $|Z_{max}|, R_{peri}, R_{apo}, e, E_j$, and $E_{char}$. Therefore, the bar pattern velocity does not have an effect on the dynamical behavior of the structures detected, and we conclude that its role seems to be absent or undetectable at the error level of the data.

\section{Conclusions}\label{sec_conclusion}
From this investigation, we conclude that:
\begin{itemize}

\item We have used high-quality APOGEE abundances and Gaia astrometric data for 1742 red giant stars located within 5 kpc of the Sun, in order to detect coherent structures associated in the chemical-kinematic-dynamical space. We limited our study to stars not rotating with the disc, i.e, $V_{\phi}<100$ km s$^{-1}$.

\item We determined orbital parameters using the non-axisymmetric galactic potential model {\tt GravPot16}, which, together with the stellar components for the disc and halo of the MW, also includes a rotating bar {{it boxy/peanut}}. The bar pattern velocity adopted was 41 km s$^{-1}$ kpc. They were also determined for velocities 31 and 51 km s$^{-1}$ kpc in order to measure the impact on the orbital parameters, since these stars may spend a significant portion of their orbits close to the Galactic plane. We did not detect significant variations in this regard.

\item The search for structures was performed with the nonlinear algorithm of dimensional reduction  known as t-SNE, using ten chemical-abundance ratios as input data: [Fe/H], [O/Fe], [Mg/Fe], [Si/Fe], [Ca/Fe], [C/Fe], [N/Fe], [Al/Fe], [Mn/Fe], and [Ni/Fe]. 

\item Seven structures were detected, of which the following were traced to previously identified populations in the literature: G1 $\sim$ Splash, G2 $\sim$ GSE, G3 $\sim$ high-$\alpha$ heated-disc population, G4 $\sim$ N-C-O peculiar stars, and G7 $\sim$ inner disk-like stars. In the many tests performed, t-SNE occasionally separated the metal-poor high-$\alpha$ plateau portion of GSE from the metal-rich lower-$\alpha$ declining section at the so-called {\it knee}. 

\item No known structures were detected to be unambiguously similar to groups Galileo 5 (G5) and Galileo 6 (G6), therefore we posit them as new structures found in this work. These groups have low values of [Ni/Fe] but [Mn/Fe] with thin disc-like chemical abundances, which hints at an ex-situ origin, i.e, an accreted satellite galaxy. Both G5 and G6 have Splash-like dynamics. G6 may also correspond to Eos, which was proposed by \citet{2022ApJ...938...21M} to be formed in-situ from gas contaminated by GSE. 

\item All groups except G4 (N-C-O peculiar stars) as detected by t-SNE in this work occupy distinctive loci in the [Mg/Fe], [Si/Fe], and [Al/Fe] versus [Fe/H] planes, with little overlap between the groups. We recommend using these abundances in future searches for these structures.

\item We report three new N-rich stars and probable GC debris: APOGEE IDs 2M06572697$+$5543115, 2M21372238$+$1244305, and 2M16254515-2620462.

\item The t-SNE algorithm was thoroughly tested on several issues. 1) A wide range of parameters were tested for the parameters {\tt perplexity, early exaggeration}, and {\tt number of iterations} to find those that produced the greatest visible separation in the t-SNE plane. 2) In those cases where the groups separation was not completely clean in the t-SNE plane, several procedures were applied: (i) t-SNE was applied iteratively, i.e., the outputs t-SNE X and t-SNE Y were introduced as two additional inputs for a second run of the method; (ii) t-SNE was applied to sub-samples in which two or more groups were suspected to coexist, to check if t-SNE separated them or not; and (iii) t-SNE was applied to the larger individual groups to determine whether or not they remained as a single entity. 3) Through Monte Carlo realizations considering the uncertainties in the abundances used, it was determined that the errors in these do not appreciably affect the groups detected by t-SNE.

\item The t-SNE algorithm proved to be very useful for exploring high-dimensional data sets, objectively separating real structures at various scales present in a large number of data, e.g., small hidden groups with a very dominant feature, or large ensembles with more dispersed properties that overlap in some variables with other structures and whose boundaries are less evident. In any case, it is recommended to cross-check t-SNE results with independent data (e.g., dynamics) or previous findings, and also gauge the stability of the structures detected, as described in the previous item.

\item A table with all the relevant data for the full dataset, including a tag for each of the seven groups detected, is available online. A sample is shown in Table \ref{ta:morse} in Appendix \ref{online_data}.
\end{itemize}

\begin{acknowledgements}  
This work has received funding from the grant support provided by Agencia Nacional de Investigaci\'on y Desarrollo de Chile (ANID) under the Proyecto Fondecyt Iniciaci\'on 2022 Agreement No. 11220340 (PI: Jos\'e G. Fern\'andez-Trincado) and from ANID under the Concurso de Fomento a la Vinculaci\'on Internacional para Instituciones de Investigaci\'on Regionales (Modalidad corta duraci\'on) Agreement No. FOVI210020 (PI: Jos\'e G. Fern\'andez-Trincado) and from the Joint Committee ESO-Government of Chile 2021 under the Agreement No. ORP 023/2021 (PI: Jos\'e G. Fern\'andez-Trincado) and from Becas Santander Movilidad Internacional Profesores 2022, Banco Santander Chile (PI: Jos\'e G. Fern\'andez-Trincado).\\

T.C.B. acknowledges partial support for this work from grant PHY 14-30152; Physics
Frontier Center/JINA Center for the Evolution of the Elements
(JINA-CEE), and from OISE-1927130: The International Research
Network for Nuclear Astrophysics (IReNA), awarded by the U.S. National Science
Foundation.\\

B.T. gratefully acknowledges support from the Natural Science Foundation of Guangdong Province under grant No. 2022A1515010732, the National Natural Science Foundation of China through grants No. 12233013, and the China Manned Space Project No. CMS-CSST-2021-B03.\\

Funding for the Sloan Digital Sky Survey IV has been provided by the Alfred P. Sloan Foundation, the U.S. Department of Energy Office of Science, and the Participating Institutions. SDSS-IV acknowledges support and resources from the Center for High-Performance Computing at the University of Utah. The SDSS website is www.sdss.org. SDSS-IV is managed by the Astrophysical Research Consortium for the Participating Institutions of the SDSS Collaboration including the Brazilian Participation Group, the Carnegie Institution for Science, Carnegie Mellon University, the Chilean Participation Group, the French Participation Group, Harvard-Smithsonian Center for Astrophysics, Instituto de Astrof\`{i}sica de Canarias, The Johns Hopkins University, Kavli Institute for the Physics and Mathematics of the Universe (IPMU) / University of Tokyo, Lawrence Berkeley National Laboratory, Leibniz Institut f\"{u}r Astrophysik Potsdam (AIP), Max-Planck-Institut f\"{u}r Astronomie (MPIA Heidelberg), Max-Planck-Institut f\"{u}r Astrophysik (MPA Garching), Max-Planck-Institut f\"{u}r Extraterrestrische Physik (MPE), National Astronomical Observatory of China, New Mexico State University, New York University, University of Notre Dame, Observat\'{o}rio Nacional / MCTI, The Ohio State University, Pennsylvania State University, Shanghai Astronomical Observatory, United Kingdom Participation Group, Universidad Nacional Aut\'{o}noma de M\'{e}xico, University of Arizona, University of Colorado Boulder, University of Oxford, University of Portsmouth, University of Utah, University of Virginia, University of Washington, University of Wisconsin, Vanderbilt University, and Yale University.\\

This work has made use of data from the European Space Agency (ESA) mission \textit{Gaia} (\url{http://www.cosmos.esa.int/gaia}), processed by the \textit{Gaia} Data Processing and Analysis Consortium (DPAC, \url{http://www.cosmos.esa.int/web/gaia/dpac/consortium}). Funding for the DPAC has been provided by national institutions, in particular, the institutions participating in the \textit{Gaia} Multilateral Agreement.\\\end{acknowledgements}
	
\begin{appendix}

\section{t-SNE tests}\label{sec_tests}
\begin{figure*}
\centering
\includegraphics[width=\textwidth]{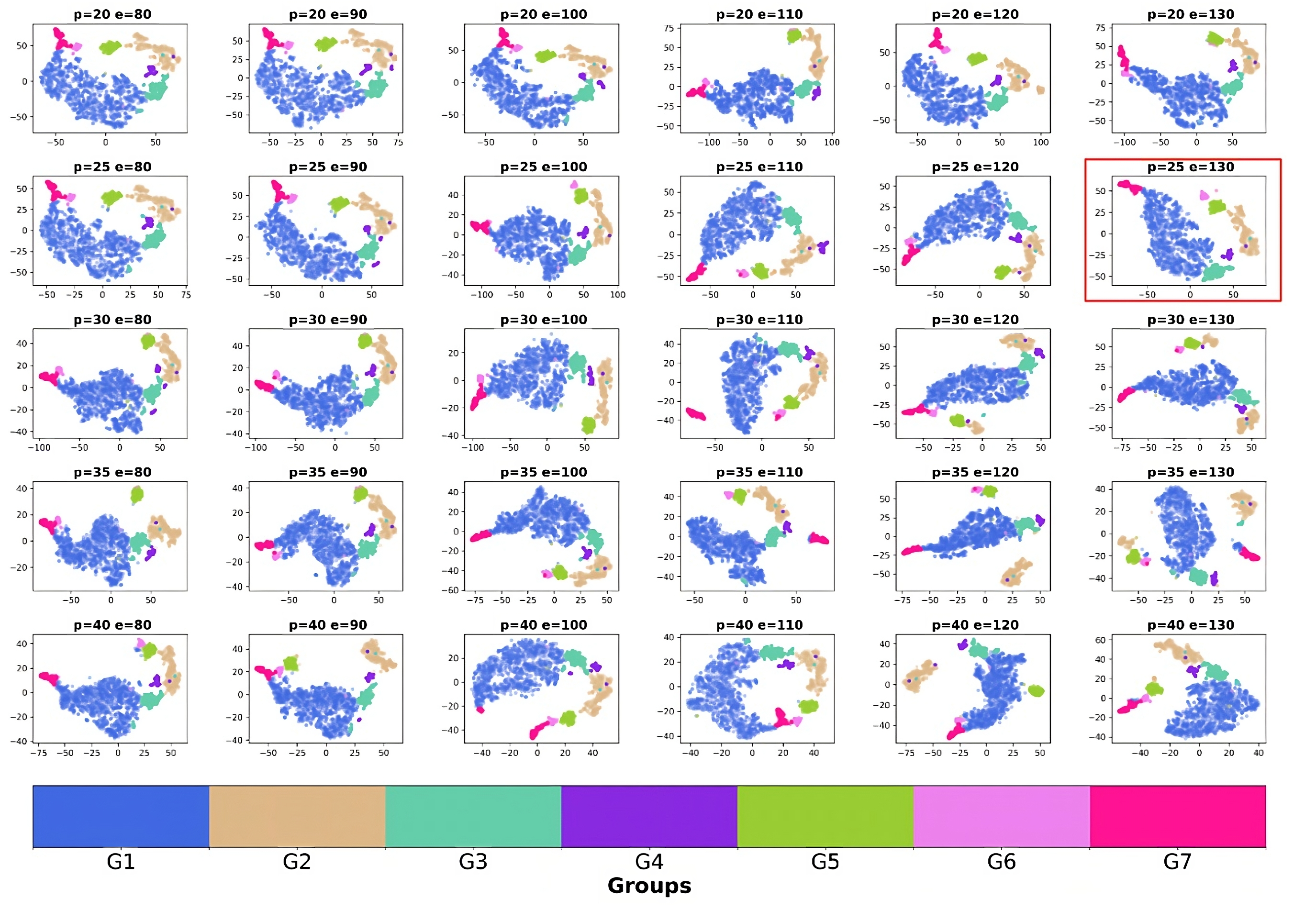}\caption{The t-SNE plane for different input parameter values, as indicated in each panel. Horizontal and vertical axes correspond, respectively, to the t-SNE X and Y dimensions.} The panel marked in red is the one chosen as final -- i.e. Figure \ref{fig_main_tsne} -- and best to select the structures found in this investigation.
\label{fig_tsne_params}
\end{figure*}

\begin{figure}
\centering
\includegraphics[scale=0.07]{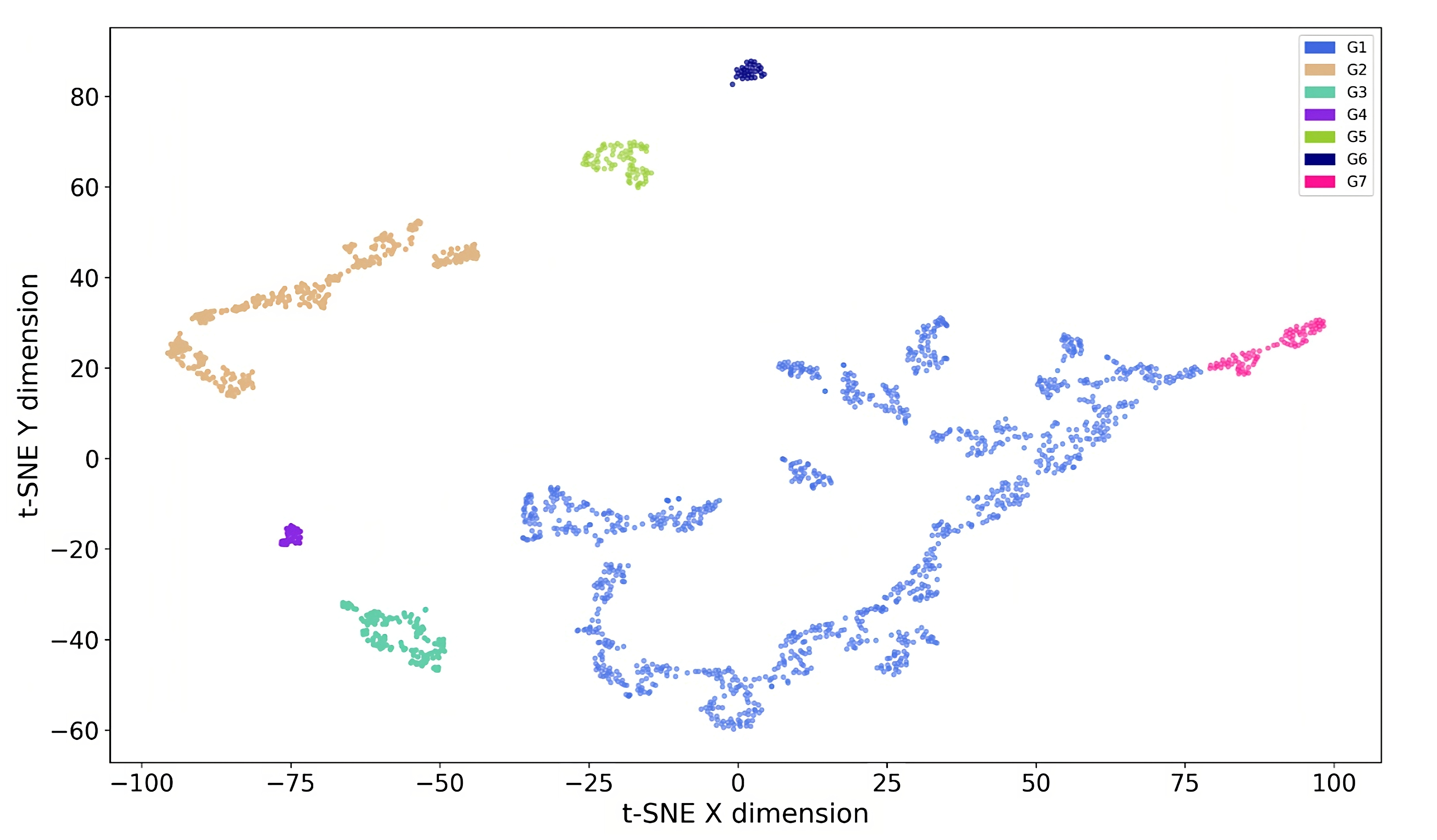}
\caption{The iterated t-SNE plane. Values of t-SNE X and t-SNE Y from Figure \ref{fig_main_tsne}-central panel were fed as additional data to another run of t-SNE, to check if adjacent groups in the original t-SNE plane were separated or not, and where the borders occur. Notice how the overall shape is kept, but in fact in this case t-SNE flipped the orientation of the points in the plane.}
\label{fig_tsne_iter}
\end{figure}

\begin{figure*}
\centering
\includegraphics[width=\textwidth]{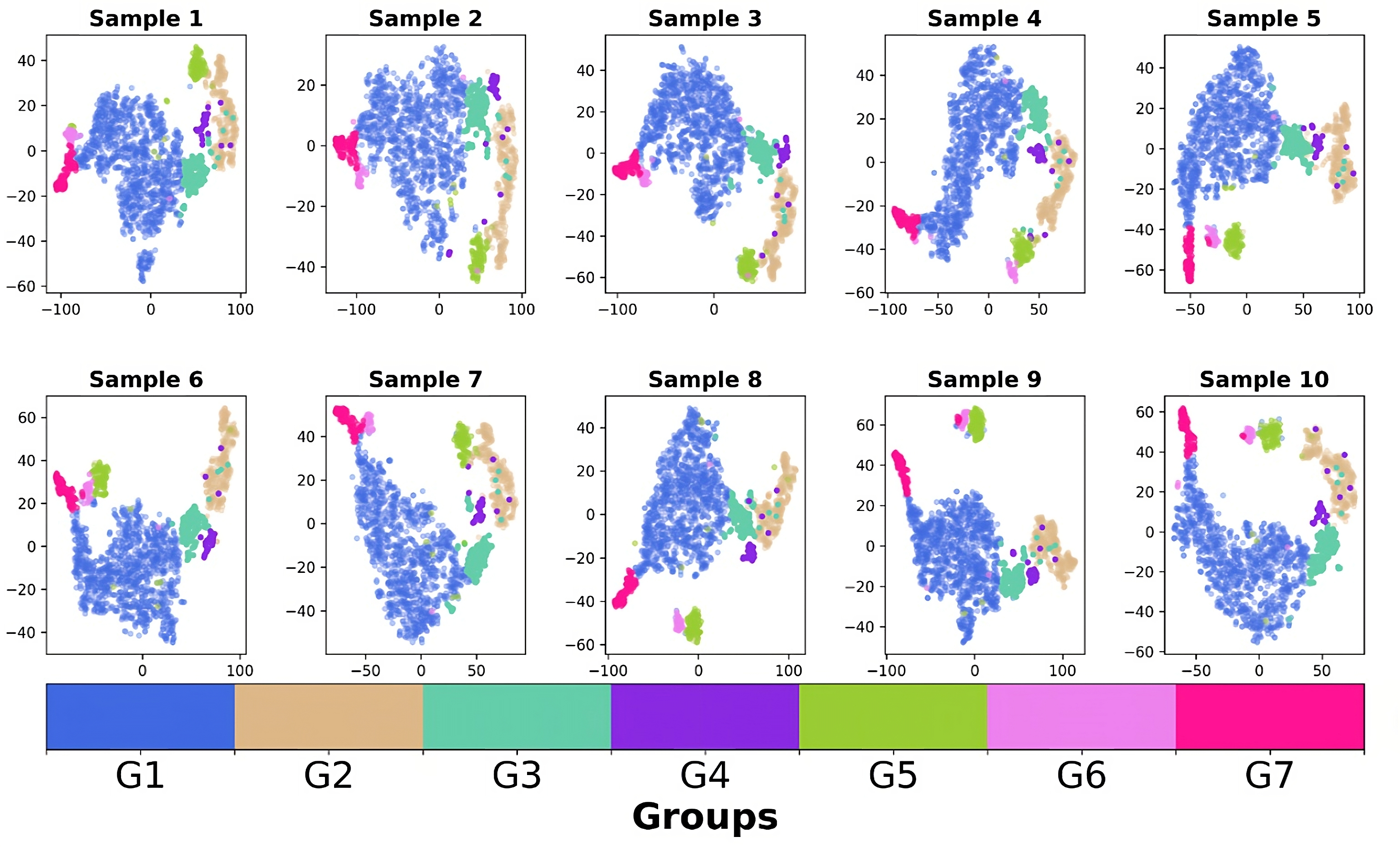}
\caption{Effect of abundance errors on the t-SNE plane. Horizontal and vertical axes correspond, respectively, to the t-SNE X and Y dimensions.} Monte-Carlo realizations of the abundances used were fed into the t-SNE algorithm to check if the groups as selected in Figure \ref{fig_main_tsne} were still associated and kept as coherent structures.
\label{fig_tsne_errors}
\end{figure*}

The following figures show the various tests performed with t-SNE to assess its performance, and to make the final selection of adjacent groups that were not clearly separated in the t-SNE plane chosen, as described in Section \ref{sec_strategy}. Figure \ref{fig_tsne_params} shows the effect of varying the input parameters: $25<$\texttt{perplexity}$<40$, and $80<$\texttt{early\_exageration}$<130$. Figure \ref{fig_tsne_iter} shows the {\it iterated} t-SNE plane, and Figure \ref{fig_tsne_errors} shows the effect of abundance errors. These tests helped to identify truly stable groups that t-SNE kept finding regardless of the changes introduced.

\section{Summary of chemical abundances and dynamical parameters of each detected group}\label{app_median}

Table \ref{tab:median_values} contains the median value and median absolute deviation Median$(X-\mbox{Median}(X))$ of each the abundances and dynamical parameters studied in this work. Table \ref{tab:orbit_percent} shows the percentage of prograde, P-R, and retrograde orbits per group. See Section \ref{sec_results} and subsections within for details.

\begin{table*}
\renewcommand*{\arraystretch}{1.25}
    \centering
    \resizebox{18cm}{!} {
    \begin{tabular}{c|r|r|r|r|r|r|r}
\hline
Item & G1\hspace{0.8cm} & G2\hspace{0.8cm} & G3\hspace{0.8cm} & G4\hspace{0.8cm} & G5\hspace{0.8cm} & G6\hspace{0.8cm} & G7\hspace{0.8cm} \\ \hline 
       [C/Fe] & +0.087 $\pm$0.035 & -0.366 $\pm$0.077 & -0.124 $\pm$0.095 & -0.239 $\pm$0.115 & -0.097 $\pm$0.037 
  & -0.019 $\pm$0.037 & +0.034 $\pm$0.034 \\
   $[$N/Fe$]$ & +0.120 $\pm$0.044 & +0.155 $\pm$0.099 & +0.195 $\pm$0.082 & +0.772 $\pm$0.153 & +0.134 $\pm$0.066 
  & +0.114 $\pm$0.027 & +0.235 $\pm$0.063 \\
   $[$O/Fe$]$ & +0.349 $\pm$0.040 & +0.305 $\pm$0.067 & +0.429 $\pm$0.071 & +0.212 $\pm$0.122 & +0.241 $\pm$0.046
  & +0.159 $\pm$0.028 & +0.068 $\pm$0.043 \\
  $[$Mg/Fe$]$ & +0.339 $\pm$0.026 & +0.195 $\pm$0.053 & +0.327 $\pm$0.028 & +0.176 $\pm$0.065 & +0.162 $\pm$0.040
  & +0.129 $\pm$0.018 & +0.052 $\pm$0.049 \\
  $[$Al/Fe$]$ & +0.265 $\pm$0.039 & -0.187 $\pm$0.062 & +0.170 $\pm$0.088 & +0.106 $\pm$0.215 & +0.031 $\pm$0.041
  & +0.052 $\pm$0.044 & +0.007 $\pm$0.056 \\
  $[$Si/Fe$]$ & +0.240 $\pm$0.027 & +0.201 $\pm$0.041 & +0.334 $\pm$0.045 & +0.178 $\pm$0.056 & +0.109 $\pm$0.020
  & +0.069 $\pm$0.014 & +0.031 $\pm$0.031 \\
  $[$Ca/Fe$]$ & +0.172 $\pm$0.034 & +0.186 $\pm$0.049 & +0.258 $\pm$0.029 & +0.229 $\pm$0.058 & +0.098 $\pm$0.033
  & +0.041 $\pm$0.031 & -0.025 $\pm$0.035 \\
  $[$Mn/Fe$]$ & -0.136 $\pm$0.034 & -0.297 $\pm$0.055 & -0.279 $\pm$0.054 & -0.321 $\pm$0.081 & -0.166 $\pm$0.025
  & -0.093 $\pm$0.015 & +0.026 $\pm$0.079 \\
  $[$Fe/H$]$  & -0.570 $\pm$0.103 & -1.230 $\pm$0.197 & -1.105 $\pm$0.105 & -1.027 $\pm$0.136 & -0.692 $\pm$0.042
  & -0.455 $\pm$0.031 & +0.058 $\pm$0.170 \\
  $[$Ni/Fe$]$ & +0.079 $\pm$0.017 & -0.040 $\pm$0.033 & +0.028 $\pm$0.030 & -0.007 $\pm$0.039 & -0.016 $\pm$0.020
  & -0.009 $\pm$0.019 & +0.039 $\pm$0.034 \\
  $[$Mg/Mn$]$ & +0.479 $\pm$0.050 & +0.497 $\pm$0.087 & +0.619 $\pm$0.055 & +0.541 $\pm$0.085 & +0.311 $\pm$0.048
  & +0.218 $\pm$0.022 & -0.007 $\pm$0.157 \\
   $R_{peri}$ & 0.808 $\pm$0.529 & 0.457  $\pm$0.360 & 0.688 $\pm$0.542 & 0.471 $\pm$0.406 & 0.414 $\pm$0.315 
   & 0.372 $\pm$0.211 & 0.391 $\pm$0.228 \\
    $R_{apo}$ & 8.673 $\pm$1.268 & 11.759 $\pm$2.724 & 8.890 $\pm$1.153 & 9.889 $\pm$2.340 & 10.212 $\pm$1.354
    & 9.116 $\pm$0.974 & 5.955 $\pm$2.256 \\
          $e$ & 0.811 $\pm$0.093 & 0.933 $\pm$0.053 & 0.830 $\pm$0.089 & 0.920 $\pm$0.070 & 0.923 $\pm$0.061
          & 0.922 $\pm$0.052 & 0.835 $\pm$0.059 \\
  $|Z_{max}|$ & 3.303 $\pm$0.914 & 5.065 $\pm$2.075 & 3.857 $\pm$1.202 & 3.681 $\pm$1.046 & 4.183 $\pm$1.183
    & 3.534 $\pm$0.998 & 1.894 $\pm$1.431 \\
       $E_j$ & -2.115 $\pm$0.100 & -1.789 $\pm$0.166 & -2.003 $\pm$0.154 & -1.934 $\pm$0.207 & -1.867 $\pm$0.116 
    & -1.877 $\pm$0.077 & -2.284 $\pm$0.218 \\
  $E_{char}$ & -1.927 $\pm$0.097 & -1.747 $\pm$0.125 & -1.887 $\pm$0.092 & -1.838 $\pm$0.132 & -1.813 $\pm$0.077
    & -1.898 $\pm$0.071 & -2.117 $\pm$0.211 \\
\hline
\end{tabular}
}
    \caption{Median and median absolute deviation of each of the chemical abundances and dynamical parameters investigated in this work, and plotted in Figs. \ref{fig_main_tsne} to \ref{fig_niquel}, for each detected group. $R_{peri}$, $R_{apo}$, and $|Z_{max}|$ are measured in kpc; $E_j$ and $E_{char}$ in  $10^5$ km$^2$ s$^{-2}$. The first column (Item) indicates the variable evaluated.}
    \label{tab:median_values}
\end{table*}

\begin{table}
\renewcommand*{\arraystretch}{1.25}
    \centering
    \begin{tabular}{c|c|c|c|c|c}
    G & Name & \# of P & \# of P-R & \# of R  & Total\\ \hline 
    1 & Splash &    851 (76\%) &   175 (16\%) &  99 (8\%) & 1125   \\
    2 & GSE &    136 (49\%) &   55 (20\%)  &  89 (31\%) & 280 \\
    3 & High-$\alpha$  &    61  (54\%) &   19 (17\%)  &  34 (29\%) & 114   \\ 
    4 & N-rich  &    13  (46\%) &   9  (32\%)  &  6 (22\%)  &  28 \\ 
    5 & Galileo 5 &    39  (49\%) &   17 (21\%)  &  24 (30\%) & 80    \\
    6 & Galileo 6 &    14  (44\%) &   6 (19\%)   &  12 (37\%) & 32   \\
    7 & Inner Disk &    64  (77\%) &   10 (12\%)   &  9 (11\%) &  83  \\
    \end{tabular}
    \caption{Number and percentage of stars in each group detected according to their orbit orientation. G$=$group number assigned in this work, P$=$prograde, P-R$=$prograde-retrograde, R$=$retrograde.}
    \label{tab:orbit_percent}
\end{table}

\section{Supplementary online data}\label{online_data}
A catalog with all the relevant information, including a tag for each detected group, is available online. A sample table showing the column labels is shown in Table \ref{ta:morse}.

\begin{table*}
\caption{Relevant information for each star used in this investigation, including a tag for each detected group.}
\resizebox{19cm}{!} {
\scriptsize{\begin{tabular}{clll}

\hline
ID \# & Column name & Units & Column Description \\
\hline
1 & APOGEE\_ID                    &                              & APOGEE id\\
2 & RA                            & deg                          & $\alpha(J2000)$\\
3 & DEC                           & deg                          & $\delta(J2000)$\\
4 & SNR                           & pixel$^{-1}$                 & Spectral signal-to-noise\\
5 & RV                            & km s$^{-1}$                  & APOGEE-2 radial velocity\\
6 & VSCATTER                      & km s$^{-1}$                  & APOGEE-2 radial velocity scatter\\
7 & LOGG                          & [cgs]                        & Surface gravity\\
8 & ERROR\_LOGG                   & [cgs]                        & Uncertainty in LOGG\\
9 & C\_FE                         &                              & [C/Fe] from ASPCAP\\
10 & ERROR\_C\_FE                 &                              & Uncertainty in [C/Fe]\\
11 & N\_FE                        &                              & [N/Fe] from ASPCAP\\
12 & ERROR\_N\_FE                 &                              & Uncertainty in [N/Fe]\\
13 & O\_FE                        &                              & [O/Fe] from ASPCAP\\
14 & ERROR\_O\_FE                 &                              & Uncertainty in [O/Fe]\\
15 & MG\_FE                       &                              & [Mg/Fe] from ASPCAP\\
16 & ERROR\_MG\_FE                &                              & Uncertainty in [Mg/Fe]\\
17 & AL\_FE                       &                              & [Al/Fe] from ASPCAP\\
18 & ERROR\_AL\_FE                &                              & Uncertainty in [Al/Fe]\\
19 & SI\_FE                       &                              & [Si/Fe] from ASPCAP\\
20 & ERROR\_SI\_FE                &                              & Uncertainty in [Si/Fe]\\
21 & CA\_FE                       &                              & [Ca/Fe] from ASPCAP\\
22 & ERROR\_CA\_FE                &                              & Uncertainty in [Ca/Fe]\\
23 & MN\_FE                       &                              & [Mn/Fe] from ASPCAP\\
24 & ERROR\_MN\_FE                &                              & Uncertainty in [Mn/Fe]\\
25 & FE\_H                        &                              & [Fe/H] from ASPCAP\\
26 & ERROR\_FE\_H                 &                              & Uncertainty in [Fe/H]\\
27 & NI\_FE                       &                              & [Ni/Fe] from ASPCAP\\
28 & ERROR\_NI\_FE                &                              & Uncertainty in [Ni/Fe]\\
29 & VR                           & km s$^{-1}$                  & Galactocentric radial velocity\\
30 & VPHI                         & km s$^{-1}$                  & Galactocentric azimuthal velocity\\
31 & PERIGALACTICON               & kpc                          & Perigalactocentric distance\\
32 & ERROR\_PERIGALACTICON        & kpc                          & Uncertainty in PERIGALACTICON\\
33 & APOGALACTICON                & kpc                          & Apogalactocentric distance\\
34 & ERROR\_APOGALACTICON         & kpc                          & Uncertainty in APOGALACTICON\\
35 & Eccentricity                 &                              & Orbital eccentricity\\
36 & ERROR\_Eccentricity          &                              & Uncertainty in eccentricity\\
37 & $|Zmax|$                     & kpc                          & Maximum vertical excursion from the Galactic plane\\
38 & ERROR\_Zmax                  & kpc                          & Uncertainty in Zmax\\
39 & $E_{j}$                      & $10^5$ km$^2$ s$^{-2}$       & Jacobi energy\\
40 & ERROR\_$E_{j}$               & $10^5$ km$^2$ s$^{-2}$       & Uncertainty in Jacobi energy\\
41 & $E_{char}$                   & $10^5$ km$^2$ s$^{-2}$       & Characteristic orbital energy\\
42 & ERROR\_$E_{char}$            & $10^5$ km$^2$ s$^{-2}$       & Uncertainty in  characteristic orbital energy\\
43 & $L_{zmin}$                   & $10^1$ km$^2$ s$^{-2}$ kpc   & Minimum $z$-component of angular momentum\\
44 & ERROR\_$L_{zmin}$            & $10^1$ km$^2$ s$^{-2}$ kpc   & Uncertainty in minimum $z$-component of angular momentum\\
45 & $L_{zmax}$                   & $10^1$ km$^2$ s$^{-2}$ kpc   & Maximum $z$-component of angular momentum\\
46 & ERROR\_$L_{zmax}$            & $10^1$ km$^2$ s$^{-2}$ kpc   & Uncertainty in maximum $z$-component of angular momentum\\
47 & GAIADR3\_SOURCE\_ID          &                              & GAIA DR3 source id\\
48 & ruwe                         &                              & Gaia DR3 renormalized unit weight error\\
49 & d\_STARHORSE                 & kpc                          & Bayesian StarHorse distance, 50th percentile\\
50 & ERROR\_d\_STARHORSE          & kpc                          & Uncertainty in d\_STARHORSE, ERROR = (84th-16th )/2 percentile\\
51 & pmRA\_GaiaDR3                & mas yr$^{-1}$                & $\mu_{\alpha}$ cos ($\delta$) from Gaia DR3\\
52 & ERROR\_pmRA\_GaiaDR3         & mas yr$^{-1}$                & Uncertainty in $\mu\alpha$ cos ($\delta$) from Gaia EDR3\\
53 & pmDEC\_GaiaDR3               & mas yr$^{-1}$                & $\mu_{\alpha}$ from Gaia EDR3\\
54 & ERROR\_pmDEC\_GaiaDR3        & mas yr$^{-1}$                & Uncertainty in $\mu\alpha$ from Gaia EDR3\\
55 & Group                        &                              & Group as determined by t-SNE     
\\
\hline
\\
\end{tabular}}
}
\label{ta:morse}
\end{table*}

\end{appendix}

\begin{thebibliography}{110}
	\expandafter\ifx\csname natexlab\endcsname\relax\def\natexlab#1{#1}\fi
	
	\bibitem[{{Abdurro'uf} {et~al.}(2022){Abdurro'uf}, {Accetta}, {Aerts}, {Silva
			Aguirre}, {Ahumada}, {Ajgaonkar}, {Filiz Ak}, {Alam}, {Allende Prieto},
		{Almeida}, {Anders}, {Anderson}, {Andrews}, {Anguiano}, {Aquino-Ort{\'\i}z},
		{Arag{\'o}n-Salamanca}, {Argudo-Fern{\'a}ndez}, {Ata}, {Aubert},
		{Avila-Reese}, {Badenes}, {Barb{\'a}}, {Barger}, {Barrera-Ballesteros},
		{Beaton}, {Beers}, {Belfiore}, {Bender}, {Bernardi}, {Bershady}, {Beutler},
		{Bidin}, {Bird}, {Bizyaev}, {Blanc}, {Blanton}, {Boardman}, {Bolton},
		{Boquien}, {Borissova}, {Bovy}, {Brandt}, {Brown}, {Brownstein}, {Brusa},
		{Buchner}, {Bundy}, {Burchett}, {Bureau}, {Burgasser}, {Cabang}, {Campbell},
		{Cappellari}, {Carlberg}, {Wanderley}, {Carrera}, {Cash}, {Chen}, {Chen},
		{Cherinka}, {Chiappini}, {Choi}, {Chojnowski}, {Chung}, {Clerc}, {Cohen},
		{Comerford}, {Comparat}, {da Costa}, {Covey}, {Crane}, {Cruz-Gonzalez},
		{Culhane}, {Cunha}, {Dai}, {Damke}, {Darling}, {Davidson}, {Davies},
		{Dawson}, {De Lee}, {Diamond-Stanic}, {Cano-D{\'\i}az}, {S{\'a}nchez},
		{Donor}, {Duckworth}, {Dwelly}, {Eisenstein}, {Elsworth}, {Emsellem},
		{Eracleous}, {Escoffier}, {Fan}, {Farr}, {Feng}, {Fern{\'a}ndez-Trincado},
		{Feuillet}, {Filipp}, {Fillingham}, {Frinchaboy}, {Fromenteau}, {Galbany},
		{Garc{\'\i}a}, {Garc{\'\i}a-Hern{\'a}ndez}, {Ge}, {Geisler}, {Gelfand},
		{G{\'e}ron}, {Gibson}, {Goddy}, {Godoy-Rivera}, {Grabowski}, {Green},
		{Greener}, {Grier}, {Griffith}, {Guo}, {Guy}, {Hadjara}, {Harding},
		{Hasselquist}, {Hayes}, {Hearty}, {Hern{\'a}ndez}, {Hill}, {Hogg},
		{Holtzman}, {Horta}, {Hsieh}, {Hsu}, {Hsu}, {Huber}, {Huertas-Company},
		{Hutchinson}, {Hwang}, {Ibarra-Medel}, {Chitham}, {Ilha}, {Imig}, {Jaekle},
		{Jayasinghe}, {Ji}, {Johnson}, {Jones}, {J{\"o}nsson}, {Katkov}, {Khalatyan},
		{Kinemuchi}, {Kisku}, {Knapen}, {Kneib}, {Kollmeier}, {Kong}, {Kounkel},
		{Kreckel}, {Krishnarao}, {Lacerna}, {Lane}, {Langgin}, {Lavender}, {Law},
		{Lazarz}, {Leung}, {Leung}, {Lewis}, {Li}, {Li}, {Lian}, {Liang}, {Lin},
		{Lin}, {Lin}, {Lintott}, {Long}, {Longa-Pe{\~n}a}, {L{\'o}pez-Cob{\'a}},
		{Lu}, {Lundgren}, {Luo}, {Mackereth}, {de la Macorra}, {Mahadevan},
		{Majewski}, {Manchado}, {Mandeville}, {Maraston}, {Margalef-Bentabol},
		{Masseron}, {Masters}, {Mathur}, {McDermid}, {Mckay}, {Merloni},
		{Merrifield}, {Meszaros}, {Miglio}, {Di Mille}, {Minniti}, {Minsley},
		{Monachesi}, {Moon}, {Mosser}, {Mulchaey}, {Muna}, {Mu{\~n}oz}, {Myers},
		{Myers}, {Nadathur}, {Nair}, {Nandra}, {Neumann}, {Newman}, {Nidever},
		{Nikakhtar}, {Nitschelm}, {O'Connell}, {Garma-Oehmichen}, {Luan Souza de
			Oliveira}, {Olney}, {Oravetz}, {Ortigoza-Urdaneta}, {Osorio}, {Otter},
		{Pace}, {Padilla}, {Pan}, {Pan}, {Parikh}, {Parker}, {Peirani}, {Pe{\~n}a
			Ram{\'\i}rez}, {Penny}, {Percival}, {Perez-Fournon}, {Pinsonneault},
		{Poidevin}, {Poovelil}, {Price-Whelan}, {B{\'a}rbara de Andrade Queiroz},
		{Raddick}, {Ray}, {Rembold}, {Riddle}, {Riffel}, {Riffel}, {Rix}, {Robin},
		{Rodr{\'\i}guez-Puebla}, {Roman-Lopes}, {Rom{\'a}n-Z{\'u}{\~n}iga}, {Rose},
		{Ross}, {Rossi}, {Rubin}, {Salvato}, {S{\'a}nchez}, {S{\'a}nchez-Gallego},
		{Sanderson}, {Santana Rojas}, {Sarceno}, {Sarmiento}, {Sayres}, {Sazonova},
		{Schaefer}, {Schiavon}, {Schlegel}, {Schneider}, {Schultheis}, {Schwope},
		{Serenelli}, {Serna}, {Shao}, {Shapiro}, {Sharma}, {Shen}, {Shetrone}, {Shu},
		{Simon}, {Skrutskie}, {Smethurst}, {Smith}, {Sobeck}, {Spoo}, {Sprague},
		{Stark}, {Stassun}, {Steinmetz}, {Stello}, {Stone-Martinez},
		{Storchi-Bergmann}, {Stringfellow}, {Stutz}, {Su}, {Taghizadeh-Popp},
		{Talbot}, {Tayar}, {Telles}, {Teske}, {Thakar}, {Theissen}, {Tkachenko},
		{Thomas}, {Tojeiro}, {Hernandez Toledo}, {Troup}, {Trump}, {Trussler},
		{Turner}, {Tuttle}, {Unda-Sanzana}, {V{\'a}zquez-Mata}, {Valentini},
		{Valenzuela}, {Vargas-Gonz{\'a}lez}, {Vargas-Maga{\~n}a}, {Alfaro},
		{Villanova}, {Vincenzo}, {Wake}, {Warfield}, {Washington}, {Weaver},
		{Weijmans}, {Weinberg}, {Weiss}, {Westfall}, {Wild}, {Wilde}, {Wilson},
		{Wilson}, {Wilson}, {Wolf}, {Wood-Vasey}, {Yan}, {Zamora}, {Zasowski},
		{Zhang}, {Zhao}, {Zheng}, {Zheng}, \& {Zhu}}]{Abdurro2022}
	{Abdurro'uf}, {Accetta}, K., {Aerts}, C., {et~al.} 2022, \apjs, 259, 35
	
	\bibitem[{{Allende Prieto} {et~al.}(2006){Allende Prieto}, {Beers}, {Wilhelm},
		{Newberg}, {Rockosi}, {Yanny}, \& {Lee}}]{Allende2006}
	{Allende Prieto}, C., {Beers}, T.~C., {Wilhelm}, R., {et~al.} 2006, \apj, 636,
	804
	
	\bibitem[{{Amarante} {et~al.}(2020){Amarante}, {Beraldo e Silva}, {Debattista},
		\& {Smith}}]{2020ApJ...891L..30A}
	{Amarante}, J. A.~S., {Beraldo e Silva}, L., {Debattista}, V.~P., \& {Smith},
	M.~C. 2020, \apjl, 891, L30
	
	\bibitem[{{Anders} {et~al.}(2018){Anders}, {Chiappini}, {Santiago},
		{Matijevi{\v{c}}}, {Queiroz}, {Steinmetz}, \& {Guiglion}}]{Anders2018}
	{Anders}, F., {Chiappini}, C., {Santiago}, B.~X., {et~al.} 2018, \aap, 619,
	A125
	
	\bibitem[{{Anders} {et~al.}(2019){Anders}, {Khalatyan}, {Chiappini}, {Queiroz},
		{Santiago}, {Jordi}, {Girardi}, {Brown}, {Matijevi{\v{c}}}, {Monari},
		{Cantat-Gaudin}, {Weiler}, {Khan}, {Miglio}, {Carrillo}, {Romero-G{\'o}mez},
		{Minchev}, {de Jong}, {Antoja}, {Ramos}, {Steinmetz}, \& {Enke}}]{Anders2019}
	{Anders}, F., {Khalatyan}, A., {Chiappini}, C., {et~al.} 2019, \aap, 628, A94
	
	\bibitem[{{Barb{\'a}} {et~al.}(2019){Barb{\'a}}, {Minniti}, {Geisler},
		{Alonso-Garc{\'\i}a}, {Hempel}, {Monachesi}, {Arias}, \&
		{G{\'o}mez}}]{Barba2019}
	{Barb{\'a}}, R.~H., {Minniti}, D., {Geisler}, D., {et~al.} 2019, \apjl, 870,
	L24
	
	\bibitem[{{Baumgardt} \& {Vasiliev}(2021)}]{2021-Baumgardt-Vasiliev}
	{Baumgardt}, H. \& {Vasiliev}, E. 2021, \mnras, 505, 5957
	
	\bibitem[{{Beaton} {et~al.}(2021){Beaton}, {Oelkers}, {Hayes}, {Covey},
		{Chojnowski}, {De Lee}, {Sobeck}, {Majewski}, {Cohen},
		{Fern{\'a}ndez-Trincado}, {Longa-Pe{\~n}a}, {O'Connell}, {Santana},
		{Stringfellow}, {Zasowski}, {Aerts}, {Anguiano}, {Bender}, {Ca{\~n}as},
		{Cunha}, {Donor}, {Fleming}, {Frinchaboy}, {Feuillet}, {Harding},
		{Hasselquist}, {Holtzman}, {Johnson}, {Kollmeier}, {Kounkel}, {Mahadevan},
		{Price-Whelan}, {Rojas-Arriagada}, {Rom{\'a}n-Z{\'u}{\~n}iga}, {Schlafly},
		{Schultheis}, {Shetrone}, {Simon}, {Stassun}, {Stutz}, {Tayar}, {Teske},
		{Tkachenko}, {Troup}, {Albareti}, {Bizyaev}, {Bovy}, {Burgasser}, {Comparat},
		{Downes}, {Geisler}, {Inno}, {Manchado}, {Ness}, {Pinsonneault}, {Prada},
		{Roman-Lopes}, {Simonian}, {Smith}, {Yan}, \& {Zamora}}]{Beaton2021}
	{Beaton}, R.~L., {Oelkers}, R.~J., {Hayes}, C.~R., {et~al.} 2021, \aj, 162, 302
	
	\bibitem[{{Beers} {et~al.}(2012){Beers}, {Carollo}, {Ivezi{\'c}}, {An},
		{Chiba}, {Norris}, {Freeman}, {Lee}, {Munn}, {Re Fiorentin}, {Sivarani},
		{Wilhelm}, {Yanny}, \& {York}}]{Beers2012d}
	{Beers}, T.~C., {Carollo}, D., {Ivezi{\'c}}, {\v{Z}}., {et~al.} 2012, \apj,
	746, 34
	
	\bibitem[{{Beers} \& {Christlieb}(2005)}]{2005ARA&A..43..531B}
	{Beers}, T.~C. \& {Christlieb}, N. 2005, \araa, 43, 531
	
	\bibitem[{{Belokurov} {et~al.}(2018){Belokurov}, {Erkal}, {Evans}, {Koposov},
		\& {Deason}}]{Belokurov2018}
	{Belokurov}, V., {Erkal}, D., {Evans}, N.~W., {Koposov}, S.~E., \& {Deason},
	A.~J. 2018, \mnras, 478, 611
	
	\bibitem[{{Belokurov} {et~al.}(2020){Belokurov}, {Sanders}, {Fattahi}, {Smith},
		{Deason}, {Evans}, \& {Grand}}]{2020MNRAS.494.3880B}
	{Belokurov}, V., {Sanders}, J.~L., {Fattahi}, A., {et~al.} 2020, \mnras, 494,
	3880
	
	\bibitem[{{Blanton} {et~al.}(2017){Blanton}, {Bershady}, {Abolfathi},
		{Albareti}, {Allende Prieto}, {Almeida}, {Alonso-Garc{\'\i}a}, {Anders},
		{Anderson}, {Andrews}, {Aquino-Ort{\'\i}z}, {Arag{\'o}n-Salamanca},
		{Argudo-Fern{\'a}ndez}, {Armengaud}, {Aubourg}, {Avila-Reese}, {Badenes},
		{Bailey}, {Barger}, {Barrera-Ballesteros}, {Bartosz}, {Bates}, {Baumgarten},
		{Bautista}, {Beaton}, {Beers}, {Belfiore}, {Bender}, {Berlind}, {Bernardi},
		{Beutler}, {Bird}, {Bizyaev}, {Blanc}, {Blomqvist}, {Bolton}, {Boquien},
		{Borissova}, {van den Bosch}, {Bovy}, {Brandt}, {Brinkmann}, {Brownstein},
		{Bundy}, {Burgasser}, {Burtin}, {Busca}, {Cappellari}, {Delgado Carigi},
		{Carlberg}, {Carnero Rosell}, {Carrera}, {Chanover}, {Cherinka}, {Cheung},
		{G{\'o}mez Maqueo Chew}, {Chiappini}, {Choi}, {Chojnowski}, {Chuang},
		{Chung}, {Cirolini}, {Clerc}, {Cohen}, {Comparat}, {da Costa}, {Cousinou},
		{Covey}, {Crane}, {Croft}, {Cruz-Gonzalez}, {Garrido Cuadra}, {Cunha},
		{Damke}, {Darling}, {Davies}, {Dawson}, {de la Macorra}, {Dell'Agli}, {De
			Lee}, {Delubac}, {Di Mille}, {Diamond-Stanic}, {Cano-D{\'\i}az}, {Donor},
		{Downes}, {Drory}, {du Mas des Bourboux}, {Duckworth}, {Dwelly}, {Dyer},
		{Ebelke}, {Eigenbrot}, {Eisenstein}, {Emsellem}, {Eracleous}, {Escoffier},
		{Evans}, {Fan}, {Fern{\'a}ndez-Alvar}, {Fernandez-Trincado}, {Feuillet},
		{Finoguenov}, {Fleming}, {Font-Ribera}, {Fredrickson}, {Freischlad},
		{Frinchaboy}, {Fuentes}, {Galbany}, {Garcia-Dias},
		{Garc{\'\i}a-Hern{\'a}ndez}, {Gaulme}, {Geisler}, {Gelfand},
		{Gil-Mar{\'\i}n}, {Gillespie}, {Goddard}, {Gonzalez-Perez}, {Grabowski},
		{Green}, {Grier}, {Gunn}, {Guo}, {Guy}, {Hagen}, {Hahn}, {Hall}, {Harding},
		{Hasselquist}, {Hawley}, {Hearty}, {Gonzalez Hern{\'a}ndez}, {Ho}, {Hogg},
		{Holley-Bockelmann}, {Holtzman}, {Holzer}, {Huehnerhoff}, {Hutchinson},
		{Hwang}, {Ibarra-Medel}, {da Silva Ilha}, {Ivans}, {Ivory}, {Jackson},
		{Jensen}, {Johnson}, {Jones}, {J{\"o}nsson}, {Jullo}, {Kamble}, {Kinemuchi},
		{Kirkby}, {Kitaura}, {Klaene}, {Knapp}, {Kneib}, {Kollmeier}, {Lacerna},
		{Lane}, {Lang}, {Law}, {Lazarz}, {Lee}, {Le Goff}, {Liang}, {Li}, {Li},
		{Lian}, {Lima}, {Lin}, {Lin}, {Bertran de Lis}, {Liu}, {de Icaza Lizaola},
		{Long}, {Lucatello}, {Lundgren}, {MacDonald}, {Deconto Machado}, {MacLeod},
		{Mahadevan}, {Geimba Maia}, {Maiolino}, {Majewski}, {Malanushenko},
		{Malanushenko}, {Manchado}, {Mao}, {Maraston}, {Marques-Chaves}, {Masseron},
		{Masters}, {McBride}, {McDermid}, {McGrath}, {McGreer}, {Medina Pe{\~n}a},
		{Melendez}, {Merloni}, {Merrifield}, {Meszaros}, {Meza}, {Minchev},
		{Minniti}, {Miyaji}, {More}, {Mulchaey}, {M{\"u}ller-S{\'a}nchez}, {Muna},
		{Munoz}, {Myers}, {Nair}, {Nandra}, {Correa do Nascimento}, {Negrete},
		{Ness}, {Newman}, {Nichol}, {Nidever}, {Nitschelm}, {Ntelis}, {O'Connell},
		{Oelkers}, {Oravetz}, {Oravetz}, {Pace}, {Padilla}, {Palanque-Delabrouille},
		{Alonso Palicio}, {Pan}, {Parejko}, {Parikh}, {P{\^a}ris}, {Park}, {Patten},
		{Peirani}, {Pellejero-Ibanez}, {Penny}, {Percival}, {Perez-Fournon},
		{Petitjean}, {Pieri}, {Pinsonneault}, {Pisani}, {Poleski}, {Prada},
		{Prakash}, {Queiroz}, {Raddick}, {Raichoor}, {Barboza Rembold}, {Richstein},
		{Riffel}, {Riffel}, {Rix}, {Robin}, {Rockosi}, {Rodr{\'\i}guez-Torres},
		{Roman-Lopes}, {Rom{\'a}n-Z{\'u}{\~n}iga}, {Rosado}, {Ross}, {Rossi}, {Ruan},
		{Ruggeri}, {Rykoff}, {Salazar-Albornoz}, {Salvato}, {S{\'a}nchez}, {Aguado},
		{S{\'a}nchez-Gallego}, {Santana}, {Santiago}, {Sayres}, {Schiavon}, {da Silva
			Schimoia}, {Schlafly}, {Schlegel}, {Schneider}, {Schultheis}, {Schuster},
		{Schwope}, {Seo}, {Shao}, {Shen}, {Shetrone}, {Shull}, {Simon}, {Skinner},
		{Skrutskie}, {Slosar}, {Smith}, {Sobeck}, {Sobreira}, {Somers}, {Souto},
		{Stark}, {Stassun}, {Stauffer}, {Steinmetz}, {Storchi-Bergmann},
		{Streblyanska}, {Stringfellow}, {Su{\'a}rez}, {Sun}, {Suzuki}, {Szigeti},
		{Taghizadeh-Popp}, {Tang}, {Tao}, {Tayar}, {Tembe}, {Teske}, {Thakar},
		{Thomas}, {Thompson}, {Tinker}, {Tissera}, {Tojeiro}, {Hernandez Toledo}, {de
			la Torre}, {Tremonti}, {Troup}, {Valenzuela}, {Martinez Valpuesta},
		{Vargas-Gonz{\'a}lez}, {Vargas-Maga{\~n}a}, {Vazquez}, {Villanova}, {Vivek},
		{Vogt}, {Wake}, {Walterbos}, {Wang}, {Weaver}, {Weijmans}, {Weinberg},
		{Westfall}, {Whelan}, {Wild}, {Wilson}, {Wood-Vasey}, {Wylezalek}, {Xiao},
		{Yan}, {Yang}, {Ybarra}, {Y{\`e}che}, {Zakamska}, {Zamora}, {Zarrouk},
		{Zasowski}, {Zhang}, {Zhao}, {Zheng}, {Zheng}, {Zhou}, {Zhou}, {Zhu},
		{Zoccali}, \& {Zou}}]{Blanton2017}
	{Blanton}, M.~R., {Bershady}, M.~A., {Abolfathi}, B., {et~al.} 2017, \aj, 154,
	28
	
	\bibitem[{{Bonaca} {et~al.}(2020){Bonaca}, {Conroy}, {Cargile}, {Naidu},
		{Johnson}, {Zaritsky}, {Ting}, {Caldwell}, {Han}, \& {van
			Dokkum}}]{2020ApJ...897L..18B}
	{Bonaca}, A., {Conroy}, C., {Cargile}, P.~A., {et~al.} 2020, \apjl, 897, L18
	
	\bibitem[{{Bonaca} {et~al.}(2017){Bonaca}, {Conroy}, {Wetzel}, {Hopkins}, \&
		{Kere{\v{s}}}}]{2017ApJ...845..101B}
	{Bonaca}, A., {Conroy}, C., {Wetzel}, A., {Hopkins}, P.~F., \& {Kere{\v{s}}},
	D. 2017, \apj, 845, 101
	
	\bibitem[{{Bovy} {et~al.}(2019){Bovy}, {Leung}, {Hunt}, {Mackereth},
		{Garc{\'\i}a-Hern{\'a}ndez}, \& {Roman-Lopes}}]{Bovy2019}
	{Bovy}, J., {Leung}, H.~W., {Hunt}, J. A.~S., {et~al.} 2019, \mnras, 490, 4740
	
	\bibitem[{{Bowen} \& {Vaughan}(1973)}]{Bowen1973}
	{Bowen}, I.~S. \& {Vaughan}, A.~H., J. 1973, \ao, 12, 1430
	
	\bibitem[{{Brunthaler} {et~al.}(2011){Brunthaler}, {Reid}, {Menten}, {Zheng},
		{Bartkiewicz}, {Choi}, {Dame}, {Hachisuka}, {Immer}, {Moellenbrock},
		{Moscadelli}, {Rygl}, {Sanna}, {Sato}, {Wu}, {Xu}, \&
		{Zhang}}]{Brunthaler2011}
	{Brunthaler}, A., {Reid}, M.~J., {Menten}, K.~M., {et~al.} 2011, Astronomische
	Nachrichten, 332, 461
	
	\bibitem[{{Buder} {et~al.}(2022){Buder}, {Lind}, {Ness}, {Feuillet}, {Horta},
		{Monty}, {Buck}, {Nordlander}, {Bland-Hawthorn}, {Casey}, {de Silva},
		{D'Orazi}, {Freeman}, {Hayden}, {Kos}, {Martell}, {Lewis}, {Lin},
		{Schlesinger}, {Sharma}, {Simpson}, {Stello}, {Zucker}, {Zwitter},
		{Ciuc{\u{a}}}, {Horner}, {Kobayashi}, {Ting}, {Wyse}, \& {Wyse}}]{Buder2022}
	{Buder}, S., {Lind}, K., {Ness}, M.~K., {et~al.} 2022, \mnras, 510, 2407
	
	\bibitem[{{Carollo} {et~al.}(2010){Carollo}, {Beers}, {Chiba}, {Norris},
		{Freeman}, {Lee}, {Ivezi{\'c}}, {Rockosi}, \& {Yanny}}]{Carollo2010}
	{Carollo}, D., {Beers}, T.~C., {Chiba}, M., {et~al.} 2010, \apj, 712, 692
	
	\bibitem[{{Carollo} {et~al.}(2007){Carollo}, {Beers}, {Lee}, {Chiba}, {Norris},
		{Wilhelm}, {Sivarani}, {Marsteller}, {Munn}, {Bailer-Jones}, {Fiorentin}, \&
		{York}}]{Carollo2007}
	{Carollo}, D., {Beers}, T.~C., {Lee}, Y.~S., {et~al.} 2007, \nat, 450, 1020
	
	\bibitem[{{Chiba} \& {Beers}(2000)}]{2000AJ....119.2843C}
	{Chiba}, M. \& {Beers}, T.~C. 2000, \aj, 119, 2843
	
	\bibitem[{{Combes}(2017)}]{2017sf2a.conf..223C}
	{Combes}, F. 2017, in SF2A-2017: Proceedings of the Annual meeting of the
	French Society of Astronomy and Astrophysics, ed. C.~{Reyl{\'e}}, P.~{Di
		Matteo}, F.~{Herpin}, E.~{Lagadec}, A.~{Lan{\c{c}}on}, Z.~{Meliani}, \&
	F.~{Royer}, Di
	
	\bibitem[{{Cunha} {et~al.}(2017){Cunha}, {Smith}, {Hasselquist}, {Souto},
		{Shetrone}, {Allende Prieto}, {Bizyaev}, {Frinchaboy},
		{Garc{\'\i}a-Hern{\'a}ndez}, {Holtzman}, {Johnson}, {J{\H{o}}nsson},
		{Majewski}, {M{\'e}sz{\'a}ros}, {Nidever}, {Pinsonneault}, {Schiavon},
		{Sobeck}, {Skrutskie}, {Zamora}, {Zasowski}, \&
		{Fern{\'a}ndez-Trincado}}]{Cunha2017}
	{Cunha}, K., {Smith}, V.~V., {Hasselquist}, S., {et~al.} 2017, \apj, 844, 145
	
	\bibitem[{{Das} {et~al.}(2020){Das}, {Hawkins}, \& {Jofr{\'e}}}]{Das2020}
	{Das}, P., {Hawkins}, K., \& {Jofr{\'e}}, P. 2020, \mnras, 493, 5195
	
	\bibitem[{{Di Matteo} {et~al.}(2019){Di Matteo}, {Haywood}, {Lehnert}, {Katz},
		{Khoperskov}, {Snaith}, {G{\'o}mez}, \& {Robichon}}]{2019A&A...632A...4D}
	{Di Matteo}, P., {Haywood}, M., {Lehnert}, M.~D., {et~al.} 2019, \aap, 632, A4
	
	\bibitem[{{Einasto}(1979)}]{Einasto1979}
	{Einasto}, J. 1979, in The Large-Scale Characteristics of the Galaxy, ed. W.~B.
	{Burton}, Vol.~84, 451
	
	\bibitem[{{Fehlberg}(1968)}]{fehlberg68}
	{Fehlberg}, E. 1968, NASA Technical Report, 315
	
	\bibitem[{{Fern{\'a}ndez-Trincado} {et~al.}(2022){Fern{\'a}ndez-Trincado},
		{Beers}, {Barbuy}, {Minniti}, {Chiappini}, {Garro}, {Tang}, {Alves-Brito},
		{Villanova}, {Geisler}, {Lane}, \& {Diaz}}]{Trincado11}
	{Fern{\'a}ndez-Trincado}, J.~G., {Beers}, T.~C., {Barbuy}, B., {et~al.} 2022,
	\aap, 663, A126
	
	\bibitem[{{Fern{\'a}ndez-Trincado}
		{et~al.}(2020{\natexlab{a}}){Fern{\'a}ndez-Trincado}, {Beers}, {Minniti},
		{Carigi}, {Barbuy}, {Placco}, {Moni Bidin}, {Villanova}, {Roman-Lopes}, \&
		{Nitschelm}}]{Trincado7}
	{Fern{\'a}ndez-Trincado}, J.~G., {Beers}, T.~C., {Minniti}, D., {et~al.}
	2020{\natexlab{a}}, \apjl, 903, L17
	
	\bibitem[{{Fern{\'a}ndez-Trincado}
		{et~al.}(2021{\natexlab{a}}){Fern{\'a}ndez-Trincado}, {Beers}, {Minniti},
		{Carigi}, {Placco}, {Chun}, {Lane}, {Geisler}, {Villanova}, {Souza},
		{Barbuy}, {P{\'e}rez-Villegas}, {Chiappini}, {Queiroz}, {Tang},
		{Alonso-Garc{\'\i}a}, {Piatti}, {Palma}, {Alves-Brito}, {Moni Bidin},
		{Roman-Lopes}, {Mu{\~n}oz}, {Singh}, {Kundu}, {Chaves-Velasquez},
		{Romero-Colmenares}, {Longa-Pe{\~n}a}, {Soto}, \& {Vieira}}]{Trincado8}
	{Fern{\'a}ndez-Trincado}, J.~G., {Beers}, T.~C., {Minniti}, D., {et~al.}
	2021{\natexlab{a}}, \aap, 647, A64
	
	\bibitem[{{Fern{\'a}ndez-Trincado}
		{et~al.}(2021{\natexlab{b}}){Fern{\'a}ndez-Trincado}, {Beers}, {Minniti},
		{Moni Bidin}, {Barbuy}, {Villanova}, {Geisler}, {Lane}, {Roman-Lopes}, \&
		{Bizyaev}}]{Trincado9}
	{Fern{\'a}ndez-Trincado}, J.~G., {Beers}, T.~C., {Minniti}, D., {et~al.}
	2021{\natexlab{b}}, \aap, 648, A70
	
	\bibitem[{{Fern{\'a}ndez-Trincado}
		{et~al.}(2020{\natexlab{b}}){Fern{\'a}ndez-Trincado}, {Beers}, {Minniti},
		{Tang}, {Villanova}, {Geisler}, {P{\'e}rez-Villegas}, \&
		{Vieira}}]{Trincado6}
	{Fern{\'a}ndez-Trincado}, J.~G., {Beers}, T.~C., {Minniti}, D., {et~al.}
	2020{\natexlab{b}}, \aap, 643, L4
	
	\bibitem[{{Fern{\'a}ndez-Trincado}
		{et~al.}(2021{\natexlab{c}}){Fern{\'a}ndez-Trincado}, {Beers}, {Queiroz},
		{Chiappini}, {Minniti}, {Barbuy}, {Majewski}, {Ortigoza-Urdaneta}, {Moni
			Bidin}, {Robin}, {Moreno}, {Chaves-Velasquez}, {Villanova}, {Lane}, {Pan}, \&
		{Bizyaev}}]{Trincado10}
	{Fern{\'a}ndez-Trincado}, J.~G., {Beers}, T.~C., {Queiroz}, A. B.~A., {et~al.}
	2021{\natexlab{c}}, \apjl, 918, L37
	
	\bibitem[{{Fern{\'a}ndez-Trincado}
		{et~al.}(2019{\natexlab{a}}){Fern{\'a}ndez-Trincado}, {Beers}, {Tang},
		{Moreno}, {P{\'e}rez-Villegas}, \& {Ortigoza-Urdaneta}}]{Trincado3}
	{Fern{\'a}ndez-Trincado}, J.~G., {Beers}, T.~C., {Tang}, B., {et~al.}
	2019{\natexlab{a}}, \mnras, 488, 2864
	
	\bibitem[{{Fern{\'a}ndez-Trincado}
		{et~al.}(2020{\natexlab{c}}){Fern{\'a}ndez-Trincado}, {Chaves-Velasquez},
		{P{\'e}rez-Villegas}, {Vieira}, {Moreno}, {Ortigoza-Urdaneta}, \&
		{Vega-Neme}}]{Trincado5}
	{Fern{\'a}ndez-Trincado}, J.~G., {Chaves-Velasquez}, L., {P{\'e}rez-Villegas},
	A., {et~al.} 2020{\natexlab{c}}, \mnras, 495, 4113
	
	\bibitem[{{Fern{\'a}ndez-Trincado}
		{et~al.}(2019{\natexlab{b}}){Fern{\'a}ndez-Trincado}, {Mennickent},
		{Cabezas}, {Zamora}, {Martell}, {Beers}, {Placco}, {Nataf},
		{M{\'e}sz{\'a}ros}, {Minniti}, {Schleicher}, {Tang}, {P{\'e}rez-Villegas},
		{Robin}, \& {Reyl{\'e}}}]{Trincado4}
	{Fern{\'a}ndez-Trincado}, J.~G., {Mennickent}, R., {Cabezas}, M., {et~al.}
	2019{\natexlab{b}}, \aap, 631, A97
	
	\bibitem[{{Fern{\'a}ndez-Trincado}
		{et~al.}(2020{\natexlab{d}}){Fern{\'a}ndez-Trincado}, {Minniti}, {Beers},
		{Villanova}, {Geisler}, {Souza}, {Smith}, {Placco}, {Vieira},
		{P{\'e}rez-Villegas}, {Barbuy}, {Alves-Brito}, {Bidin}, {Alonso-Garc{\'\i}a},
		{Tang}, \& {Palma}}]{Fernandez-Trincado2020}
	{Fern{\'a}ndez-Trincado}, J.~G., {Minniti}, D., {Beers}, T.~C., {et~al.}
	2020{\natexlab{d}}, \aap, 643, A145
	
	\bibitem[{{Fern{\'a}ndez-Trincado}
		{et~al.}(2021{\natexlab{d}}){Fern{\'a}ndez-Trincado}, {Minniti}, {Souza},
		{Beers}, {Geisler}, {Moni Bidin}, {Villanova}, {Majewski}, {Barbuy},
		{P{\'e}rez-Villegas}, {Henao}, {Romero-Colmenares}, {Roman-Lopes}, \&
		{Lane}}]{2021ApJ...908L..42F}
	{Fern{\'a}ndez-Trincado}, J.~G., {Minniti}, D., {Souza}, S.~O., {et~al.}
	2021{\natexlab{d}}, \apjl, 908, L42
	
	\bibitem[{{Fern{\'a}ndez-Trincado} {et~al.}(2016){Fern{\'a}ndez-Trincado},
		{Robin}, {Moreno}, {Schiavon}, {Garc{\'\i}a P{\'e}rez}, {Vieira}, {Cunha},
		{Zamora}, {Sneden}, {Souto}, {Carrera}, {Johnson}, {Shetrone}, {Zasowski},
		{Garc{\'\i}a-Hern{\'a}ndez}, {Majewski}, {Reyl{\'e}}, {Blanco-Cuaresma},
		{Martinez-Medina}, {P{\'e}rez-Villegas}, {Valenzuela}, {Pichardo}, {Meza},
		{M{\'e}sz{\'a}ros}, {Sobeck}, {Geisler}, {Anders}, {Schultheis}, {Tang},
		{Roman-Lopes}, {Mennickent}, {Pan}, {Nitschelm}, \& {Allard}}]{Trincado1}
	{Fern{\'a}ndez-Trincado}, J.~G., {Robin}, A.~C., {Moreno}, E., {et~al.} 2016,
	\apj, 833, 132
	
	\bibitem[{{Fern{\'a}ndez-Trincado} {et~al.}(2017){Fern{\'a}ndez-Trincado},
		{Zamora}, {Garc{\'\i}a-Hern{\'a}ndez}, {Souto}, {Dell'Agli}, {Schiavon},
		{Geisler}, {Tang}, {Villanova}, {Hasselquist}, {Mennickent}, {Cunha},
		{Shetrone}, {Allende Prieto}, {Vieira}, {Zasowski}, {Sobeck}, {Hayes},
		{Majewski}, {Placco}, {Beers}, {Schleicher}, {Robin}, {M{\'e}sz{\'a}ros},
		{Masseron}, {Garc{\'\i}a P{\'e}rez}, {Anders}, {Meza}, {Alves-Brito},
		{Carrera}, {Minniti}, {Lane}, {Fern{\'a}ndez-Alvar}, {Moreno}, {Pichardo},
		{P{\'e}rez-Villegas}, {Schultheis}, {Roman-Lopes}, {Fuentes}, {Nitschelm},
		{Harding}, {Bizyaev}, {Pan}, {Oravetz}, {Simmons}, {Ivans},
		{Blanco-Cuaresma}, {Hern{\'a}ndez}, {Alonso-Garc{\'\i}a}, {Valenzuela}, \&
		{Chanam{\'e}}}]{Trincado2}
	{Fern{\'a}ndez-Trincado}, J.~G., {Zamora}, O., {Garc{\'\i}a-Hern{\'a}ndez},
	D.~A., {et~al.} 2017, \apjl, 846, L2
	
	\bibitem[{{Feuillet} {et~al.}(2021){Feuillet}, {Sahlholdt}, {Feltzing}, \&
		{Casagrande}}]{Feuillet2021}
	{Feuillet}, D.~K., {Sahlholdt}, C.~L., {Feltzing}, S., \& {Casagrande}, L.
	2021, \mnras, 508, 1489
	
	\bibitem[{{Fiteni} {et~al.}(2021){Fiteni}, {Caruana}, {Amarante}, {Debattista},
		\& {Beraldo e Silva}}]{2021MNRAS.503.1418F}
	{Fiteni}, K., {Caruana}, J., {Amarante}, J. A.~S., {Debattista}, V.~P., \&
	{Beraldo e Silva}, L. 2021, \mnras, 503, 1418
	
	\bibitem[{{Freeman} \& {Bland-Hawthorn}(2002)}]{Freeman2002}
	{Freeman}, K. \& {Bland-Hawthorn}, J. 2002, \araa, 40, 487
	
	\bibitem[{{Gaia Collaboration} {et~al.}(2021){Gaia Collaboration}, {Brown},
		{Vallenari}, {Prusti}, {de Bruijne}, {Babusiaux}, {Biermann}, {Creevey},
		{Evans}, {Eyer}, {Hutton}, {Jansen}, {Jordi}, {Klioner}, {Lammers},
		{Lindegren}, {Luri}, {Mignard}, {Panem}, {Pourbaix}, {Randich}, {Sartoretti},
		{Soubiran}, {Walton}, {Arenou}, {Bailer-Jones}, {Bastian}, {Cropper},
		{Drimmel}, {Katz}, {Lattanzi}, {van Leeuwen}, {Bakker}, {Cacciari},
		{Casta{\~n}eda}, {De Angeli}, {Ducourant}, {Fabricius}, {Fouesneau},
		{Fr{\'e}mat}, {Guerra}, {Guerrier}, {Guiraud}, {Jean-Antoine Piccolo},
		{Masana}, {Messineo}, {Mowlavi}, {Nicolas}, {Nienartowicz}, {Pailler},
		{Panuzzo}, {Riclet}, {Roux}, {Seabroke}, {Sordo}, {Tanga}, {Th{\'e}venin},
		{Gracia-Abril}, {Portell}, {Teyssier}, {Altmann}, {Andrae}, {Bellas-Velidis},
		{Benson}, {Berthier}, {Blomme}, {Brugaletta}, {Burgess}, {Busso}, {Carry},
		{Cellino}, {Cheek}, {Clementini}, {Damerdji}, {Davidson}, {Delchambre},
		{Dell'Oro}, {Fern{\'a}ndez-Hern{\'a}ndez}, {Galluccio}, {Garc{\'\i}a-Lario},
		{Garcia-Reinaldos}, {Gonz{\'a}lez-N{\'u}{\~n}ez}, {Gosset}, {Haigron},
		{Halbwachs}, {Hambly}, {Harrison}, {Hatzidimitriou}, {Heiter},
		{Hern{\'a}ndez}, {Hestroffer}, {Hodgkin}, {Holl}, {Jan{\ss}en}, {Jevardat de
			Fombelle}, {Jordan}, {Krone-Martins}, {Lanzafame}, {L{\"o}ffler}, {Lorca},
		{Manteiga}, {Marchal}, {Marrese}, {Moitinho}, {Mora}, {Muinonen}, {Osborne},
		{Pancino}, {Pauwels}, {Petit}, {Recio-Blanco}, {Richards}, {Riello},
		{Rimoldini}, {Robin}, {Roegiers}, {Rybizki}, {Sarro}, {Siopis}, {Smith},
		{Sozzetti}, {Ulla}, {Utrilla}, {van Leeuwen}, {van Reeven}, {Abbas}, {Abreu
			Aramburu}, {Accart}, {Aerts}, {Aguado}, {Ajaj}, {Altavilla}, {{\'A}lvarez},
		{{\'A}lvarez Cid-Fuentes}, {Alves}, {Anderson}, {Anglada Varela}, {Antoja},
		{Audard}, {Baines}, {Baker}, {Balaguer-N{\'u}{\~n}ez}, {Balbinot}, {Balog},
		{Barache}, {Barbato}, {Barros}, {Barstow}, {Bartolom{\'e}}, {Bassilana},
		{Bauchet}, {Baudesson-Stella}, {Becciani}, {Bellazzini}, {Bernet}, {Bertone},
		{Bianchi}, {Blanco-Cuaresma}, {Boch}, {Bombrun}, {Bossini}, {Bouquillon},
		{Bragaglia}, {Bramante}, {Breedt}, {Bressan}, {Brouillet}, {Bucciarelli},
		{Burlacu}, {Busonero}, {Butkevich}, {Buzzi}, {Caffau}, {Cancelliere},
		{C{\'a}novas}, {Cantat-Gaudin}, {Carballo}, {Carlucci}, {Carnerero},
		{Carrasco}, {Casamiquela}, {Castellani}, {Castro-Ginard}, {Castro Sampol},
		{Chaoul}, {Charlot}, {Chemin}, {Chiavassa}, {Cioni}, {Comoretto}, {Cooper},
		{Cornez}, {Cowell}, {Crifo}, {Crosta}, {Crowley}, {Dafonte}, {Dapergolas},
		{David}, {David}, {de Laverny}, {De Luise}, {De March}, {De Ridder}, {de
			Souza}, {de Teodoro}, {de Torres}, {del Peloso}, {del Pozo}, {Delbo},
		{Delgado}, {Delgado}, {Delisle}, {Di Matteo}, {Diakite}, {Diener},
		{Distefano}, {Dolding}, {Eappachen}, {Edvardsson}, {Enke}, {Esquej}, {Fabre},
		{Fabrizio}, {Faigler}, {Fedorets}, {Fernique}, {Fienga}, {Figueras},
		{Fouron}, {Fragkoudi}, {Fraile}, {Franke}, {Gai}, {Garabato},
		{Garcia-Gutierrez}, {Garc{\'\i}a-Torres}, {Garofalo}, {Gavras}, {Gerlach},
		{Geyer}, {Giacobbe}, {Gilmore}, {Girona}, {Giuffrida}, {Gomel}, {Gomez},
		{Gonzalez-Santamaria}, {Gonz{\'a}lez-Vidal}, {Granvik},
		{Guti{\'e}rrez-S{\'a}nchez}, {Guy}, {Hauser}, {Haywood}, {Helmi}, {Hidalgo},
		{Hilger}, {H{\l}adczuk}, {Hobbs}, {Holland}, {Huckle}, {Jasniewicz},
		{Jonker}, {Juaristi Campillo}, {Julbe}, {Karbevska}, {Kervella}, {Khanna},
		{Kochoska}, {Kontizas}, {Kordopatis}, {Korn}, {Kostrzewa-Rutkowska},
		{Kruszy{\'n}ska}, {Lambert}, {Lanza}, {Lasne}, {Le Campion}, {Le Fustec},
		{Lebreton}, {Lebzelter}, {Leccia}, {Leclerc}, {Lecoeur-Taibi}, {Liao},
		{Licata}, {Lindstr{\o}m}, {Lister}, {Livanou}, {Lobel}, {Madrero Pardo},
		{Managau}, {Mann}, {Marchant}, {Marconi}, {Marcos Santos}, {Marinoni},
		{Marocco}, {Marshall}, {Martin Polo}, {Mart{\'\i}n-Fleitas}, {Masip},
		{Massari}, {Mastrobuono-Battisti}, {Mazeh}, {McMillan}, {Messina},
		{Michalik}, {Millar}, {Mints}, {Molina}, {Molinaro}, {Moln{\'a}r},
		{Montegriffo}, {Mor}, {Morbidelli}, {Morel}, {Morris}, {Mulone}, {Munoz},
		{Muraveva}, {Murphy}, {Musella}, {Noval}, {Ord{\'e}novic}, {Orr{\`u}},
		{Osinde}, {Pagani}, {Pagano}, {Palaversa}, {Palicio}, {Panahi}, {Pawlak},
		{Pe{\~n}alosa Esteller}, {Penttil{\"a}}, {Piersimoni}, {Pineau}, {Plachy},
		{Plum}, {Poggio}, {Poretti}, {Poujoulet}, {Pr{\v{s}}a}, {Pulone}, {Racero},
		{Ragaini}, {Rainer}, {Raiteri}, {Rambaux}, {Ramos}, {Ramos-Lerate}, {Re
			Fiorentin}, {Regibo}, {Reyl{\'e}}, {Ripepi}, {Riva}, {Rixon}, {Robichon},
		{Robin}, {Roelens}, {Rohrbasser}, {Romero-G{\'o}mez}, {Rowell}, {Royer},
		{Rybicki}, {Sadowski}, {Sagrist{\`a} Sell{\'e}s}, {Sahlmann}, {Salgado},
		{Salguero}, {Samaras}, {Sanchez Gimenez}, {Sanna}, {Santove{\~n}a},
		{Sarasso}, {Schultheis}, {Sciacca}, {Segol}, {Segovia}, {S{\'e}gransan},
		{Semeux}, {Shahaf}, {Siddiqui}, {Siebert}, {Siltala}, {Slezak}, {Smart},
		{Solano}, {Solitro}, {Souami}, {Souchay}, {Spagna}, {Spoto}, {Steele},
		{Steidelm{\"u}ller}, {Stephenson}, {S{\"u}veges}, {Szabados}, {Szegedi-Elek},
		{Taris}, {Tauran}, {Taylor}, {Teixeira}, {Thuillot}, {Tonello}, {Torra},
		{Torra}, {Turon}, {Unger}, {Vaillant}, {van Dillen}, {Vanel}, {Vecchiato},
		{Viala}, {Vicente}, {Voutsinas}, {Weiler}, {Wevers}, {Wyrzykowski}, {Yoldas},
		{Yvard}, {Zhao}, {Zorec}, {Zucker}, {Zurbach}, \& {Zwitter}}]{Brown2021}
	{Gaia Collaboration}, {Brown}, A.~G.~A., {Vallenari}, A., {et~al.} 2021, \aap,
	649, A1
	
	\bibitem[{{Gaia Collaboration} {et~al.}(2022){Gaia Collaboration}, {Vallenari},
		{Brown}, {Prusti}, {de Bruijne}, {Arenou}, {Babusiaux}, {Biermann},
		{Creevey}, {Ducourant}, {Evans}, {Eyer}, {Guerra}, {Hutton}, {Jordi},
		{Klioner}, {Lammers}, {Lindegren}, {Luri}, {Mignard}, {Panem}, {Pourbaix},
		{Randich}, {Sartoretti}, {Soubiran}, {Tanga}, {Walton}, {Bailer-Jones},
		{Bastian}, {Drimmel}, {Jansen}, {Katz}, {Lattanzi}, {van Leeuwen}, {Bakker},
		{Cacciari}, {Casta{\~n}eda}, {De Angeli}, {Fabricius}, {Fouesneau},
		{Fr{\'e}mat}, {Galluccio}, {Guerrier}, {Heiter}, {Masana}, {Messineo},
		{Mowlavi}, {Nicolas}, {Nienartowicz}, {Pailler}, {Panuzzo}, {Riclet}, {Roux},
		{Seabroke}, {Sordo{\o}rcit}, {Th{\'e}venin}, {Gracia-Abril}, {Portell},
		{Teyssier}, {Altmann}, {Andrae}, {Audard}, {Bellas-Velidis}, {Benson},
		{Berthier}, {Blomme}, {Burgess}, {Busonero}, {Busso}, {C{\'a}novas}, {Carry},
		{Cellino}, {Cheek}, {Clementini}, {Damerdji}, {Davidson}, {de Teodoro},
		{Nu{\~n}ez Campos}, {Delchambre}, {Dell'Oro}, {Esquej},
		{Fern{\'a}ndez-Hern{\'a}ndez}, {Fraile}, {Garabato}, {Garc{\'\i}a-Lario},
		{Gosset}, {Haigron}, {Halbwachs}, {Hambly}, {Harrison}, {Hern{\'a}ndez},
		{Hestroffer}, {Hodgkin}, {Holl}, {Jan{\ss}en}, {Jevardat de Fombelle},
		{Jordan}, {Krone-Martins}, {Lanzafame}, {L{\"o}ffler}, {Marchal}, {Marrese},
		{Moitinho}, {Muinonen}, {Osborne}, {Pancino}, {Pauwels}, {Recio-Blanco},
		{Reyl{\'e}}, {Riello}, {Rimoldini}, {Roegiers}, {Rybizki}, {Sarro}, {Siopis},
		{Smith}, {Sozzetti}, {Utrilla}, {van Leeuwen}, {Abbas}, {{\'A}brah{\'a}m},
		{Abreu Aramburu}, {Aerts}, {Aguado}, {Ajaj}, {Aldea-Montero}, {Altavilla},
		{{\'A}lvarez}, {Alves}, {Anders}, {Anderson}, {Anglada Varela}, {Antoja},
		{Baines}, {Baker}, {Balaguer-N{\'u}{\~n}ez}, {Balbinot}, {Balog}, {Barache},
		{Barbato}, {Barros}, {Barstow}, {Bartolom{\'e}}, {Bassilana}, {Bauchet},
		{Becciani}, {Bellazzini}, {Berihuete}, {Bernet}, {Bertone}, {Bianchi},
		{Binnenfeld}, {Blanco-Cuaresma}, {Blazere}, {Boch}, {Bombrun}, {Bossini},
		{Bouquillon}, {Bragaglia}, {Bramante}, {Breedt}, {Bressan}, {Brouillet},
		{Brugaletta}, {Bucciarelli}, {Burlacu}, {Butkevich}, {Buzzi}, {Caffau},
		{Cancelliere}, {Cantat-Gaudin}, {Carballo}, {Carlucci}, {Carnerero},
		{Carrasco}, {Casamiquela}, {Castellani}, {Castro-Ginard}, {Chaoul},
		{Charlot}, {Chemin}, {Chiaramida}, {Chiavassa}, {Chornay}, {Comoretto},
		{Contursi}, {Cooper}, {Cornez}, {Cowell}, {Crifo}, {Cropper}, {Crosta},
		{Crowley}, {Dafonte}, {Dapergolas}, {David}, {David}, {de Laverny}, {De
			Luise}, {De March}, {De Ridder}, {de Souza}, {de Torres}, {del Peloso}, {del
			Pozo}, {Delbo}, {Delgado}, {Delisle}, {Demouchy}, {Dharmawardena}, {Di
			Matteo}, {Diakite}, {Diener}, {Distefano}, {Dolding}, {Edvardsson}, {Enke},
		{Fabre}, {Fabrizio}, {Faigler}, {Fedorets}, {Fernique}, {Fienga}, {Figueras},
		{Fournier}, {Fouron}, {Fragkoudi}, {Gai}, {Garcia-Gutierrez},
		{Garcia-Reinaldos}, {Garc{\'\i}a-Torres}, {Garofalo}, {Gavel}, {Gavras},
		{Gerlach}, {Geyer}, {Giacobbe}, {Gilmore}, {Girona}, {Giuffrida}, {Gomel},
		{Gomez}, {Gonz{\'a}lez-N{\'u}{\~n}ez}, {Gonz{\'a}lez-Santamar{\'\i}a},
		{Gonz{\'a}lez-Vidal}, {Granvik}, {Guillout}, {Guiraud},
		{Guti{\'e}rrez-S{\'a}nchez}, {Guy}, {Hatzidimitriou}, {Hauser}, {Haywood},
		{Helmer}, {Helmi}, {Sarmiento}, {Hidalgo}, {Hilger}, {H{\l}adczuk}, {Hobbs},
		{Holland}, {Huckle}, {Jardine}, {Jasniewicz}, {Jean-Antoine Piccolo},
		{Jim{\'e}nez-Arranz}, {Jorissen}, {Juaristi Campillo}, {Julbe}, {Karbevska},
		{Kervella}, {Khanna}, {Kontizas}, {Kordopatis}, {Korn}, {K{\'o}sp{\'a}l},
		{Kostrzewa-Rutkowska}, {Kruszy{\'n}ska}, {Kun}, {Laizeau}, {Lambert},
		{Lanza}, {Lasne}, {Le Campion}, {Lebreton}, {Lebzelter}, {Leccia}, {Leclerc},
		{Lecoeur-Taibi}, {Liao}, {Licata}, {Lindstr{\o}m}, {Lister}, {Livanou},
		{Lobel}, {Lorca}, {Loup}, {Madrero Pardo}, {Magdaleno Romeo}, {Managau},
		{Mann}, {Manteiga}, {Marchant}, {Marconi}, {Marcos}, {Marcos Santos},
		{Mar{\'\i}n Pina}, {Marinoni}, {Marocco}, {Marshall}, {Polo},
		{Mart{\'\i}n-Fleitas}, {Marton}, {Mary}, {Masip}, {Massari},
		{Mastrobuono-Battisti}, {Mazeh}, {McMillan}, {Messina}, {Michalik}, {Millar},
		{Mints}, {Molina}, {Molinaro}, {Moln{\'a}r}, {Monari}, {Mongui{\'o}},
		{Montegriffo}, {Montero}, {Mor}, {Mora}, {Morbidelli}, {Morel}, {Morris},
		{Muraveva}, {Murphy}, {Musella}, {Nagy}, {Noval}, {Oca{\~n}a}, {Ogden},
		{Ordenovic}, {Osinde}, {Pagani}, {Pagano}, {Palaversa}, {Palicio},
		{Pallas-Quintela}, {Panahi}, {Payne-Wardenaar}, {Pe{\~n}alosa Esteller},
		{Penttil{\"a}}, {Pichon}, {Piersimoni}, {Pineau}, {Plachy}, {Plum}, {Poggio},
		{Pr{\v{s}}a}, {Pulone}, {Racero}, {Ragaini}, {Rainer}, {Raiteri}, {Rambaux},
		{Ramos}, {Ramos-Lerate}, {Re Fiorentin}, {Regibo}, {Richards}, {Rios Diaz},
		{Ripepi}, {Riva}, {Rix}, {Rixon}, {Robichon}, {Robin}, {Robin}, {Roelens},
		{Rogues}, {Rohrbasser}, {Romero-G{\'o}mez}, {Rowell}, {Royer}, {Ruz Mieres},
		{Rybicki}, {Sadowski}, {S{\'a}ez N{\'u}{\~n}ez}, {Sagrist{\`a} Sell{\'e}s},
		{Sahlmann}, {Salguero}, {Samaras}, {Sanchez Gimenez}, {Sanna},
		{Santove{\~n}a}, {Sarasso}, {Schultheis}, {Sciacca}, {Segol}, {Segovia},
		{S{\'e}gransan}, {Semeux}, {Shahaf}, {Siddiqui}, {Siebert}, {Siltala},
		{Silvelo}, {Slezak}, {Slezak}, {Smart}, {Snaith}, {Solano}, {Solitro},
		{Souami}, {Souchay}, {Spagna}, {Spina}, {Spoto}, {Steele},
		{Steidelm{\"u}ller}, {Stephenson}, {S{\"u}veges}, {Surdej}, {Szabados},
		{Szegedi-Elek}, {Taris}, {Taylo}, {Teixeira}, {Tolomei}, {Tonello}, {Torra},
		{Torra}, {Torralba Elipe}, {Trabucchi}, {Tsounis}, {Turon}, {Ulla}, {Unger},
		{Vaillant}, {van Dillen}, {van Reeven}, {Vanel}, {Vecchiato}, {Viala},
		{Vicente}, {Voutsinas}, {Weiler}, {Wevers}, {Wyrzykowski}, {Yoldas}, {Yvard},
		{Zhao}, {Zorec}, {Zucker}, \& {Zwitter}}]{GaiaDR3}
	{Gaia Collaboration}, {Vallenari}, A., {Brown}, A.~G.~A., {et~al.} 2022, arXiv
	e-prints, arXiv:2208.00211
	
	\bibitem[{{Garc{\'\i}a P{\'e}rez} {et~al.}(2016){Garc{\'\i}a P{\'e}rez},
		{Allende Prieto}, {Holtzman}, {Shetrone}, {M{\'e}sz{\'a}ros}, {Bizyaev},
		{Carrera}, {Cunha}, {Garc{\'\i}a-Hern{\'a}ndez}, {Johnson}, {Majewski},
		{Nidever}, {Schiavon}, {Shane}, {Smith}, {Sobeck}, {Troup}, {Zamora},
		{Weinberg}, {Bovy}, {Eisenstein}, {Feuillet}, {Frinchaboy}, {Hayden},
		{Hearty}, {Nguyen}, {O'Connell}, {Pinsonneault}, {Wilson}, \&
		{Zasowski}}]{Garcia2016}
	{Garc{\'\i}a P{\'e}rez}, A.~E., {Allende Prieto}, C., {Holtzman}, J.~A.,
	{et~al.} 2016, \aj, 151, 144
	
	\bibitem[{{GRAVITY Collaboration} {et~al.}(2019){GRAVITY Collaboration},
		{Abuter}, {Amorim}, {Baub{\"o}ck}, {Berger}, {Bonnet}, {Brandner},
		{Cl{\'e}net}, {Coud{\'e} Du Foresto}, {de Zeeuw}, {Dexter}, {Duvert},
		{Eckart}, {Eisenhauer}, {F{\"o}rster Schreiber}, {Garcia}, {Gao}, {Gendron},
		{Genzel}, {Gerhard}, {Gillessen}, {Habibi}, {Haubois}, {Henning}, {Hippler},
		{Horrobin}, {Jim{\'e}nez-Rosales}, {Jocou}, {Kervella}, {Lacour},
		{Lapeyr{\`e}re}, {Le Bouquin}, {L{\'e}na}, {Ott}, {Paumard}, {Perraut},
		{Perrin}, {Pfuhl}, {Rabien}, {Rodriguez Coira}, {Rousset}, {Scheithauer},
		{Sternberg}, {Straub}, {Straubmeier}, {Sturm}, {Tacconi}, {Vincent}, {von
			Fellenberg}, {Waisberg}, {Widmann}, {Wieprecht}, {Wiezorrek}, {Woillez}, \&
		{Yazici}}]{Gravity2019}
	{GRAVITY Collaboration}, {Abuter}, R., {Amorim}, A., {et~al.} 2019, \aap, 625,
	L10
	
	\bibitem[{{Gunn} {et~al.}(2006){Gunn}, {Siegmund}, {Mannery}, {Owen}, {Hull},
		{Leger}, {Carey}, {Knapp}, {York}, {Boroski}, {Kent}, {Lupton}, {Rockosi},
		{Evans}, {Waddell}, {Anderson}, {Annis}, {Barentine}, {Bartoszek}, {Bastian},
		{Bracker}, {Brewington}, {Briegel}, {Brinkmann}, {Brown}, {Carr},
		{Czarapata}, {Drennan}, {Dombeck}, {Federwitz}, {Gillespie}, {Gonzales},
		{Hansen}, {Harvanek}, {Hayes}, {Jordan}, {Kinney}, {Klaene}, {Kleinman},
		{Kron}, {Kresinski}, {Lee}, {Limmongkol}, {Lindenmeyer}, {Long}, {Loomis},
		{McGehee}, {Mantsch}, {Neilsen}, {Neswold}, {Newman}, {Nitta}, {Peoples},
		{Pier}, {Prieto}, {Prosapio}, {Rivetta}, {Schneider}, {Snedden}, \&
		{Wang}}]{Gunn2006}
	{Gunn}, J.~E., {Siegmund}, W.~A., {Mannery}, E.~J., {et~al.} 2006, \aj, 131,
	2332
	
	\bibitem[{{Hasselquist} {et~al.}(2021){Hasselquist}, {Hayes}, {Lian},
		{Weinberg}, {Zasowski}, {Horta}, {Beaton}, {Feuillet}, {Garro}, {Gallart},
		{Smith}, {Holtzman}, {Minniti}, {Lacerna}, {Shetrone}, {J{\"o}nsson},
		{Cioni}, {Fillingham}, {Cunha}, {O'Connell}, {Fern{\'a}ndez-Trincado},
		{Mu{\~n}oz}, {Schiavon}, {Almeida}, {Anguiano}, {Beers}, {Bizyaev},
		{Brownstein}, {Cohen}, {Frinchaboy}, {Garc{\'\i}a-Hern{\'a}ndez}, {Geisler},
		{Lane}, {Majewski}, {Nidever}, {Nitschelm}, {Povick}, {Price-Whelan},
		{Roman-Lopes}, {Rosado}, {Sobeck}, {Stringfellow}, {Valenzuela}, {Villanova},
		\& {Vincenzo}}]{2021ApJ...923..172H}
	{Hasselquist}, S., {Hayes}, C.~R., {Lian}, J., {et~al.} 2021, \apj, 923, 172
	
	\bibitem[{{Hasselquist} {et~al.}(2016){Hasselquist}, {Shetrone}, {Cunha},
		{Smith}, {Holtzman}, {Lawler}, {Allende Prieto}, {Beers}, {Chojnowski},
		{Fern{\'a}ndez-Trincado}, {Garc{\'\i}a-Hern{\'a}ndez}, {Hearty}, {Majewski},
		{Pereira}, {Placco}, {Villanova}, \& {Zamora}}]{Hasselquist2016}
	{Hasselquist}, S., {Shetrone}, M., {Cunha}, K., {et~al.} 2016, \apj, 833, 81
	
	\bibitem[{{Hawkins} {et~al.}(2015){Hawkins}, {Jofr{\'e}}, {Masseron}, \&
		{Gilmore}}]{2015MNRAS.453..758H}
	{Hawkins}, K., {Jofr{\'e}}, P., {Masseron}, T., \& {Gilmore}, G. 2015, \mnras,
	453, 758
	
	\bibitem[{{Hayes} {et~al.}(2018){Hayes}, {Majewski}, {Shetrone},
		{Fern{\'a}ndez-Alvar}, {Allende Prieto}, {Schuster}, {Carigi}, {Cunha},
		{Smith}, {Sobeck}, {Almeida}, {Beers}, {Carrera}, {Fern{\'a}ndez-Trincado},
		{Garc{\'\i}a-Hern{\'a}ndez}, {Geisler}, {Lane}, {Lucatello}, {Matthews},
		{Minniti}, {Nitschelm}, {Tang}, {Tissera}, \& {Zamora}}]{2018ApJ...852...49H}
	{Hayes}, C.~R., {Majewski}, S.~R., {Shetrone}, M., {et~al.} 2018, \apj, 852, 49
	
	\bibitem[{{Haywood} {et~al.}(2018){Haywood}, {Di Matteo}, {Lehnert}, {Snaith},
		{Khoperskov}, \& {G{\'o}mez}}]{2018ApJ...863..113H}
	{Haywood}, M., {Di Matteo}, P., {Lehnert}, M.~D., {et~al.} 2018, \apj, 863, 113
	
	\bibitem[{{Helmi} {et~al.}(2018{\natexlab{a}}){Helmi}, {Babusiaux},
		{Koppelman}, {Massari}, {Veljanoski}, \& {Brown}}]{2018Natur.563...85H}
	{Helmi}, A., {Babusiaux}, C., {Koppelman}, H.~H., {et~al.} 2018{\natexlab{a}},
	\nat, 563, 85
	
	\bibitem[{{Helmi} {et~al.}(2018{\natexlab{b}}){Helmi}, {Babusiaux},
		{Koppelman}, {Massari}, {Veljanoski}, \& {Brown}}]{Helmi2018Natur}
	{Helmi}, A., {Babusiaux}, C., {Koppelman}, H.~H., {et~al.} 2018{\natexlab{b}},
	\nat, 563, 85
	
	\bibitem[{{Helmi} {et~al.}(1999){Helmi}, {White}, {de Zeeuw}, \&
		{Zhao}}]{Helmi1999Natur}
	{Helmi}, A., {White}, S. D.~M., {de Zeeuw}, P.~T., \& {Zhao}, H. 1999, \nat,
	402, 53
	
	\bibitem[{Hinton \& Roweis(2003)}]{tSNE2003}
	Hinton, G. \& Roweis, S. 2003, Advances in Neural Information Processing
	Systems, 15, 857
	
	\bibitem[{{Holtzman} {et~al.}(2018){Holtzman}, {Hasselquist}, {Shetrone},
		{Cunha}, {Allende Prieto}, {Anguiano}, {Bizyaev}, {Bovy}, {Casey},
		{Edvardsson}, {Johnson}, {J{\"o}nsson}, {Meszaros}, {Smith}, {Sobeck},
		{Zamora}, {Chojnowski}, {Fernandez-Trincado}, {Garcia-Hernandez}, {Majewski},
		{Pinsonneault}, {Souto}, {Stringfellow}, {Tayar}, {Troup}, \&
		{Zasowski}}]{Holtzman2018}
	{Holtzman}, J.~A., {Hasselquist}, S., {Shetrone}, M., {et~al.} 2018, \aj, 156,
	125
	
	\bibitem[{{Holtzman} {et~al.}(2015){Holtzman}, {Shetrone}, {Johnson}, {Allende
			Prieto}, {Anders}, {Andrews}, {Beers}, {Bizyaev}, {Blanton}, {Bovy},
		{Carrera}, {Chojnowski}, {Cunha}, {Eisenstein}, {Feuillet}, {Frinchaboy},
		{Galbraith-Frew}, {Garc{\'\i}a P{\'e}rez}, {Garc{\'\i}a-Hern{\'a}ndez},
		{Hasselquist}, {Hayden}, {Hearty}, {Ivans}, {Majewski}, {Martell},
		{Meszaros}, {Muna}, {Nidever}, {Nguyen}, {O'Connell}, {Pan}, {Pinsonneault},
		{Robin}, {Schiavon}, {Shane}, {Sobeck}, {Smith}, {Troup}, {Weinberg},
		{Wilson}, {Wood-Vasey}, {Zamora}, \& {Zasowski}}]{Holtzman2015}
	{Holtzman}, J.~A., {Shetrone}, M., {Johnson}, J.~A., {et~al.} 2015, \aj, 150,
	148
	
	\bibitem[{{Horta} {et~al.}(2021){Horta}, {Schiavon}, {Mackereth}, {Pfeffer},
		{Mason}, {Kisku}, {Fragkoudi}, {Allende Prieto}, {Cunha}, {Hasselquist},
		{Holtzman}, {Majewski}, {Nataf}, {O'Connell}, {Schultheis}, \&
		{Smith}}]{2021MNRAS.500.1385H}
	{Horta}, D., {Schiavon}, R.~P., {Mackereth}, J.~T., {et~al.} 2021, \mnras, 500,
	1385
	
	\bibitem[{{Horta} {et~al.}(2023){Horta}, {Schiavon}, {Mackereth}, {Weinberg},
		{Hasselquist}, {Feuillet}, {O'Connell}, {Anguiano}, {Allende-Prieto},
		{Beaton}, {Bizyaev}, {Cunha}, {Geisler}, {Garc{\'\i}a-Hern{\'a}ndez},
		{Holtzman}, {J{\"o}nsson}, {Lane}, {Majewski}, {M{\'e}sz{\'a}ros}, {Minniti},
		{Nitschelm}, {Shetrone}, {Smith}, \& {Zasowski}}]{2023MNRAS.520.5671H}
	{Horta}, D., {Schiavon}, R.~P., {Mackereth}, J.~T., {et~al.} 2023, \mnras, 520,
	5671
	
	\bibitem[{{J{\"o}nsson} {et~al.}(2018){J{\"o}nsson}, {Allende Prieto},
		{Holtzman}, {Feuillet}, {Hawkins}, {Cunha}, {M{\'e}sz{\'a}ros},
		{Hasselquist}, {Fern{\'a}ndez-Trincado}, {Garc{\'\i}a-Hern{\'a}ndez},
		{Bizyaev}, {Carrera}, {Majewski}, {Pinsonneault}, {Shetrone}, {Smith},
		{Sobeck}, {Souto}, {Stringfellow}, {Teske}, \& {Zamora}}]{Henrik2018}
	{J{\"o}nsson}, H., {Allende Prieto}, C., {Holtzman}, J.~A., {et~al.} 2018, \aj,
	156, 126
	
	\bibitem[{{J{\"o}nsson} {et~al.}(2020){J{\"o}nsson}, {Holtzman}, {Allende
			Prieto}, {Cunha}, {Garc{\'\i}a-Hern{\'a}ndez}, {Hasselquist}, {Masseron},
		{Osorio}, {Shetrone}, {Smith}, {Stringfellow}, {Bizyaev}, {Edvardsson},
		{Majewski}, {M{\'e}sz{\'a}ros}, {Souto}, {Zamora}, {Beaton}, {Bovy}, {Donor},
		{Pinsonneault}, {Poovelil}, \& {Sobeck}}]{Henrik2020}
	{J{\"o}nsson}, H., {Holtzman}, J.~A., {Allende Prieto}, C., {et~al.} 2020, \aj,
	160, 120
	
	\bibitem[{{Juri{\'c}} {et~al.}(2008){Juri{\'c}}, {Ivezi{\'c}}, {Brooks},
		{Lupton}, {Schlegel}, {Finkbeiner}, {Padmanabhan}, {Bond}, {Sesar},
		{Rockosi}, {Knapp}, {Gunn}, {Sumi}, {Schneider}, {Barentine}, {Brewington},
		{Brinkmann}, {Fukugita}, {Harvanek}, {Kleinman}, {Krzesinski}, {Long},
		{Neilsen}, {Nitta}, {Snedden}, \& {York}}]{Juric2008}
	{Juri{\'c}}, M., {Ivezi{\'c}}, {\v{Z}}., {Brooks}, A., {et~al.} 2008, \apj,
	673, 864
	
	\bibitem[{{Koppelman} {et~al.}(2018){Koppelman}, {Helmi}, \&
		{Veljanoski}}]{2018ApJ...860L..11K}
	{Koppelman}, H., {Helmi}, A., \& {Veljanoski}, J. 2018, \apjl, 860, L11
	
	\bibitem[{{Koppelman} {et~al.}(2019{\natexlab{a}}){Koppelman}, {Helmi},
		{Massari}, {Price-Whelan}, \& {Starkenburg}}]{2019A&A...631L...9K}
	{Koppelman}, H.~H., {Helmi}, A., {Massari}, D., {Price-Whelan}, A.~M., \&
	{Starkenburg}, T.~K. 2019{\natexlab{a}}, \aap, 631, L9
	
	\bibitem[{{Koppelman} {et~al.}(2019{\natexlab{b}}){Koppelman}, {Helmi},
		{Massari}, {Roelenga}, \& {Bastian}}]{2019A&A...625A...5K}
	{Koppelman}, H.~H., {Helmi}, A., {Massari}, D., {Roelenga}, S., \& {Bastian},
	U. 2019{\natexlab{b}}, \aap, 625, A5
	
	\bibitem[{{Kruijssen} {et~al.}(2020){Kruijssen}, {Pfeffer}, {Chevance},
		{Bonaca}, {Trujillo-Gomez}, {Bastian}, {Reina-Campos}, {Crain}, \&
		{Hughes}}]{Kruijssen2020}
	{Kruijssen}, J.~M.~D., {Pfeffer}, J.~L., {Chevance}, M., {et~al.} 2020, \mnras,
	498, 2472
	
	\bibitem[{{Lindegren} {et~al.}(2018){Lindegren}, {Hern{\'a}ndez}, {Bombrun},
		{Klioner}, {Bastian}, {Ramos-Lerate}, {de Torres}, {Steidelm{\"u}ller},
		{Stephenson}, {Hobbs}, {Lammers}, {Biermann}, {Geyer}, {Hilger}, {Michalik},
		{Stampa}, {McMillan}, {Casta{\~n}eda}, {Clotet}, {Comoretto}, {Davidson},
		{Fabricius}, {Gracia}, {Hambly}, {Hutton}, {Mora}, {Portell}, {van Leeuwen},
		{Abbas}, {Abreu}, {Altmann}, {Andrei}, {Anglada}, {Balaguer-N{\'u}{\~n}ez},
		{Barache}, {Becciani}, {Bertone}, {Bianchi}, {Bouquillon}, {Bourda},
		{Br{\"u}semeister}, {Bucciarelli}, {Busonero}, {Buzzi}, {Cancelliere},
		{Carlucci}, {Charlot}, {Cheek}, {Crosta}, {Crowley}, {de Bruijne}, {de
			Felice}, {Drimmel}, {Esquej}, {Fienga}, {Fraile}, {Gai}, {Garralda},
		{Gonz{\'a}lez-Vidal}, {Guerra}, {Hauser}, {Hofmann}, {Holl}, {Jordan},
		{Lattanzi}, {Lenhardt}, {Liao}, {Licata}, {Lister}, {L{\"o}ffler},
		{Marchant}, {Martin-Fleitas}, {Messineo}, {Mignard}, {Morbidelli}, {Poggio},
		{Riva}, {Rowell}, {Salguero}, {Sarasso}, {Sciacca}, {Siddiqui}, {Smart},
		{Spagna}, {Steele}, {Taris}, {Torra}, {van Elteren}, {van Reeven}, \&
		{Vecchiato}}]{Lindegren2018}
	{Lindegren}, L., {Hern{\'a}ndez}, J., {Bombrun}, A., {et~al.} 2018, \aap, 616,
	A2
	
	\bibitem[{{Linderman} \& {Steinerberger}(2017)}]{Linderman2017EfficientAF}
	{Linderman}, G.~C. \& {Steinerberger}, S. 2017, arXiv e-prints,
	arXiv:1706.02582
	
	\bibitem[{{Mackereth} \& {Bovy}(2020)}]{Mackereth&Bovy2020}
	{Mackereth}, J.~T. \& {Bovy}, J. 2020, \mnras, 492, 3631
	
	\bibitem[{{Mackereth} {et~al.}(2019){Mackereth}, {Schiavon}, {Pfeffer},
		{Hayes}, {Bovy}, {Anguiano}, {Allende Prieto}, {Hasselquist}, {Holtzman},
		{Johnson}, {Majewski}, {O'Connell}, {Shetrone}, {Tissera}, \&
		{Fern{\'a}ndez-Trincado}}]{Mackereth}
	{Mackereth}, J.~T., {Schiavon}, R.~P., {Pfeffer}, J., {et~al.} 2019, \mnras,
	482, 3426
	
	\bibitem[{{Majewski} {et~al.}(2017){Majewski}, {Schiavon}, {Frinchaboy},
		{Allende Prieto}, {Barkhouser}, {Bizyaev}, {Blank}, {Brunner}, {Burton},
		{Carrera}, {Chojnowski}, {Cunha}, {Epstein}, {Fitzgerald}, {Garc{\'\i}a
			P{\'e}rez}, {Hearty}, {Henderson}, {Holtzman}, {Johnson}, {Lam}, {Lawler},
		{Maseman}, {M{\'e}sz{\'a}ros}, {Nelson}, {Nguyen}, {Nidever}, {Pinsonneault},
		{Shetrone}, {Smee}, {Smith}, {Stolberg}, {Skrutskie}, {Walker}, {Wilson},
		{Zasowski}, {Anders}, {Basu}, {Beland}, {Blanton}, {Bovy}, {Brownstein},
		{Carlberg}, {Chaplin}, {Chiappini}, {Eisenstein}, {Elsworth}, {Feuillet},
		{Fleming}, {Galbraith-Frew}, {Garc{\'\i}a}, {Garc{\'\i}a-Hern{\'a}ndez},
		{Gillespie}, {Girardi}, {Gunn}, {Hasselquist}, {Hayden}, {Hekker}, {Ivans},
		{Kinemuchi}, {Klaene}, {Mahadevan}, {Mathur}, {Mosser}, {Muna}, {Munn},
		{Nichol}, {O'Connell}, {Parejko}, {Robin}, {Rocha-Pinto}, {Schultheis},
		{Serenelli}, {Shane}, {Silva Aguirre}, {Sobeck}, {Thompson}, {Troup},
		{Weinberg}, \& {Zamora}}]{Majewski2017}
	{Majewski}, S.~R., {Schiavon}, R.~P., {Frinchaboy}, P.~M., {et~al.} 2017, \aj,
	154, 94
	
	\bibitem[{{M{\'e}sz{\'a}ros} {et~al.}(2020){M{\'e}sz{\'a}ros}, {Masseron},
		{Garc{\'\i}a-Hern{\'a}ndez}, {Allende Prieto}, {Beers}, {Bizyaev},
		{Chojnowski}, {Cohen}, {Cunha}, {Dell'Agli}, {Ebelke},
		{Fern{\'a}ndez-Trincado}, {Frinchaboy}, {Geisler}, {Hasselquist}, {Hearty},
		{Holtzman}, {Johnson}, {Lane}, {Lacerna}, {Longa-Pe{\~n}a}, {Majewski},
		{Martell}, {Minniti}, {Nataf}, {Nidever}, {Pan}, {Schiavon}, {Shetrone},
		{Smith}, {Sobeck}, {Stringfellow}, {Szigeti}, {Tang}, {Wilson}, \&
		{Zamora}}]{Szalbolcs2020}
	{M{\'e}sz{\'a}ros}, S., {Masseron}, T., {Garc{\'\i}a-Hern{\'a}ndez}, D.~A.,
	{et~al.} 2020, \mnras, 492, 1641
	
	\bibitem[{{Montalb{\'a}n} {et~al.}(2021){Montalb{\'a}n}, {Mackereth}, {Miglio},
		{Vincenzo}, {Chiappini}, {Buldgen}, {Mosser}, {Noels}, {Scuflaire}, {Vrard},
		{Willett}, {Davies}, {Hall}, {Nielsen}, {Khan}, {Rendle}, {van Rossem},
		{Ferguson}, \& {Chaplin}}]{2021NatAs...5..640M}
	{Montalb{\'a}n}, J., {Mackereth}, J.~T., {Miglio}, A., {et~al.} 2021, Nature
	Astronomy, 5, 640
	
	\bibitem[{{Moreno} {et~al.}(2015){Moreno}, {Pichardo}, \&
		{Schuster}}]{Moreno2015}
	{Moreno}, E., {Pichardo}, B., \& {Schuster}, W.~J. 2015, \mnras, 451, 705
	
	\bibitem[{{Myeong} {et~al.}(2022){Myeong}, {Belokurov}, {Aguado}, {Evans},
		{Caldwell}, \& {Bradley}}]{2022ApJ...938...21M}
	{Myeong}, G.~C., {Belokurov}, V., {Aguado}, D.~S., {et~al.} 2022, \apj, 938, 21
	
	\bibitem[{{Myeong} {et~al.}(2018){Myeong}, {Evans}, {Belokurov}, {Sanders}, \&
		{Koposov}}]{2018ApJ...863L..28M}
	{Myeong}, G.~C., {Evans}, N.~W., {Belokurov}, V., {Sanders}, J.~L., \&
	{Koposov}, S.~E. 2018, \apjl, 863, L28
	
	\bibitem[{{Naidu} {et~al.}(2020){Naidu}, {Conroy}, {Bonaca}, {Johnson}, {Ting},
		{Caldwell}, {Zaritsky}, \& {Cargile}}]{Naidu2020}
	{Naidu}, R.~P., {Conroy}, C., {Bonaca}, A., {et~al.} 2020, \apj, 901, 48
	
	\bibitem[{{Naidu} {et~al.}(2022){Naidu}, {Ji}, {Conroy}, {Bonaca}, {Ting},
		{Zaritsky}, {van Son}, {Broekgaarden}, {Tacchella}, {Chandra}, {Caldwell},
		{Cargile}, \& {Speagle}}]{2022ApJ...926L..36N}
	{Naidu}, R.~P., {Ji}, A.~P., {Conroy}, C., {et~al.} 2022, \apjl, 926, L36
	
	\bibitem[{{Nidever} {et~al.}(2015){Nidever}, {Holtzman}, {Allende Prieto},
		{Beland}, {Bender}, {Bizyaev}, {Burton}, {Desphande}, {Fleming}, {Garc{\'\i}a
			P{\'e}rez}, {Hearty}, {Majewski}, {M{\'e}sz{\'a}ros}, {Muna}, {Nguyen},
		{Schiavon}, {Shetrone}, {Skrutskie}, {Sobeck}, \& {Wilson}}]{Nidever2015}
	{Nidever}, D.~L., {Holtzman}, J.~A., {Allende Prieto}, C., {et~al.} 2015, \aj,
	150, 173
	
	\bibitem[{{Nissen} \& {Schuster}(2010)}]{2010A&A...511L..10N}
	{Nissen}, P.~E. \& {Schuster}, W.~J. 2010, \aap, 511, L10
	
	\bibitem[{Pedregosa {et~al.}(2011)Pedregosa, Varoquaux, Gramfort, Michel,
		Thirion, Grisel, Blondel, Prettenhofer, Weiss, Dubourg,
		{et~al.}}]{pedregosa2011scikit}
	Pedregosa, F., Varoquaux, G., Gramfort, A., {et~al.} 2011, Journal of machine
	learning research, 12, 2825
	
	\bibitem[{{Queiroz} {et~al.}(2023){Queiroz}, {Anders}, {Chiappini},
		{Khalatyan}, {Santiago}, {Nepal}, {Steinmetz}, {Gallart}, {Valentini}, {Dal
			Ponte}, {Barbuy}, {P{\'e}rez-Villegas}, {Masseron}, {Fern{\'a}ndez-Trincado},
		{Khoperskov}, {Minchev}, {Fern{\'a}ndez-Alvar}, {Lane}, \&
		{Nitschelm}}]{Queiroz2023}
	{Queiroz}, A.~B.~A., {Anders}, F., {Chiappini}, C., {et~al.} 2023, \aap, 673,
	A155
	
	\bibitem[{{Queiroz} {et~al.}(2020){Queiroz}, {Anders}, {Chiappini},
		{Khalatyan}, {Santiago}, {Steinmetz}, {Valentini}, {Miglio}, {Bossini},
		{Barbuy}, {Minchev}, {Minniti}, {Garc{\'\i}a Hern{\'a}ndez}, {Schultheis},
		{Beaton}, {Beers}, {Bizyaev}, {Brownstein}, {Cunha},
		{Fern{\'a}ndez-Trincado}, {Frinchaboy}, {Lane}, {Majewski}, {Nataf},
		{Nitschelm}, {Pan}, {Roman-Lopes}, {Sobeck}, {Stringfellow}, \&
		{Zamora}}]{Queiroz2020}
	{Queiroz}, A.~B.~A., {Anders}, F., {Chiappini}, C., {et~al.} 2020, \aap, 638,
	A76
	
	\bibitem[{{Queiroz} {et~al.}(2018){Queiroz}, {Anders}, {Santiago}, {Chiappini},
		{Steinmetz}, {Dal Ponte}, {Stassun}, {da Costa}, {Maia}, {Crestani}, {Beers},
		{Fern{\'a}ndez-Trincado}, {Garc{\'\i}a-Hern{\'a}ndez}, {Roman-Lopes}, \&
		{Zamora}}]{Queiroz2018}
	{Queiroz}, A.~B.~A., {Anders}, F., {Santiago}, B.~X., {et~al.} 2018, \mnras,
	476, 2556
	
	\bibitem[{Rebonato \& J{\"a}ckel(2011)}]{rebonato2011most}
	Rebonato, R. \& J{\"a}ckel, P. 2011, Available at SSRN 1969689
	
	\bibitem[{{Reid} \& {Brunthaler}(2020)}]{Reid2020}
	{Reid}, M.~J. \& {Brunthaler}, A. 2020, \apj, 892, 39
	
	\bibitem[{{Rim} {et~al.}(2022){Rim}, {Steinhardt}, {Clark}, {Diaconu},
		{Rusakov}, \& {Sneppen}}]{Rim2022}
	{Rim}, P., {Steinhardt}, C., {Clark}, T., {et~al.} 2022, in American
	Astronomical Society Meeting Abstracts, Vol.~54, American Astronomical
	Society Meeting Abstracts, 241.38
	
	\bibitem[{{Robin} {et~al.}(2012){Robin}, {Marshall}, {Schultheis}, \&
		{Reyl{\'e}}}]{Robin2012}
	{Robin}, A.~C., {Marshall}, D.~J., {Schultheis}, M., \& {Reyl{\'e}}, C. 2012,
	\aap, 538, A106
	
	\bibitem[{{Robin} {et~al.}(2003){Robin}, {Reyl{\'e}}, {Derri{\`e}re}, \&
		{Picaud}}]{Robin2003}
	{Robin}, A.~C., {Reyl{\'e}}, C., {Derri{\`e}re}, S., \& {Picaud}, S. 2003,
	\aap, 409, 523
	
	\bibitem[{{Robin} {et~al.}(2014){Robin}, {Reyl{\'e}}, {Fliri}, {Czekaj},
		{Robert}, \& {Martins}}]{Robin2014}
	{Robin}, A.~C., {Reyl{\'e}}, C., {Fliri}, J., {et~al.} 2014, \aap, 569, A13
	
	\bibitem[{{Sanders} {et~al.}(2019){Sanders}, {Smith}, \& {Evans}}]{Sanders2019}
	{Sanders}, J.~L., {Smith}, L., \& {Evans}, N.~W. 2019, \mnras, 488, 4552
	
	\bibitem[{{Santana} {et~al.}(2021){Santana}, {Beaton}, {Covey}, {Saito},
		{Hempel}, {Pietrukowicz}, {Ahumada}, {Alonso}, {Alonso-Garcia}, {Arias},
		{Bandyopadhyay}, {Barb{\'a}}, {Barbuy}, {Bedin}, {Bica}, {Borissova},
		{Bronfman}, {Carraro}, {Catelan}, {Clari{\'a}}, {Cross}, {de Grijs},
		{D{\'e}k{\'a}ny}, {Drew}, {Fari{\~n}a}, {Feinstein}, {Fern{\'a}ndez
			Laj{\'u}s}, {Gamen}, {Geisler}, {Gieren}, {Goldman}, {Gonzalez}, {Gunthardt},
		{Gurovich}, {Hambly}, {Irwin}, {Ivanov}, {Jord{\'a}n}, {Kerins}, {Kinemuchi},
		{Kurtev}, {L{\'o}pez-Corredoira}, {Maccarone}, {Masetti}, {Merlo},
		{Messineo}, {Mirabel}, {Monaco}, {Morelli}, {Padilla}, {Palma}, {Parisi},
		{Pignata}, {Rejkuba}, {Roman-Lopes}, {Sale}, {Schreiber}, {Schr{\"o}der},
		{Smith}, {}, {Soto}, {Tamura}, {Tappert}, {Thompson}, {Toledo}, {Zoccali}, \&
		{Pietrzynski}}]{Santana2021}
	{Santana}, F.~A., {Beaton}, R.~L., {Covey}, K., {et~al.} 2021, \apj, in
	preparation
	
	\bibitem[{{Santiago} {et~al.}(2016){Santiago}, {Brauer}, {Anders}, {Chiappini},
		{Queiroz}, {Girardi}, {Rocha-Pinto}, {Balbinot}, {da Costa}, {Maia},
		{Schultheis}, {Steinmetz}, {Miglio}, {Montalb{\'a}n}, {Schneider}, {Beers},
		{Frinchaboy}, {Lee}, \& {Zasowski}}]{Santiago2016}
	{Santiago}, B.~X., {Brauer}, D.~E., {Anders}, F., {et~al.} 2016, \aap, 585, A42
	
	\bibitem[{{Schiavon} {et~al.}(2017){Schiavon}, {Zamora}, {Carrera},
		{Lucatello}, {Robin}, {Ness}, {Martell}, {Smith},
		{Garc{\'\i}a-Hern{\'a}ndez}, {Manchado}, {Sch{\"o}nrich}, {Bastian},
		{Chiappini}, {Shetrone}, {Mackereth}, {Williams}, {M{\'e}sz{\'a}ros},
		{Allende Prieto}, {Anders}, {Bizyaev}, {Beers}, {Chojnowski}, {Cunha},
		{Epstein}, {Frinchaboy}, {Garc{\'\i}a P{\'e}rez}, {Hearty}, {Holtzman},
		{Johnson}, {Kinemuchi}, {Majewski}, {Muna}, {Nidever}, {Nguyen}, {O'Connell},
		{Oravetz}, {Pan}, {Pinsonneault}, {Schneider}, {Schultheis}, {Simmons},
		{Skrutskie}, {Sobeck}, {Wilson}, \& {Zasowski}}]{Schiavon2017Nrich}
	{Schiavon}, R.~P., {Zamora}, O., {Carrera}, R., {et~al.} 2017, \mnras, 465, 501
	
	\bibitem[{{Shetrone} {et~al.}(2015){Shetrone}, {Bizyaev}, {Lawler}, {Allende
			Prieto}, {Johnson}, {Smith}, {Cunha}, {Holtzman}, {Garc{\'\i}a P{\'e}rez},
		{M{\'e}sz{\'a}ros}, {Sobeck}, {Zamora}, {Garc{\'\i}a-Hern{\'a}ndez}, {Souto},
		{Chojnowski}, {Koesterke}, {Majewski}, \& {Zasowski}}]{Shetrone2015}
	{Shetrone}, M., {Bizyaev}, D., {Lawler}, J.~E., {et~al.} 2015, \apjs, 221, 24
	
	\bibitem[{{Smith} {et~al.}(2021){Smith}, {Bizyaev}, {Cunha}, {Shetrone},
		{Souto}, {Allende Prieto}, {Masseron}, {M{\'e}sz{\'a}ros}, {J{\"o}nsson},
		{Hasselquist}, {Osorio}, {Garc{\'\i}a-Hern{\'a}ndez}, {Plez}, {Beaton},
		{Holtzman}, {Majewski}, {Stringfellow}, \& {Sobeck}}]{2021AJ....161..254S}
	{Smith}, V.~V., {Bizyaev}, D., {Cunha}, K., {et~al.} 2021, \aj, 161, 254
	
	\bibitem[{{Tang} {et~al.}(2018){Tang}, {Fern{\'a}ndez-Trincado}, {Geisler},
		{Zamora}, {M{\'e}sz{\'a}ros}, {Masseron}, {Cohen},
		{Garc{\'\i}a-Hern{\'a}ndez}, {Dell'Agli}, {Beers}, {Schiavon}, {Sohn},
		{Hasselquist}, {Robin}, {Shetrone}, {Majewski}, {Villanova}, {Schiappacasse
			Ulloa}, {Lane}, {Minnti}, {Roman-Lopes}, {Almeida}, \& {Moreno}}]{Tang2018}
	{Tang}, B., {Fern{\'a}ndez-Trincado}, J.~G., {Geisler}, D., {et~al.} 2018,
	\apj, 855, 38
	
	\bibitem[{{Ting} {et~al.}(2012){Ting}, {Freeman}, {Kobayashi}, {De Silva}, \&
		{Bland-Hawthorn}}]{2012MNRAS.421.1231T}
	{Ting}, Y.-S., {Freeman}, K.~C., {Kobayashi}, C., {De Silva}, G.~M., \&
	{Bland-Hawthorn}, J. 2012, \mnras, 421, 1231
	
	\bibitem[{van~der Maaten \& Hinton(2008)}]{ref_tsne}
	van~der Maaten, L. \& Hinton, G. 2008, Journal of Machine Learning Research, 9,
	2579
	
	\bibitem[{{Vasiliev} \& {Baumgardt}(2021)}]{2021MNRAS.505.5978V}
	{Vasiliev}, E. \& {Baumgardt}, H. 2021, \mnras, 505, 5978
	
	\bibitem[{{Vera} {et~al.}(2016){Vera}, {Alonso}, \&
		{Coldwell}}]{2016A&A...595A..63V}
	{Vera}, M., {Alonso}, S., \& {Coldwell}, G. 2016, \aap, 595, A63
	
	\bibitem[{{Verma} {et~al.}(2021){Verma}, {Matijevi{\v{c}}}, {Denker},
		{Diercke}, {Dineva}, {Balthasar}, {Kamlah}, {Kontogiannis}, {Kuckein}, \&
		{Pal}}]{Verma2021}
	{Verma}, M., {Matijevi{\v{c}}}, G., {Denker}, C., {et~al.} 2021, \apj, 907, 54
	
	\bibitem[{Wattenberg {et~al.}(2016)Wattenberg, Viégas, \&
		Johnson}]{wattenberg_how_2016}
	Wattenberg, M., Viégas, F., \& Johnson, I. 2016, Distill, 1,
	10.23915/distill.00002
	
	\bibitem[{{Wilson} {et~al.}(2019){Wilson}, {Hearty}, {Skrutskie}, {Majewski},
		{Holtzman}, {Eisenstein}, {Gunn}, {Blank}, {Henderson}, {Smee}, {Nelson},
		{Nidever}, {Arns}, {Barkhouser}, {Barr}, {Beland}, {Bershady}, {Blanton},
		{Brunner}, {Burton}, {Carey}, {Carr}, {Colque}, {Crane}, {Damke}, {Davidson},
		{Dean}, {Di Mille}, {Don}, {Ebelke}, {Evans}, {Fitzgerald}, {Gillespie},
		{Hall}, {Harding}, {Harding}, {Hammond}, {Hancock}, {Harrison}, {Hope},
		{Horne}, {Karakla}, {Lam}, {Leger}, {MacDonald}, {Maseman}, {Matsunari},
		{Melton}, {Mitcheltree}, {O'Brien}, {O'Connell}, {Patten}, {Richardson},
		{Rieke}, {Rieke}, {Roman-Lopes}, {Schiavon}, {Sobeck}, {Stolberg}, {Stoll},
		{Tembe}, {Trujillo}, {Uomoto}, {Vernieri}, {Walker}, {Weinberg}, {Young},
		{Anthony-Brumfield}, {Bizyaev}, {Breslauer}, {De Lee}, {Downey}, {Halverson},
		{Huehnerhoff}, {Klaene}, {Leon}, {Long}, {Mahadevan}, {Malanushenko},
		{Nguyen}, {Owen}, {S{\'a}nchez-Gallego}, {Sayres}, {Shane}, {Shectman},
		{Shetrone}, {Skinner}, {Stauffer}, \& {Zhao}}]{Wilson2019}
	{Wilson}, J.~C., {Hearty}, F.~R., {Skrutskie}, M.~F., {et~al.} 2019, \pasp,
	131, 055001
	
	\bibitem[{{Zamora} {et~al.}(2015){Zamora}, {Garc{\'\i}a-Hern{\'a}ndez},
		{Allende Prieto}, {Carrera}, {Koesterke}, {Edvardsson}, {Castelli}, {Plez},
		{Bizyaev}, {Cunha}, {Garc{\'\i}a P{\'e}rez}, {Gustafsson}, {Holtzman},
		{Lawler}, {Majewski}, {Manchado}, {M{\'e}sz{\'a}ros}, {Shane}, {Shetrone},
		{Smith}, \& {Zasowski}}]{Zamora2015}
	{Zamora}, O., {Garc{\'\i}a-Hern{\'a}ndez}, D.~A., {Allende Prieto}, C.,
	{et~al.} 2015, \aj, 149, 181
	
	\bibitem[{{Zasowski} {et~al.}(2017){Zasowski}, {Cohen}, {Chojnowski},
		{Santana}, {Oelkers}, {Andrews}, {Beaton}, {Bender}, {Bird}, {Bovy},
		{Carlberg}, {Covey}, {Cunha}, {Dell'Agli}, {Fleming}, {Frinchaboy},
		{Garc{\'\i}a-Hern{\'a}ndez}, {Harding}, {Holtzman}, {Johnson}, {Kollmeier},
		{Majewski}, {M{\'e}sz{\'a}ros}, {Munn}, {Mu{\~n}oz}, {Ness}, {Nidever},
		{Poleski}, {Rom{\'a}n-Z{\'u}{\~n}iga}, {Shetrone}, {Simon}, {Smith},
		{Sobeck}, {Stringfellow}, {Szigeti{\'a}ros}, {Tayar}, \&
		{Troup}}]{Zasowski2017}
	{Zasowski}, G., {Cohen}, R.~E., {Chojnowski}, S.~D., {et~al.} 2017, \aj, 154,
	198
	
	\bibitem[{{Zasowski} {et~al.}(2013){Zasowski}, {Johnson}, {Frinchaboy},
		{Majewski}, {Nidever}, {Rocha Pinto}, {Girardi}, {Andrews}, {Chojnowski},
		{Cudworth}, {Jackson}, {Munn}, {Skrutskie}, {Beaton}, {Blake}, {Covey},
		{Deshpande}, {Epstein}, {Fabbian}, {Fleming}, {Garcia Hernandez}, {Herrero},
		{Mahadevan}, {M{\'e}sz{\'a}ros}, {Schultheis}, {Sellgren}, {Terrien}, {van
			Saders}, {Allende Prieto}, {Bizyaev}, {Burton}, {Cunha}, {da Costa},
		{Hasselquist}, {Hearty}, {Holtzman}, {Garc{\'\i}a P{\'e}rez}, {Maia},
		{O'Connell}, {O'Donnell}, {Pinsonneault}, {Santiago}, {Schiavon}, {Shetrone},
		{Smith}, \& {Wilson}}]{Zasowski2013}
	{Zasowski}, G., {Johnson}, J.~A., {Frinchaboy}, P.~M., {et~al.} 2013, \aj, 146,
	81
	
\end{thebibliography}

\end{document}